\documentclass[]{aa} 

\usepackage[]{graphicx}
\usepackage{subfig}
\usepackage{xcolor}
\usepackage[varg]{txfonts}
\usepackage{lscape}
\usepackage{natbib}
\usepackage{rotating}
\usepackage{multirow}
\usepackage[perpage,para,symbol,hang,flushmargin]{footmisc}
\usepackage{hhline}
\usepackage{makecell}
\usepackage{tabularx}
\usepackage{booktabs}
\usepackage{caption}
\usepackage{threeparttable}
\usepackage{multirow}
\usepackage[abs]{overpic}
\usepackage{enumerate}
\usepackage[]{textpos}   
\usepackage{color}

\def\deg{\ifmmode^\circ\else$^\circ$\fi}
\def\alphaTF{\ifmmode{\alpha_{\mathrm{\,{\small TF}}}}\else{$\alpha_{\mathrm{\,{\small TF}}}$}\fi}

\def\Msun{\ifmmode{\mathrm M_\odot}\else{M$_\odot$}\fi}
\newcommand{\rbreak}{\ensuremath{R_\mathrm{break }}}
\newcommand{\mubreak}{\ensuremath{\mu_\mathrm{break}}}
\newcommand{\hi}{\ensuremath{h_{\mathrm{i}}}}
\newcommand{\mui}{\ensuremath{\mu_{0,\mathrm{i}}}}
\newcommand{\ho}{\ensuremath{h_{\mathrm{o}}}}
\newcommand{\muo}{\ensuremath{\mu_{0,\mathrm{o}}}}
\newcommand{\risoph}{\ensuremath{R_\mathrm{23}}}
\newcommand{\magarc}{mag arcsec$^{\mathrm{-2}}$}
\usepackage{txfonts}
\begin{document}

\title{Evolution of the anti-truncated stellar profiles of S0 galaxies\\ 
since $z=0.6$ in the SHARDS survey}
\subtitle{II - Structural and photometric evolution}
\titlerunning{Scaling relations of anti-truncated stellar profiles on S0 galaxies at $0.2<z<0.6$}
\authorrunning{Borlaff et al.}

\author{Alejandro Borlaff\inst{1,2,3}, M.~Carmen Eliche-Moral\inst{1,2}, John E. Beckman\inst{1,3,4}, Alexandre Vazdekis\inst{1,3}, \\ Alejandro Lumbreras-Calle\inst{1,3}, Bogdan C. Ciambur\inst{5}, Pablo G. P\'{e}rez-Gonz\'{a}lez\inst{2}, Nicol\'{a}s Cardiel\inst{2}, \\ Guillermo Barro\inst{6} and Antonio Cava\inst{7}} 

\institute{
Instituto de Astrof\'{i}sica de Canarias, C/ V\'{i}a L\'actea, E-38200 La Laguna, Tenerife, Spain
\\\email{asborlaff@iac.es}
\and
Departamento de Astrof\'{\i}sica y CC.~de la Atm\'osfera, Universidad Complutense de Madrid, E-28040 Madrid, Spain
\and
Facultad de F\'{i}sica, Universidad de La Laguna, Avda. Astrof\'{i}sico Fco. S\'{a}nchez s/n, 38200, La Laguna, Tenerife, Spain
\and
Consejo Superior de Investigaciones Cient\'{i}ficas, Spain
\and 
Observatoire de Paris, LERMA, PSL Research University, 61 Avenue de l’Observatoire, 75014, Paris, France
\and
University of California, 501 Campbell Hall, Berkeley, CA 94720 Santa Cruz, USA
\and
Observatoire de Gen{\`e}ve, Universit{\'e} de Gen{\`e}ve, 51 Ch. des Maillettes, 1290 Versoix, Switzerland
}
  \abstract
   {Anti-truncated lenticular galaxies (Type-III S0s) present tight scaling relations between their surface brightness photometric and structural parameters. Although several evolutionary models have been proposed for the formation of these structures, the observations of Type-III S0 galaxies are usually limited to the local Universe.}
   {We aim to compare the properties of Type-III discs in a sample of S0 galaxies at $0.2<z<0.6$ with those of the local Universe. In this paper, we study the evolution of the photometric and structural scaling relations measured in the rest-frame $R-$band with $z$ and the possible differences between the rest-frame $(B-R)$ colours of the inner and outer disc profiles.}
   {We make use of a sample of 14 Type-III E/S0--S0 galaxies at $0.2<z<0.6$ from the GOODS-N field identified and characterised in a previous paper. We study whether or not the correlations found in local Type-III S0 galaxies were present $\sim 6$ Gyr ago. We analyse the distribution of the surface brightness characteristic parameters (\rbreak, \mubreak, \hi, \ho, \mui\ and \muo) as a function of the stellar mass and look to see if there is a significant change with $z$. We also derive their rest-frame $(B-R)$ colour profiles. Finally, we compare these results with the predictions from a grid of SSP models.}
   {We find that the inner and outer scale-lengths of Type-III S0 galaxies at $0.4<z<0.6$ follow compatible trends and scaling relations with those observed in local S0 galaxies as a function of the break radius, \rbreak. We do not detect any significant differences between the location of \rbreak\ between $z\sim0.6$ and $z\sim0$ for a fixed stellar mass of the object, whereas the surface brightness at the break radius \mubreak\ is $\sim1.5$ \magarc\ dimmer in the local Universe than at $z\sim0.6$ for a fixed stellar mass. We find no significant differences in the $(B-R)$ colour between the inner and outer profiles of the Type-III S0 galaxies at $0.2<z<0.6$.}
   {In contrast to Type-II (down-bending) profiles, the anti-truncated surface brightness profiles of S0 galaxies present compatible \rbreak\ values and scaling relations during the last 6 Gyr. This result and the similarity of the colours of the inner and outer discs point to a highly scalable and stable formation process, probably more related to gravitational and dynamical processes than to the evolution of stellar populations.}

   \maketitle
%
%

\section{Introduction}
\label{Sec:Intro}
The mechanisms of formation of lenticular (or S0) galaxies are not yet well understood. Although S0 galaxies usually have a prominent disc component, they do not show any signs of star formation or spiral arms on it. They present an intermediate apparent morphology between spiral galaxies (which also show discs but with noticeable arms and star-formation regions) and elliptical galaxies (without any clear signs of a disc or spiral arms and with no star formation). Many authors have claimed that this intermediate position in the well known Tuning Fork diagram \citep{1926ApJ....64..321H} was supposed to imply an evolutionary path. Nevertheless, \citet{1927Obs....50..276H} warned that the "late/early-type" nomenclature referred to apparent complexity and did not imply any direction of evolution in the morphological classification system \citep[see][for a detailed discussion about this]{2008A&G....49e..25B}. 


Despite this, there are signs of evolution between the different morphological types. Recent studies such as \citet{2015FrASS...2....4D} suggest that spiral galaxies have been transforming into S0 galaxies for the last $\sim 7-8$ Gyr (since $z\sim1$). This is supported by a number of observational facts, such as that the fraction of spiral galaxies at redshift $z\sim 0.5$ was two to three times higher than in the local Universe, while there is an apparent decrease in the fraction of S0 galaxies \citep{1997ApJ...490..577D, 1999MNRAS.308..947A}. \citet{1972ApJ...176....1G} pointed out that this process may have been taking place during the fall of the spirals in the early Universe into the gravitational well of galaxy clusters through ram-pressure stripping. Another process associated to the clusters that could potentially transform spiral galaxies into S0s is galaxy harassment \citep{1996Natur.379..613M,1999MNRAS.304..465M,2015A&A...576A.103B}. In this scenario, high-velocity fly-bys would quench the star formation from spirals, strip stars into the intracluster medium and ultimately transform them into S0s. Galaxy encounters (i.e. mergers) were already suggested by \citet{1951ApJ...113..413S}, who considered that galaxy mergers might be responsible for the gas depletion of spiral galaxies, creating a parallel sequence of S0 galaxies. 

Although these environmental mechanisms can successfully transform spiral galaxies into S0s, almost $\sim 50$\% of S0 galaxies reside in groups and the field, not in clusters. \citep[see][]{1982ApJ...257..423H, 2006ApJS..167....1B,2007ApJ...655..790C, 2009ApJ...692..298W}. This implies that the mechanisms leading to the formation of S0s may depend on the environment density and the stellar mass \citep{2009MNRAS.394.1991B}. Many authors claim that in low-density environments, the most important processes for the formation of S0s are galaxy interactions and the loss of gas from the disc through secular processes. Therefore, secular evolution has been proposed as an important driver of galaxy formation, especially in low-mass S0 galaxies, while the formation of higher mass S0s may have been dominated by mergers \citep{2007ApJ...661L..37B,2009MNRAS.394.1991B}. This is consistent with the prevalence of pseudobulges in low-luminosity S0s against classical bulges in high-luminosity S0s \citep{2004ARA&A..42..603K, 2013ApJ...767L..33V}. Given that observational and theoretical studies suggest that massive E-S0 galaxies have undergone at least one major merger during the last $\sim 9$\,Gyr regardless of the environment \citep{2010A&A...519A..55E,2013MNRAS.428..999P}, galaxy interactions may have had an important role in low-density environments. It has been suggested that it is in galaxy groups, and not clusters, where the merger rate is higher \citep{2012ApJ...754...26J}. Contrary to popular belief, major mergers can produce well-defined disc remnants \citep[][]{2003ApJ...597..893N,2005A&A...437...69B,2016ApJ...821...90A, 2017A&A...600A..25R}. In fact, many authors find observational evidence of a major-merger origin of many S0s at $z<1$ \citep[see][]{2009A&A...496...51P,2009A&A...501..437Y,2009A&A...507.1313H,2009A&A...496..381H,2012MPLA...2730034H,2014A&A...565A..31T,2015A&A...573A..78Q,2015A&A...579L...2Q}. In summary, the existence of multiple evolutionary pathways for the formation of S0 galaxies \citep{2010MNRAS.405.1089L,2010MNRAS.407.1231R,2010ApJ...708..841W,2013MNRAS.432.1010C,2013MNRAS.432..430B}, has triggered a heated debate about their origin and evolution.  

One of the most common methods to study the origin and evolution of galaxies is the analysis of their surface brightness profiles. The discs of spiral and S0 galaxies show surface brightness profiles that are to first order approximation well-fitted by an exponential function out to a certain radius \citep{1940BHarO.914....9P, 1958ApJ...128..465D, 1970ApJ...160..811F}. This is due to the stellar density decline as a function of galactocentric radius \citep{1963BAAA....6...41S}. Nevertheless, many spiral and lenticular galaxies do not follow a purely exponential profile along their whole observable radius \citep{1979A&AS...38...15V,2000A&A...357L...1P, 2002PhDT..........P,2002MNRAS.334..646K,2009IAUS..254..173S,2012ApJS..198....2K,2012A&AT...27..313I}, although some others do until extremely faint magnitudes \citep[NGC300,][]{2005ApJ...629..239B}. \citet{2005ApJ...626L..81E} pointed out that a significant fraction of S0 galaxies present light excesses in the outskirts of the discs, which also show an exponential decline but with a shallower slope than their inner discs \citep[see also][]{2006A&A...454..759P}. Therefore, \citet{2006A&A...454..759P} and \citet[][E08 hereafter]{2008AJ....135...20E} designed a stellar disc classification of galaxies in three main types, according to the profile structure. Type-I discs are well modelled with a single exponential profile. Type-II galaxies present a down-bending profile, that is, the profile of the disc steepens sharply beyond a certain radius with respect to the extrapolated trend of the inner regions (truncation). Type-III discs become shallower outside the break radius than the extrapolation of the exponential trend of the inner parts \citep[anti-truncation, see][]{2006A&A...454..759P}. E08 and \citet[][G11 hereafter]{2011AJ....142..145G} showed that the fraction of anti-truncated discs was higher for S0 galaxies than for any other morphological type, increasing from approximately $10$ to $20\%$ in Sc--Sd galaxies and up to $\sim 20-50\%$ in S0--Sa galaxies \citep[see also][]{2012A&AT...27..313I,2015MNRAS.447.1506M}. In the present study we focus on S0s with Type-III profiles.

Many mechanisms have been proposed to explain the origin of Type-III profiles, which are sometimes connected to the process responsible for the formation of the S0 galaxy itself. Radially varying profiles of star formation have been proposed as a cause for both Type-II and Type-III profiles \citep{2006ApJ...636..712E}. In addition to this, some authors \citep[e.g.][]{2012ApJ...759...98C} pointed out that these transitions might be caused by combinations of thin+thick discs with different radial scalelengths. Radial mass redistribution has also been proposed as a cause for the different types of profiles \citep{2015MNRAS.448L..99H, 2017MNRAS.470.4941H, 2017A&A...608A.126R}. In a recent paper, \citet{2016ApJ...830..115E} found that in galaxy disc simulations, exponential profiles appear from random scattering when there is a slight inward bias in the scatter. They also found that radial variation or thresholds of this bias are able to produce double exponential profiles. Holes and clumps of interstellar gas have been proven to be valid gravitational scattering centres in simulations of dwarf galaxy discs and they were able to generate exponential discs from an initially flat profile in $\sim 1$ Gyr \citep{2017MNRAS.469.1157S}. However, most of the proposed mechanisms for the formation of Type-III galaxies are based on some kind of gravitational interaction, such as high-eccentricity fly-bys \citep{2001MNRAS.324..685L,2006ApJ...650L..33P}, minor mergers \citep{2007ApJ...670..269Y}, major mergers \citep{2014A&A...570A.103B}, bombardment of the disc by cold dark-matter halos \citep{2009ApJ...700.1896K} or galaxy harassment \citep{2012ApJ...758...41R}. \citet{2001MNRAS.324..685L} reported that N-body simulations of M51-like pairs present shallow outer profiles. The authors found out that this was a consequence of the stripping of gas and stars from the galaxies inner discs. 

A high fraction of the anti-truncated profiles in spiral galaxies are due entirely to disc structure, although $\sim 15\%$ might be due to outer stellar haloes \citep{2012MNRAS.419..669M}. In lenticular galaxies, the fraction of anti-truncated profiles caused by the presence of a spherical component (bulge contribution, haloes) may be as high as $\sim50\%$ \citep{2015MNRAS.447.1506M}, although it is important to note that this is only an upper limit to the real fraction due to the bulge profile used on their analysis. In some cases, transitions between the inner and outer profiles are associated with structural components of the galaxy such as rings or lenses \citep{2014MNRAS.441.1992L}.

\citet[][B14 hereafter]{2014A&A...570A.103B} tested whether major mergers can produce anti-truncated stellar profiles in S0 galaxies using hydrodynamical N-body simulations. They found that Type-III profiles of S0 remnants can be produced after a major merger event, and that these profiles obey similar scaling relations to those found in real Type-III S0 galaxies. In a later paper, \citet{2015A&A...580A..33E} reported that these relations are similar to those found in Type-III spiral galaxies and are independent of the presence of bars, which suggests that fundamental processes must be responsible for the formation of these structures in disc galaxies along the whole Hubble Sequence. To our knowledge, no other study to the date has addressed the possible formation of these tight scaling relations in Type-III profiles through mechanisms other than major mergers.

Previous studies revealed that the break radius of truncated (Type-II) spiral galaxies evolves with time, increasing by a factor of $\sim1.3\pm0.1$ between $z=1$ and $z=0$ \citep{2008ApJ...684.1026A}. However, no study has analysed the properties of anti-truncated S0 and E/S0 galaxies beyond $z\sim0.2$ to learn about their possible evolution, because cosmological dimming efficiently moves these structures towards even fainter (and prohibitive) surface brightness levels. In order to shed light on the evolution of these structures, we have analysed a sample of Type-III S0 galaxies at $0.2 < z < 0.6$, identified in \citet[][Paper I hereafter]{2017A&A...604A.119B} and compared them to their local analogues from the $z\sim0$ samples by E08 and G11. We study whether the scaling relations found by B14 in a sample of both local and simulated Type-III S0 galaxies are consistent with those found in Type-III S0 galaxies at $0.2<z<0.6$.

The outline of this paper is as follows. The methodology is described in detail in Sect.\,\ref{Sec:Methods}. The results are presented in Sect.\,\ref{Sec:Results}. We discuss the results in Sect.\,\ref{Sec:Discussion}. The final conclusions can be found in Sect.\,\ref{Sec:Conclusions}. In Appendix \ref{Appendix:kcorr} we show the K-corrections applied to the F606W and F850LP band images. We compare the results of the surface brightness profile parameters (\rbreak\ and \mubreak) for the F775W and F850LP bands in Appendix \ref{Appendix:F775WvsF850LP}. We summarise all the results of the linear fits applied to the structural, photometric, stellar mass planes and the K-corrections in Appendix \ref{Appendix:fits}. Finally, we show the surface brightness and colour profiles of our sample of Type-III S0 galaxies at $0.2<z<0.6$ in Appendix \ref{Appendix:profiles}. We assume a concordance cosmology \citep[$\Omega_{\mathrm{M}} = 0.3,\Omega_{\mathrm{\Lambda}}=0.7, H_{0}=70 $ km s$^{-1}$ Mpc$^{-1}$, see][]{2007ApJS..170..377S}. All magnitudes are in the AB system \citep{1971ApJ...170..193O} unless otherwise noted.

\section{Methods and data}
\label{Sec:Methods}
In this section we describe the available data and the process to obtain surface brightness profiles from the HST/ACS data of the GOODS-N field. We summarise the results from Paper I and describe some novel work performed for the present paper. In Sect.\,\ref{Subsec:Methods_SHARDS} we describe the initial sample and the ancillary data from the SHARDS database (SFRs, stellar masses, rest-frame magnitudes) that have been used in Paper I and the present one. We describe the HST imaging data, the PSF and photometric corrections of the images and the attainment of surface brightness profiles of the Type-III S0 galaxies in our sample in Sect.\,\ref{Subsec:Methods_Data}. In this section we also detail the methods used to analyse, classify and characterise the corrected surface brightness profiles in Paper I. Finally, we extend the analysis in this paper to two additional HST/ACS filters (F606W and F850LP) in Sect.\,\ref{Subsec:Methods_Colour}.

For a full description of the morphological classification, the PSF and photometric corrections of the surface brightness profiles and their analysis for the full sample of S0 galaxies at $0.2<z<0.6$, we refer the reader to Paper I.

\subsection{SHARDS database}
\label{Subsec:Methods_SHARDS}

The sample of Type-III S0 -- E/S0 galaxies was selected from an initial sample of quiescent galaxies on the GOODS-N field. We restricted the study to this region in order to take advantage of the Survey for High-$z$ Absorption Red and Dead Sources data in this field \citep[SHARDS,][]{2013ApJ...762...46P}. This project is an ESO/GTC Large Program carried out with the OSIRIS instrument on the 10.4\,m Gran Telescopio Canarias (GTC). The SHARDS survey obtained data between 5000 \AA\ and 9500 \AA\ for galaxies in the GOODS-N field down to $m$ < 26.5 AB magnitudes, in 25 medium band filters (FWHM $\sim$ 170 \AA). SHARDS data are available through the Rainbow Database \citep{2011ApJS..193...13B, 2011ApJS..193...30B}. For the GOODS-N field (Barro et al. in
preparation) the authors used all the available photometry to build spectral energy distributions (SEDs) from UV (GALEX) to nIR (Herschel). They derived photometric redshifts and estimates of parameters such as the stellar mass, the UV- and IR-based SFRs, the stellar population age and rest-frame magnitudes in different filters from spectral template fitting of the SEDs \citep{2005ApJ...630...82P,2008ApJ...675..234P,2013ApJ...762...46P}. The accuracy of the photometric redshifts from SHARDS for the original sample is $\Delta z/(1+z) = 0.0024$. Nevertheless, we used the spectroscopic redshifts when available. We note that all the objects in the final sample of Type-III S0 galaxies at $0.2<z<0.6$ presented available spectroscopic redshifts.

We selected all sources from the GOODS-N field in the SHARDS catalogue (excluding those classified as stars or artefacts) which present redshifts in the range $0.2<z<0.6$. We used the boundaries on the $(U-V)$ versus $(V-J)$ colour-colour diagram presented in \citet{2011ApJ...735...86W, 2012ApJ...754L..29W} to select only those objects in the red sequence. The main objective of this was to isolate a sample of potentially quiescent galaxies, which must contain the analogues of local (quiescent) S0s. We excluded those objects with $z<0.2$ in the catalogue because of the high uncertainties detected in the estimates of the photometric redshifts at these distances \citep{2011ApJS..193...30B}. After this process, we obtained a sample of 150 red-sequence objects in the range $0.2<z<0.6$.

\subsection{Surface brightness profiles: Data, reduction and corrections}
\label{Subsec:Methods_Data}

We identified every object in the red-sequence sample in the ACS F435W, F606W, F775W and F850LP mosaics from 3D-HST \citep[HST Cycle 11, programme IDs 9425 and 9583,][]{2004ApJ...600L..93G} available at the 3D-HST project webpage\footnote[4]{3D-HST - A Spectroscopic Galaxy Evolution Survey with the Hubble Space Telescope: http://3dhst.research.yale.edu/Home.html} \citep{2014ApJS..214...24S}. We performed a visual morphological
classification based on the ACS images to identify E/S0 and S0 galaxies, supported by the PSF-uncorrected surface brightness profiles in some cases to avoid misclassification of ellipticals as face-on S0s. We classified the objects in seven different morphological types: 1) Ellipticals, 2) S0-E/S0 galaxies, 3) spirals, 4) ongoing mergers, 5) compact post-starbursts, 6) green peas and
7) diffuse galaxies. We also analysed the presence of active galactic nucleus (AGN) activity in the selected E/S0-S0 sample to discard AGN hosts, which would have made the PSF corrections uncertain. We checked that the specific star formation rates (sSFR) of the selected E/S0-S0 objects were coherent with being quiescent, accordingly to local S0s. These tests were done as a function of the total stellar mass derived by the SHARDS project.

During this phase, we performed the masking of the background and foreground objects on the F775W images, in order to avoid contamination of the profiles. We refer the reader to Sect.\,2.2 on Paper I for details of the morphological classification. After this, we found that 50 objects present S0-E/S0 morphologies, all presenting low SFRs, according to local S0s.

Since the main objective of this work was to identify anti-truncated stellar profiles, which appear as light excesses over the inner exponential profile at large radii, it is mandatory to correct the images and the profiles for PSF effects. We did this following a similar methodology to that of \citet{2016ApJ...823..123T}. First, we created a PSF model using two sources: 1) the star-stacked PSF provided by the 3D-HST project for the GOODS-N mosaics of each band \citep{2014ApJS..214...24S}, and 2) a model of the PSF using the Tiny Tim software \citep{2011SPIE.8127E..0JK}. We combined these to simulate the effect of the dithering pattern and rotation of the camera in the final mosaic \citep{2016ApJ...823..123T}, because the star-stacked model size is not enough for our purposes ($\sim 4$ arcsec in diameter) while, given the apparent size of our object we need at least a $\sim 25$ arcsec diameter PSF to perform any kind of PSF subtraction without the risk of underestimating their effect in the outskirts of the galaxies \citep{2014A&A...567A..97S,2015A&A...577A.106S}. After this, we used this PSF model to generate Sérsic bulge + exponential disc models for 44 objects with S0--E/S0 morphologies on the F775W band using {\tt{GALFIT3.0}} \citep{2002AJ....124..266P}. We discarded 6 objects from the total 50 S0--E/S0 sample in this step due to their small size and possible contamination from nearby objects in the field of view. The residuals of this PSF-convolved model with respect to the original image were added to the image of the two-dimensional (2D) model before convolution with the PSF.

The final result is a deconvolved image of the selected object \citep{2013MNRAS.431.1121T,2016ApJ...823..123T}. For further details on the deconvolution procedure, we refer the reader to Sect.\,2.3 of Paper I. 
Subsequently we calculated the surface brightness profiles using the PSF-corrected images. After this, we corrected the profiles for dust extinction, cosmological dimming and K-correction for each S0 and E/S0 object, obtaining the corrected surface brightness profiles in the rest-frame $R$ band, which is the one used in the local samples by E08 and G11 that we use for comparison (see Sect.\,2.5, Table A.1 and Fig. 5 in Paper I).

Finally, we performed the identification, characterisation and analysis of the structure of the components in the surface brightness profiles by using a semi-automated method called {\tt{Elbow}}\footnote[2]{{\tt{Elbow}} is publicly available at GitHub (https://github.com/Borlaff/Elbow)}: a statistically robust method to fit and classify the surface brightness profiles, and calculate the likelihood that a certain break exists (Types II and III) or not (Type I). A detailed explanation of {\tt{Elbow}} can be found in Sect.\,2.6 of Paper I. Finally, we found that 14 S0--E/S0 galaxies present Type-III profiles after PSF corrections (13 at $0.4<z<0.6$ and 1 at $z=0.247$). To our knowledge, this is the first robust sample of anti-truncated S0--E/S0s obtained at $0.2<z<0.6$.

\subsection{Data for colour profiles}
\label{Subsec:Methods_Colour}

One of the main objectives of the present paper is to look for signs of differences between the stellar populations of the inner and outer profile  of our selected Type-III S0s. To do this, we repeat the masking, 2D modelling, photometric corrections (PSF, Galactic extinction, K-correction), surface brightness profile analysis and classification with {\tt{Elbow}} for the images of the 14 Type-III S0-E/S0 galaxies in the F606W and F850LP bands, following the same procedure previously described (see below). 
In addition, we have also used the surface brightness profiles in these bands to create colour profiles. We selected the F606W and F850LP bands in order to trace regions of the SED that are as separated as possible in wavelength, but with enough depth in the GOODS-N field to have sufficient emission in the outskirts of the objects. After several tests, the F435W band was discarded because in the vast majority of cases the surface brightness profile reached the limiting magnitude before \rbreak. The median limiting magnitudes in our profiles for the selected bands are very similar: $\mu_{\mathrm{F606W,lim}} = 27.051^{+0.099}_{-0.058}$ \magarc, $\mu_{\mathrm{F775W,lim}} = 27.092^{+0.024}_{-0.032}$ \magarc\ and $\mu_{\mathrm{F850LP,lim}} = 26.75^{+0.18}_{-0.17}$ \magarc. The F606W and F850LP filter images trace \rbreak\ in all cases, although F606W reaches the limiting magnitude at lower galactocentric radial distances than the F775W and F850LP bands, and thus it dominates the uncertainties in the colour profiles.

We corrected the images in the F606W and F850LP bands for PSF effects using the same procedure as for the F775W images (see Sect.\,\ref{Subsec:Methods_Data}). We checked that the {\tt{GALFIT3.0}} models for the F606W and F850LP bands were similar to those calculated for the F775W images. After the PSF correction process, we generate surface brightness profiles from the final images for the 14 Type-III S0 galaxies selected in Paper I at $0.2<z<0.6$. We derived the surface brightness profiles by using elliptical concentric apertures with fixed position angle and axis ratio, equal to the mean values obtained with SExtractor \citep{1996A&AS..117..393B} in the galaxy disc on the original image, with the limiting threshold set to $1\sigma$. As explained in Paper I, the surface brightness profiles of the edge-on objects were analysed separately, using {\tt{ISOFIT}} \citep{2015ApJ...810..120C}. This program also uses concentric apertures to calculate the surface brightness profile but replaces the angular parameter that defines quasi-elliptical isophotes in polar coordinates with the eccentric anomaly, providing a more accurate modelling of galaxies with high inclinations. We refer the reader to this paper for details on the method and the benefits for the modelling of edge-on galaxies. In this case, the position angle and ellipticity are allowed to vary with radius. We apply this method to the unique edge-on galaxy in our Type-III S0 sample, SHARDS20000827. 

Finally, the surface brightness profiles are corrected for Milky Way dust extinction, cosmological dimming and K-correction (see Appendix \ref{Appendix:kcorr}). We must remark that although in the present work we have included two additional filters to the analysis, the results from the F606W and F850LP bands are only used for obtaining the colour profiles, while the rest of the analysis is performed with the parameters estimated from the surface brightness profiles in the F775W band in Paper I, K-corrected to rest-frame $R$ band.


\section{Results}
\label{Sec:Results}

\begin{figure*}[h]
 \begin{center}
\includegraphics[width=0.49\textwidth]{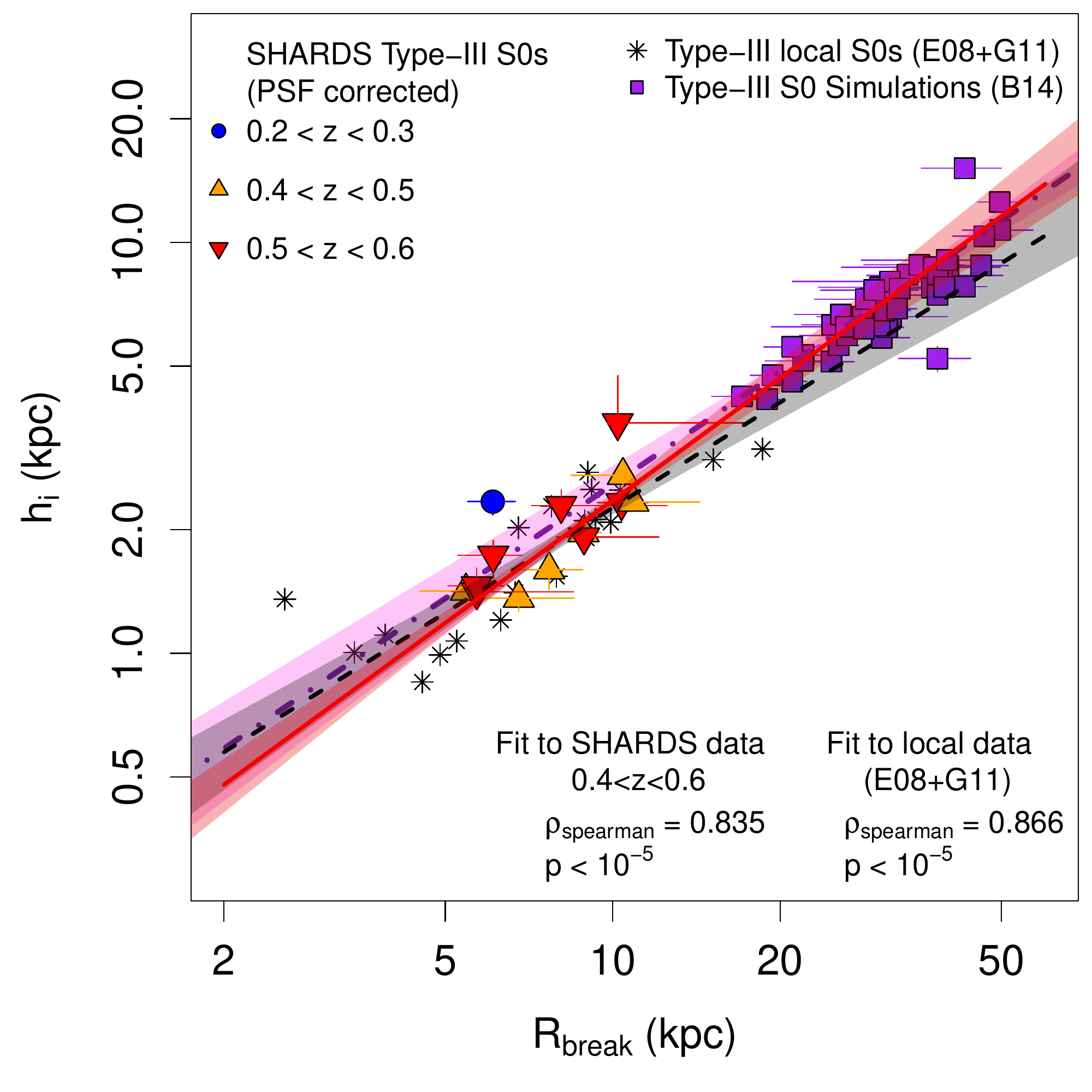}
\includegraphics[width=0.49\textwidth]{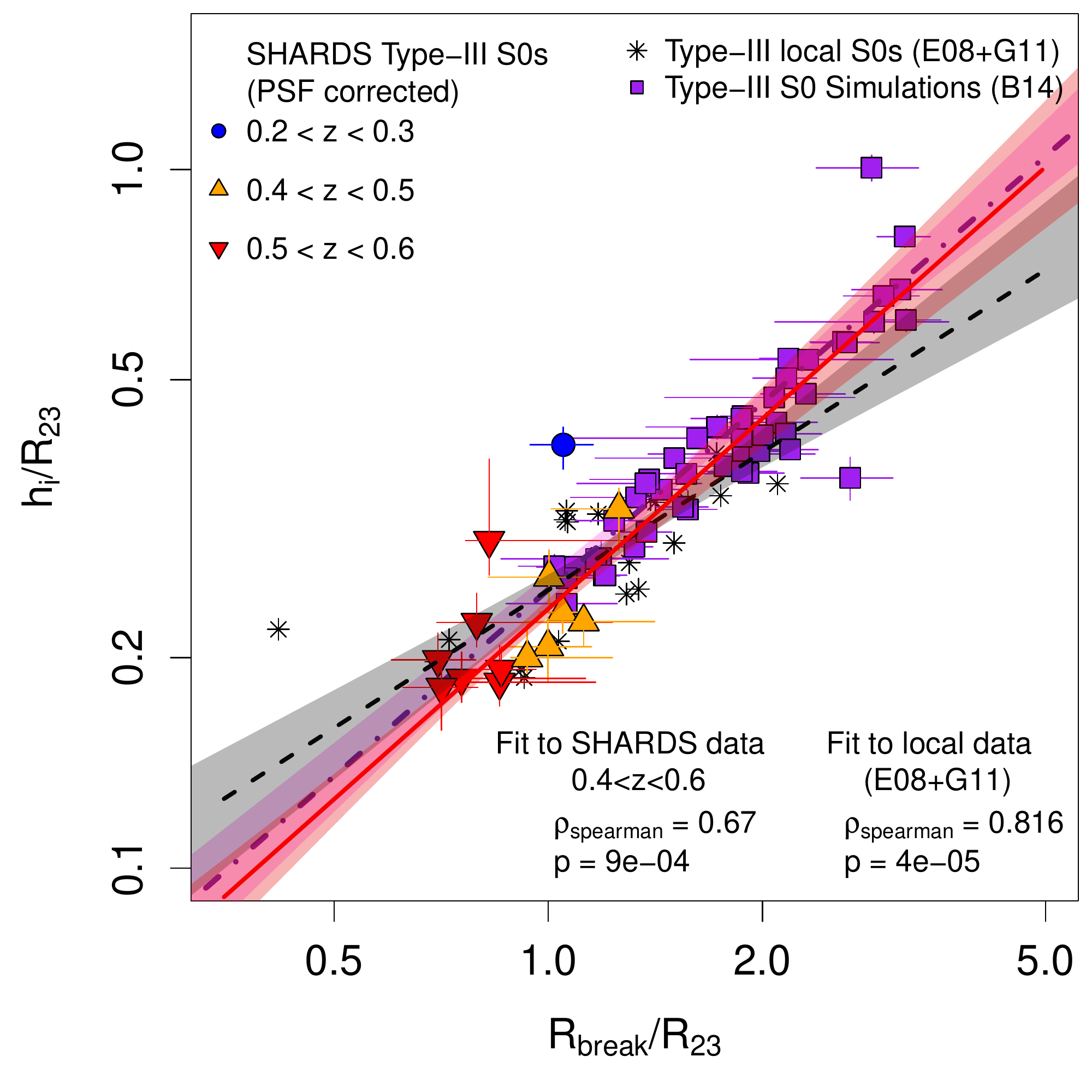}
\includegraphics[width=0.49\textwidth]{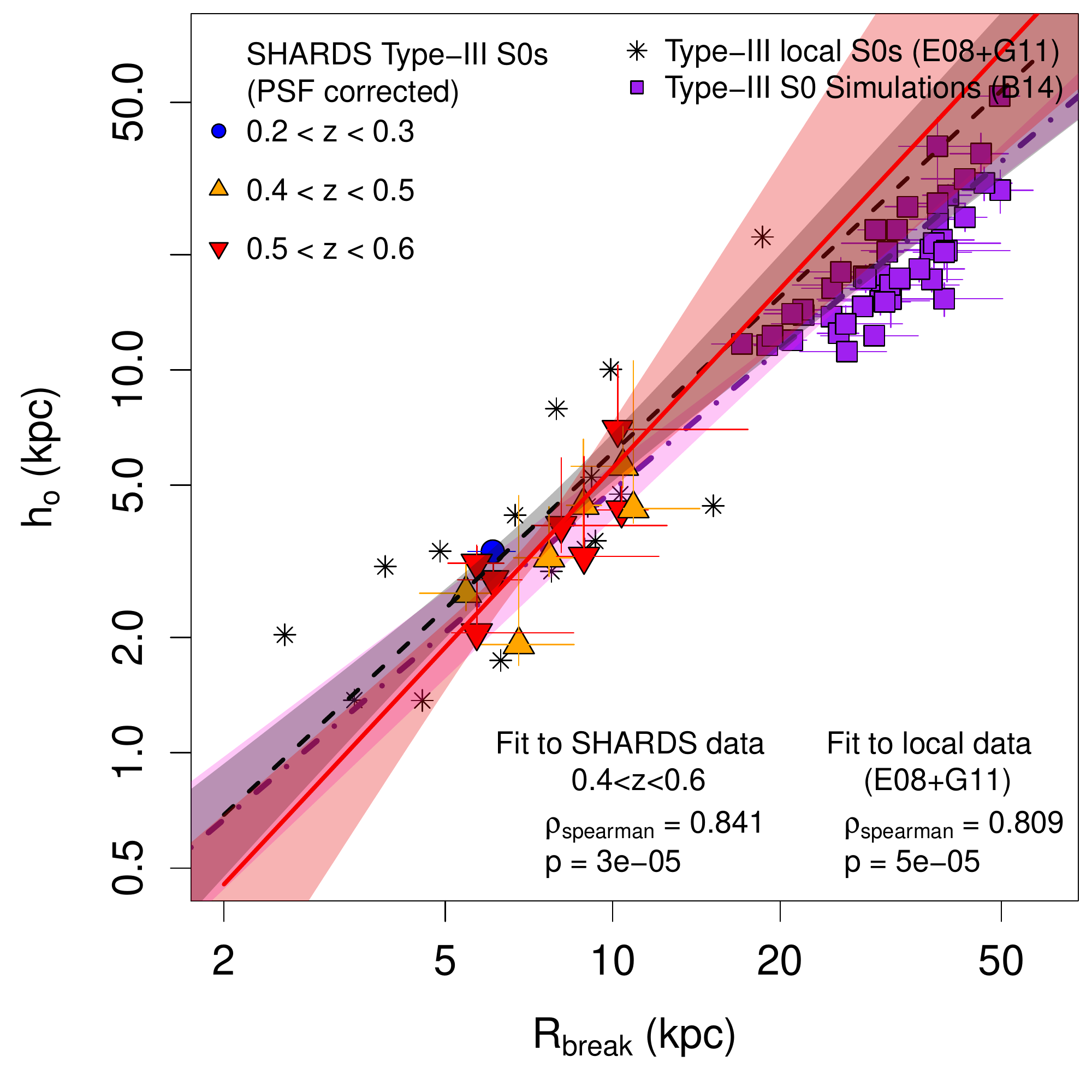}
\includegraphics[width=0.49\textwidth]{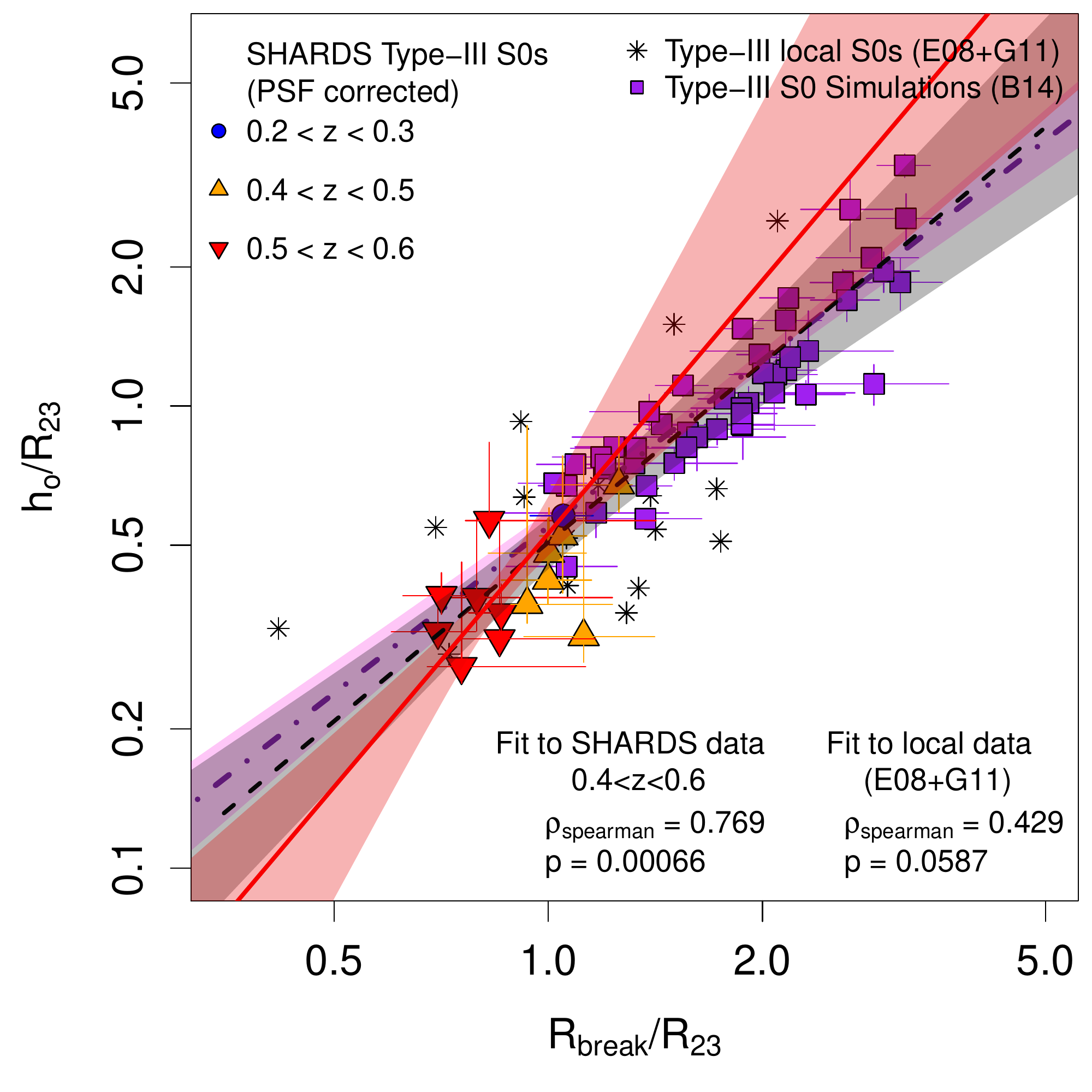}
\vspace{0.1cm}
\caption[]{Distribution of the Type-III S0 galaxies at $0.2<z<0.6$ and the local sample from E08 and G11 in the structural diagrams. \emph{Upper left panel:} \hi\ vs. \rbreak. \emph{Upper right panel:} \hi/\risoph\ vs.\rbreak/\risoph. \emph{Lower left panel:} \ho\ vs. \rbreak. \emph{Lower right panel:} \ho/\risoph\ vs. \rbreak/\risoph. \emph{Solid red line:} linear fit to the Type-III S0 galaxies at $0.4<z<0.6$. \emph{Dashed black line:} linear fit to the local Type-III S0 galaxies. \emph{Dash-dotted purple line:} linear fit to the simulations of Type-III S0 galaxies from B14. Their respective coloured regions represent the $1\sigma$ confidence level region of each linear fit. In both cases, the relations fitted were in logarithmic scale in both x and y-axis ($\log_{10}(\hi) \propto\ \log_{10}(\rbreak)$, $\log_{10}(\hi/\risoph) \propto\ \log_{10}(\rbreak/\risoph)$, $\log_{10}(\ho) \propto\ \log_{10}(\rbreak)$ and $\log_{10}(\ho/\risoph) \propto\ \log_{10}(\rbreak/\risoph)$). We refer to the legend in the figure for the redshift colour classification and correlation tests.} 
\label{fig:hiho_rbrkr23}
\end{center}
\end{figure*}

In this section we analyse the distributions of the characteristic parameters from the surface brightness profiles of Type-III S0--E/S0 galaxies. In Sect.\,\ref{Subsec:Scaling_relations} we compare the properties of our sample of 14 Type-III S0-E/S0 galaxies at $0.2<z<0.6$ with those of a sample of 20 Type-III S0 galaxies in the local Universe from E08 and G11 (local sample, hereafter) and the simulations of Type-III S0 galaxies from major mergers from B14. In Sect.\,\ref{Subsec:MassMbMk} we compare the distributions of stellar mass and rest-frame absolute magnitude in the $B$ and $K$ bands from the local and $0.2<z<0.6$ samples. We investigate if there is any significant change in the distributions of the structural parameters obtained from the surface brightness profiles with $z$ at a fixed stellar mass, in particular for \rbreak\ (Sect.\,\ref{Subsec:rbreak_evol}) and \hi, \ho\ and the ratio \ho/\hi\ (Sect.\,\ref{Subsec:hiho_evol}). We extend the analysis to the photometric parameters \mui, \muo\ and \mubreak\ in Sect.\,\ref{Subsec:mu_evol}. We study in detail the dependence with $z$ of the \mubreak\ vs. \rbreak\ relation in Sect.\,\ref{Subsec:mubreak_evol}. In addition, we analyse the colour profiles and the median profiles of the $0.4<z<0.6$ sample in Sect.\,\ref{Subsec:colour_profiles}. Finally, we compare the observed distribution of the \mubreak\ vs. \rbreak\ at $z\sim0$ and $z\sim0.5$ with the predicted evolution from a grid of single stellar population models (SSP hereafter) just to check whether the measured evolution between these two redshifts could be explained by simple passive evolution or not (see Sect.\,\ref{Subsec:stellar_pop_analysis}). 
With the exception of the colour profile analysis, where we use the F606W and F850LP bands, we again emphasise that we use the structural and photometric parameters of the antitruncations measured in the F775W band in our sample at $0.2<z<0.6$, after being transformed to rest-frame $R$ band, in order to compare with the $R$-band observations of the local sample (see Sect.\ref{Subsec:Methods_Data}).

\subsection{Scaling relations}
\label{Subsec:Scaling_relations}

\begin{figure*}[]
 \begin{center}
 \vspace{0.2cm}
\includegraphics[width=0.49\textwidth]{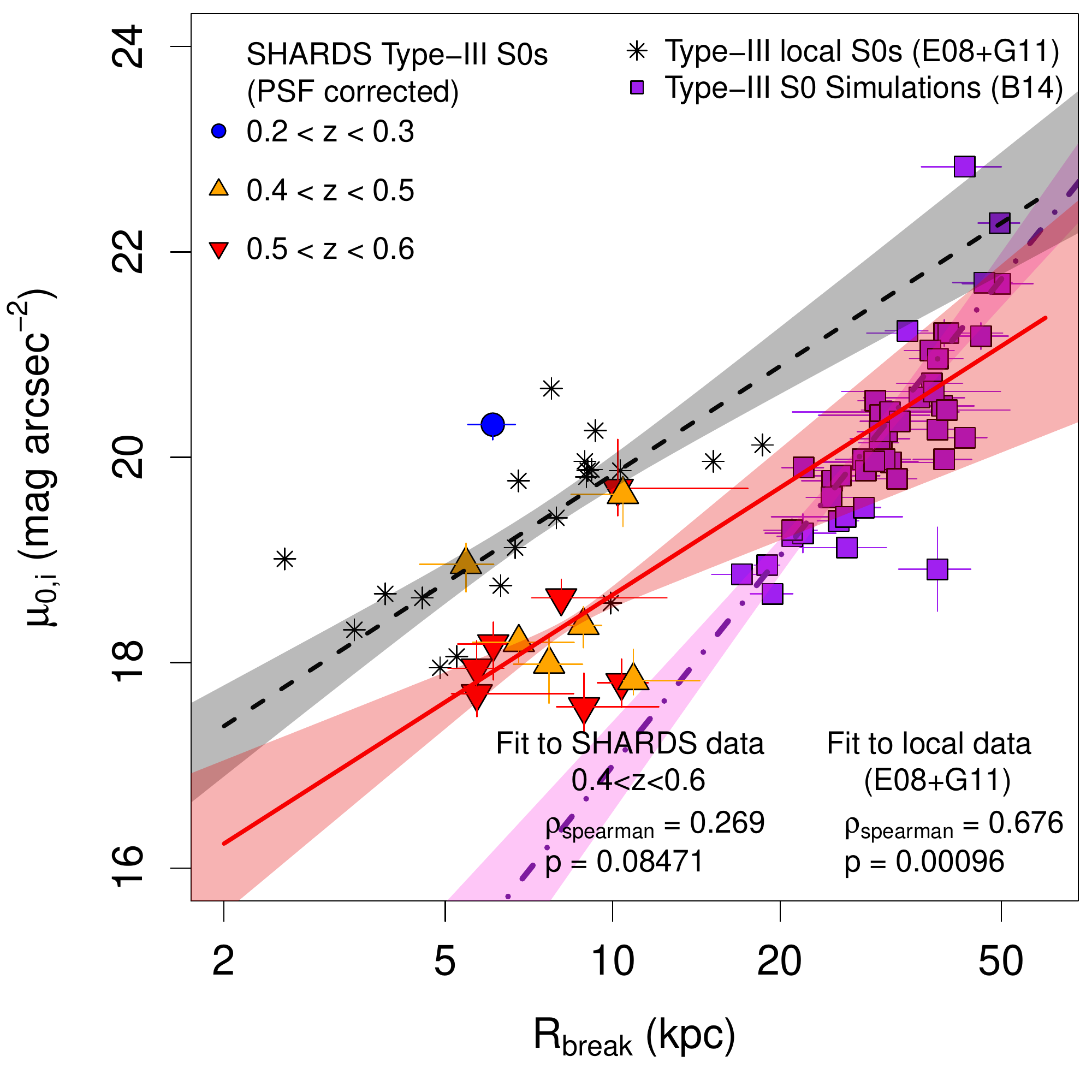}
\includegraphics[width=0.49\textwidth]{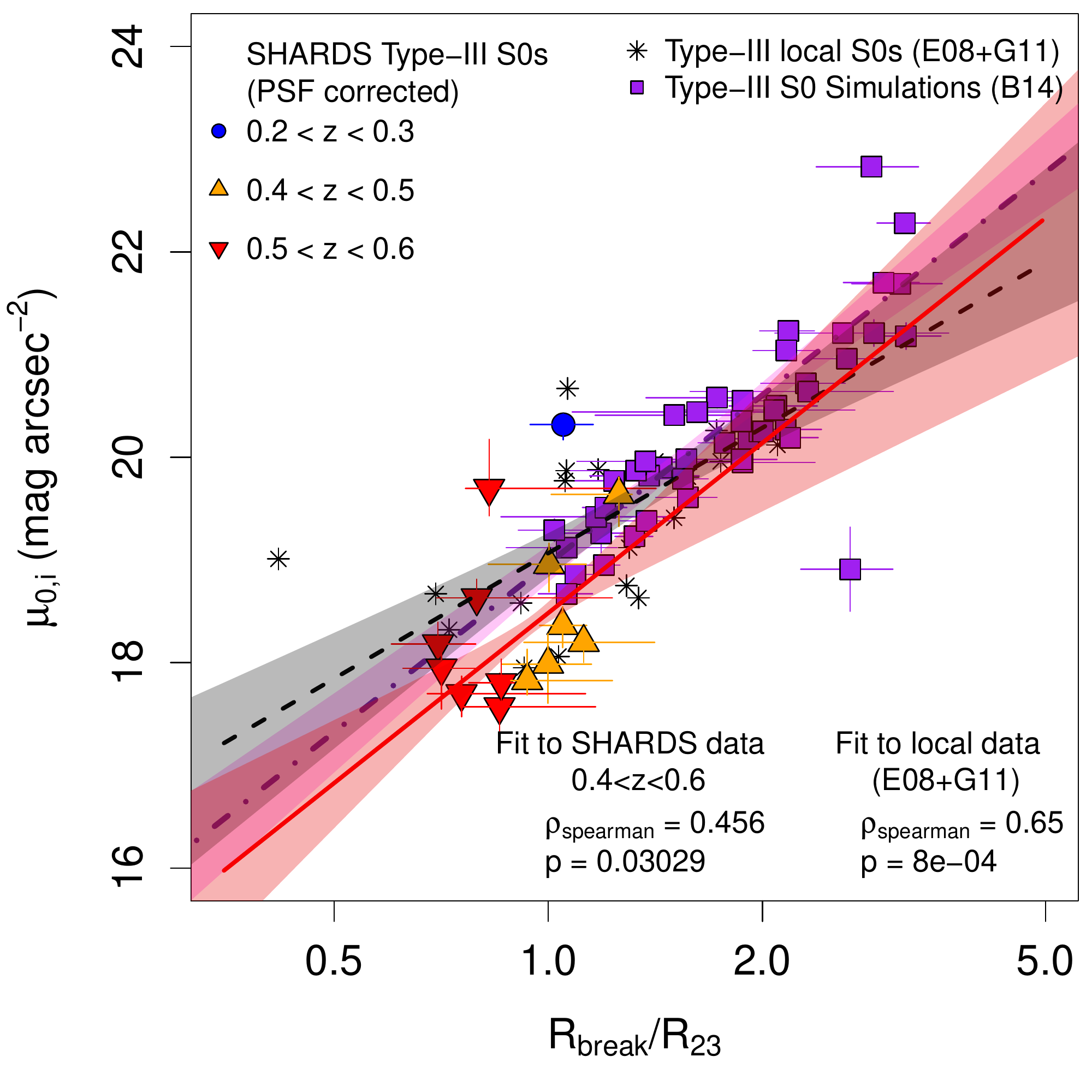}
\includegraphics[width=0.49\textwidth]{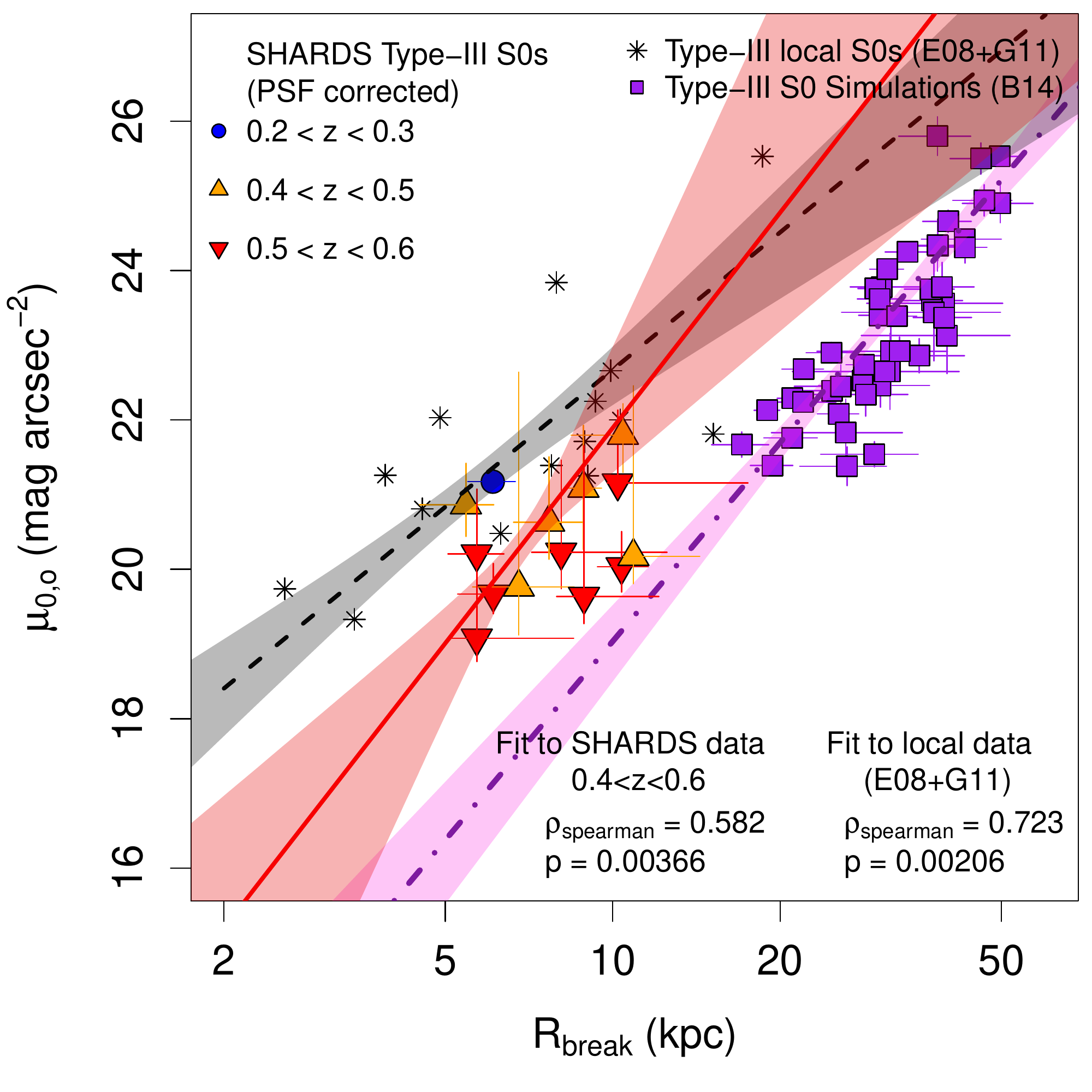}
\includegraphics[width=0.49\textwidth]{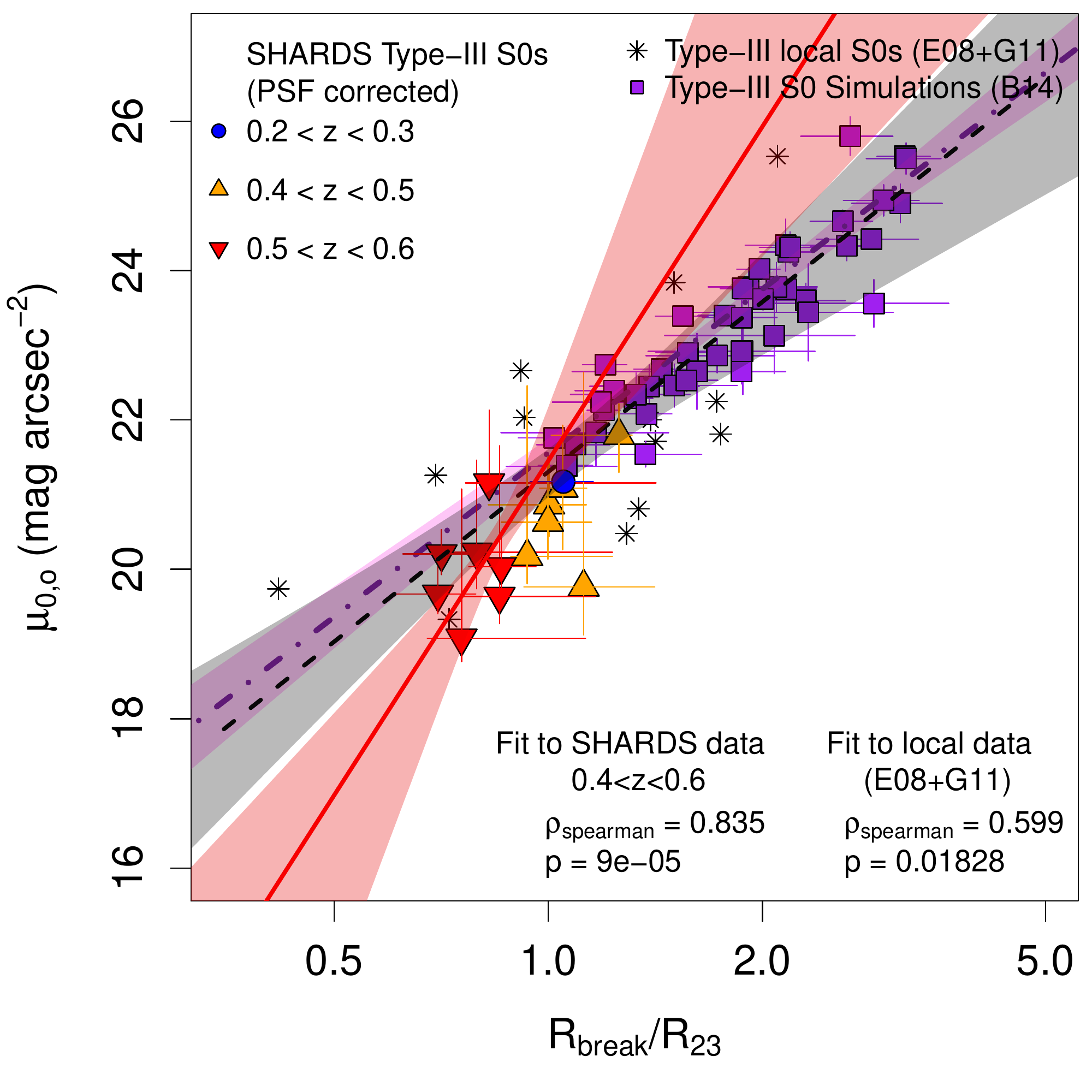}
\vspace{0.1cm}
\caption[]{Distribution of the Type-III S0 galaxies at $0.2<z<0.6$ and the local sample from E08 and G11 in the photometric diagrams. \emph{Upper left panel:} \mui\ vs. \rbreak. \emph{Upper right panel:} \mui\ vs. \rbreak/\risoph. \emph{Lower left panel:} \muo\ vs. \rbreak. \emph{Lower right panel:} \muo\ vs. \rbreak/\risoph. \emph{Solid red line:} linear fit to the Type-III S0 galaxies at $0.4<z<0.6$. \emph{Dashed black line:} linear fit to the local Type-III S0 galaxies. \emph{Dash-dotted purple line:} linear fit to the simulations of Type-III S0 galaxies from B14. Their respective coloured regions represent the $1\sigma$ confidence level region of each linear fit. In both cases, the relations fitted were in logarithmic scale for the x-axis ($\mui \propto\ \log_{10}(\rbreak)$, $\mui \propto\ \log_{10}(\rbreak/\risoph)$, $\muo \propto\ \log_{10}(\rbreak)$ and $\muo \propto\ \log_{10}(\rbreak/\risoph)$). Consult the legend in the figure for the redshift colour classification and correlation tests.} 
\label{fig:muimuo_rbrkr23}
\end{center}
\end{figure*}

The left panels of Fig.\,\ref{fig:hiho_rbrkr23} present the distribution of the inner and outer scale-lengths (\hi\ and \ho) of the Type-III S0 galaxies in the samples at $0.2<z<0.6$ and at the local Universe (E08; G11) as a function of the break radius \rbreak. In addition, we define the characteristic scale \risoph, which corresponds to the radius at which the surface brightness profile reaches 23 \magarc. For each diagram, we present another version with the scale-lengths and \rbreak\ normalised by \risoph\ in the right panels of the figure. In order to define an intermediate-redshift sample, we only include in the linear fits of the intermediate redshift sample the objects at $0.4<z<0.6$, excluding one and only Type-III S0 galaxy at $0.2<z<0.4$ that we found (SHARDS20000593, $z=0.247$). Therefore, the sample at $0.4<z<0.6$ includes 13 S0 - E/S0s with Type-III surface brightness profiles.  

As reported in Paper I, \hi, \ho\ and \rbreak\ present compatible distributions in the local and $0.4<z<0.6$ samples ($\hi \sim 1-5$ kpc, $\ho \sim 1.5 - 20$ kpc and $\rbreak \sim 2-20$ kpc, see Sect.\,3.4 on Paper I for a deeper discussion). The main result in the present work is that Type-III S0 galaxies at $0.4<z<0.6$ also present similar correlations and compatible trends to their local analogues in the structural diagrams $\log_{10}(\hi) \propto\ \log_{10}(\rbreak)$, $\log_{10}(\hi/\risoph) \propto\ \log_{10}(\rbreak/\risoph)$ and $\log_{10}(\ho) \propto\ \log_{10}(\rbreak)$, first reported in B14. We next quantify these observed correlations and trends between the parameters in each case.

We take into account the uncertainties of the parameters by performing 100,000 Bootstrapping \& Monte Carlo simulations of the Spearman correlation test. In order to perform a symmetrical treatment of the variables, we choose the OLS bisector linear regression method \citep{1990ApJ...364..104I}. Additionally, for each parameter we use the empirical probability density distribution (PDD) that resulted from the analysis performed with {\tt{Elbow}} of the surface brightness profile of each galaxy as input error (see Section 2.6 in Paper I). By doing this, we take into account any internal correlations between the variables and their uncertainties and thus we generate much more accurate Monte-Carlo simulations than from simply simulating random Gaussian noise around the central point.

We found that the structural diagrams present statistically significant ($p < 0.05$) positive correlations for both local Universe and $0.4<z<0.6$ samples (see panels on Fig.\,\ref{fig:hiho_rbrkr23}) with the exception of $\log_{10}(\ho/\risoph) \propto\ \log_{10}(\rbreak/\risoph)$ for the local sample, which does not present a significant correlation in the case of the sample at $z=0$, although it presents a low $p-$value. In contrast to this, the sample at $0.4<z<0.6$ presents a significant correlation. We performed linear fits to the distributions in those planes, which are represented with dashed black (local sample) and solid red lines ($0.4<z<0.6$ sample) in the diagrams. The results of the correlation tests and the linear fits are summarised in Table\,\ref{tab:fits_results} in Appendix \ref{Appendix:fits}. 

\begin{figure*}[ht!]
\centering
\vspace{0.1cm}
\includegraphics[width=\textwidth]{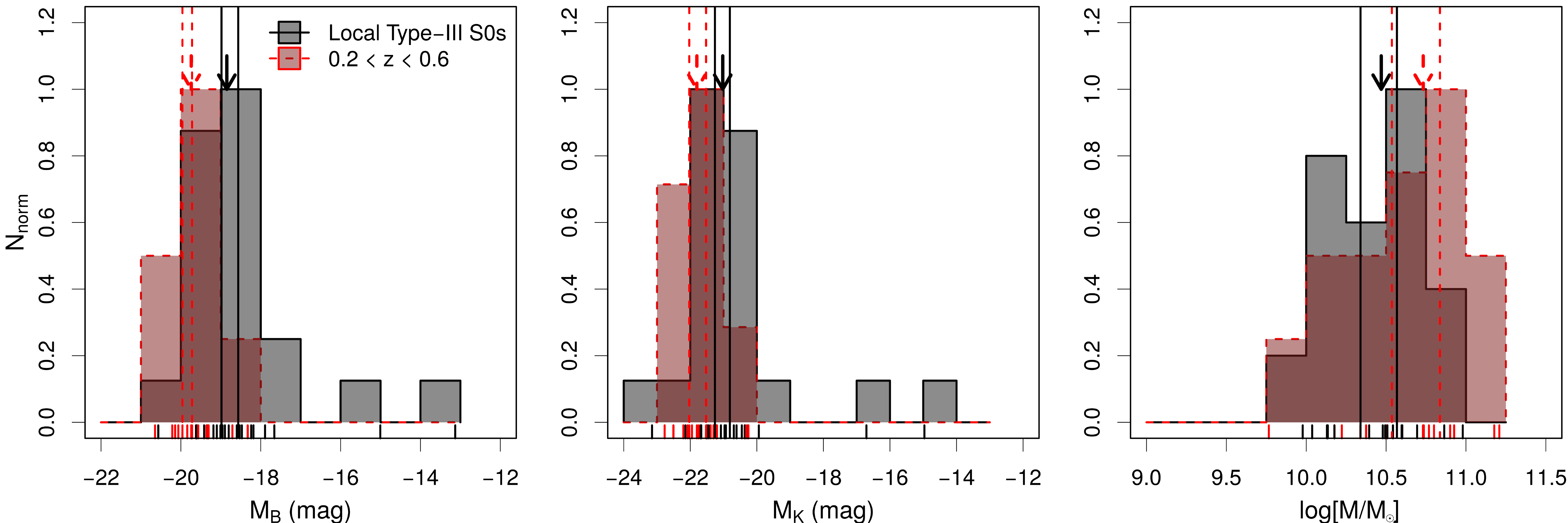}
\vspace{0.1cm}
\caption[]{Normalised distributions of the rest-frame absolute magnitude in the $B$ band (left panel), rest-frame absolute magnitude in the $K$ band (central panel) and the stellar mass (right panel) of our sample of Type-III S0 galaxies at $0.2<z<0.6$ compared to those from the local Universe (E08 and G11). \emph{Grey histogram with black solid line:} Type-III S0 galaxies in the local Universe. \emph{Red histogram with dashed line}: Type-III S0 galaxies at $0.4<z<0.6$ from the SHARDS sample. The red and black arrows represent the median value for the objects at $0.4<z<0.6$ and the local sample, respectively. The solid and dashed vertical lines represent the 1$\sigma$ confidence interval for the median for each sample.}  
\vspace{0.1cm}
\label{fig:MbMkMass_hist}
\end{figure*}

On top of the parameters of the local Universe and $0.2<z<0.6$ Type-III S0 galaxies in Fig.\,\ref{fig:hiho_rbrkr23}, we plot the results of the major merger simulations from B14 in purple squares, and the corresponding fitted linear trend with a purple dashed-dotted line. We found that in the diagrams $\log_{10}(\hi) \propto\ \log_{10}(\rbreak)$, $\log_{10}(\hi/\risoph) \propto\ \log_{10}(\rbreak/\risoph)$ and $\log_{10}(\ho) \propto\ \log_{10}(\rbreak)$, the parameters of $0.4<z<0.6$ and local Type-III S0 galaxies present compatible trends that overlap with the extrapolation towards lower values of those corresponding to the simulated mergers sample. We must remark that the simulations present higher masses ($\sim 1-3\cdot10^{11}\,\Msun$) than the real Type-III S0 galaxies samples. As a consequence, the models present an offset in the non-normalised diagrams. Despite this, the models and the observed Type-III S0 galaxies overlap in the normalised diagrams. We refer the reader to B14 for a detailed discussion on this subject. We conclude that the parameters of $0.4<z<0.6$, local and simulated Type-III S0 galaxies present positive correlations and compatible trends in the structural diagrams reported in B14.

In Fig.\,\ref{fig:muimuo_rbrkr23} we present the distribution of the central surface brightness of the inner and outer profiles (\mui\ and \muo) as a function of the break radius \rbreak\, and the normalised break radius (\rbreak/\risoph). We report several interesting results. First, in contrast with the structural parameters, the values of \mui\ and \muo\ of Type-III S0 galaxies at $0.4<z<0.6$ do not show a similar distribution to the objects in the local Universe. The objects at $0.4<z<0.6$ tend to show brighter magnitudes than the objects from the sample at $z=0$ for similar \rbreak\ values in all the photometric diagrams. When \mui\ and \muo\ are plotted against \rbreak/\risoph\ (right panels in Fig.\,\ref{fig:muimuo_rbrkr23}), the local galaxies still appear at fainter magnitudes than their $0.4<z<0.6$ analogues, but in contrast to the non-normalised diagrams, the objects tend to overlap into one trend in the normalised ones (compare with the left panels in the same figure). In the case of the sample at $0.4<z<0.6$, we find a significant correlation of the central surface brightness of the outer disc (\muo) with the break radius (\rbreak) and with the normalised break radius (\rbreak/\risoph). The $0.4<z<0.6$ sample does not show any significant correlation at the $\mui \propto\ \log_{10}(\rbreak)$ plane (see Table \ref{tab:fits_results} in Appendix \ref{Appendix:fits}), whereas it does in the normalised diagram. We also find a higher median of the \risoph\ value at $z\sim0.5$ compared to $z=0$ ($\risoph = 8.46^{+0.38}_{-0.51}$ kpc at $0.4<z<0.6$, but $\risoph = 5.89^{+0.46}_{-0.44}$ kpc in the local sample). This may be due to a general brightening of the surface brightness profiles with $z$ (we comment more on this in Sect.\,\ref{Subsec:mubreak_evol}).

Additionally, in Fig.\,\ref{fig:muimuo_rbrkr23} shows the results from the simulations from B14. As reported there, the simulations present compatible trends and positions in the $\mui \propto\ \log_{10}(\rbreak/\risoph)$ and $\muo \propto\ \log_{10}(\rbreak/\risoph)$ with the Type-III S0--E/S0 galaxies from the local sample, except that they are located in the extrapolation towards higher $\rbreak/\risoph$ values. The trends of S0 galaxies at $0.4<z<0.6$, although presenting lower \rbreak/\risoph\ values, overlap in these normalised diagrams with the global trend drawn by the local sample and the simulations at higher \rbreak/\risoph\ values.

In conclusion we find that the structural scaling relations observed in Type-III S0 galaxies at $z\sim0.5$ are compatible with those observed in their local analogues and in similar systems formed in major mergers simulations. However, our sample of Type-III S0 galaxies at $0.4<z<0.6$ present brighter values in the photometric diagrams at the same values of \rbreak\ or \rbreak/\risoph\ when compared with the local Type-III S0s.

\subsection{Stellar mass and luminosity in the $B$ and $K$ bands}

\label{Subsec:MassMbMk}
\begin{table}
{\small 
\begin{center}
\caption{Mean rest-frame absolute magnitudes in the $B$ and $K$ bands and mean stellar mass of Type-III S0 galaxies in the local and 0.4<z<0.6 samples. \emph{Columns:} (1) Parameters. (2) Median values for the samples in the local Universe (E08, G11). (3) Median values for the sample at $0.4<z<0.6$, with their corresponding $1\sigma$ uncertainty intervals. (4) Anderson-Darling test $p-$value for the null hypothesis of equal parent distributions for the $0.4<z<0.6$ and the local Universe samples. \emph{Rows from top to bottom:} Rest-frame absolute magnitudes in the $B$ and $K$ bands and stellar mass.}
\begin{tabular}{ccccccc}
\toprule
\multicolumn{1}{c}{Parameter}& &$z\simeq0$&$0.4<z<0.6$ & $p$\\
\multicolumn{1}{c}{(1)}& &(2)&(3)&(4)\\
\midrule
\multirow{1}{*}{<M$_{B}$>}&[mag]&$-18.85^{+0.28}_{-0.13}$&  $-19.74^{+0.02}_{-0.22}$ & $2.1\cdot10^{-4}$\\[1.2ex]
\multirow{1}{*}{<M$_{K}$>}&[mag]&$-21.02^{+0.21}_{-0.23}$& $-21.80^{+0.27}_{-0.23}$ & $8.2\cdot10^{-3}$\\[1.2ex]
\multirow{1}{*}{<$\log_{10}(M/M_{\odot})$>}&[dex]&$10.47^{+0.10}_{-0.13}$ & $10.73^{+0.11}_{-0.19}$ & $6.2\cdot10^{-2}$\\[1.2ex]
\hline
\bottomrule
\end{tabular}

\label{table:MbMkMass}
\end{center}
}
\end{table}

In Fig.\,\ref{fig:MbMkMass_hist} we study if the local Type-III S0 galaxies from E08 and G11 present compatible distributions in stellar mass and in the rest-frame absolute magnitudes in the $B$ and $K$ bands compared to the objects from our sample at $0.4<z<0.6$.

The M$_{B}$ and M$_{K}$ values for E08 and G11 Type-III S0 galaxies were extracted from the Leda database\footnote[3]{HyperLeda database for physics of galaxies: http://leda.univ-lyon1.fr} \citep{2003A&A...412...45P}. We use synthetic rest-frame M$_{B}$ and M$_{K}$ absolute magnitudes for the sample at $0.4<z<0.6$, available through the Rainbow Database. In the case of the $B$ band, the distribution lies in the range $-18<$ M$_{B}<-20.5$ for the sample at $0.2<z<0.6$ and $-13<$ M$_{B}<-20.5$ for the local Type-III S0s. For the $K$ band, the distribution lies in the range $-20<$ M$_{K}<-23$ for the sample at $0.2<z<0.6$ and $-15<$ M$_{K}<-23$ for the local Type-III S0s.

Total stellar masses are available through the Rainbow Database \citep{2011ApJS..193...13B,2011ApJS..193...30B} for the sample at $0.4<z<0.6$. The stellar masses of 12 objects from the local sample were available from S$^4$G \citep{2015ApJS..219....3M}, and we found mass estimates for 3 additional ones in \citep[][NGC4459 and NGC7457]{Forbes2016} and \citet[][NGC2880]{Cano-Diaz2016}. For comparison, the sample of Type-III S0 galaxies at $z\sim0$ presents total stellar masses in the range $\log_{10}(M/M_{\odot}) = [10.0 -  11.0]$, while the objects from the sample at $0.4<z<0.6$ present values in the range $\log_{10}(M/M_{\odot}) = [9.7 - 11.2]$. Again, in the case of our intermediate-redshift sample, we only include objects at $0.4<z<0.6$, which excludes the object at $z=0.247$ from our sample.

We show the results for the comparative analysis of the rest-frame absolute magnitudes in the $B$ and $K$ bands and the stellar mass in Table \ref{table:MbMkMass} and in Fig.\,\ref{fig:MbMkMass_hist}. We use the Anderson-Darling criterion \citep{10.2307/2288805} in order to test whether the parameters from both samples arose from a common unspecified distribution function (null hypothesis). We find that the median values and the distributions of stellar masses are compatible for the objects from the local Universe and our sample at $0.4<z<0.6$ ($p>0.05$ in Table \ref{table:MbMkMass}, see right panel of Fig.\,\ref{fig:MbMkMass_hist}). In contrast to that, the rest-frame absolute magnitudes on the $B$ and $K$ bands are significantly brighter at $0.4<z<0.6$ when compared to the Type-III S0 galaxies at $z\sim0$ ($p<0.05$ in Table \ref{table:MbMkMass}, see also the corresponding panels in Fig.\,\ref{fig:MbMkMass_hist}). This result confirms that our sample of Type-III S0 galaxies at $z\sim0.5$ and the objects from E08 and G11 are comparable in terms of stellar mass and enables us to perform a fair comparison in this sense. However, this analysis also suggests that Type-III S0 galaxies of similar stellar mass were brighter in the past, meaning that they have undergone a significant dimming in the last $6$ Gyr. In Sects.\,\ref{Subsec:mu_evol} and \ref{Subsec:mubreak_evol} we study whether this has affected the whole structure or the inner (\mui) and outer (\muo) profiles in a different way.

\begin{figure*}[ht!]
\centering
\vspace{0.35cm}
\includegraphics[width=0.48\textwidth]{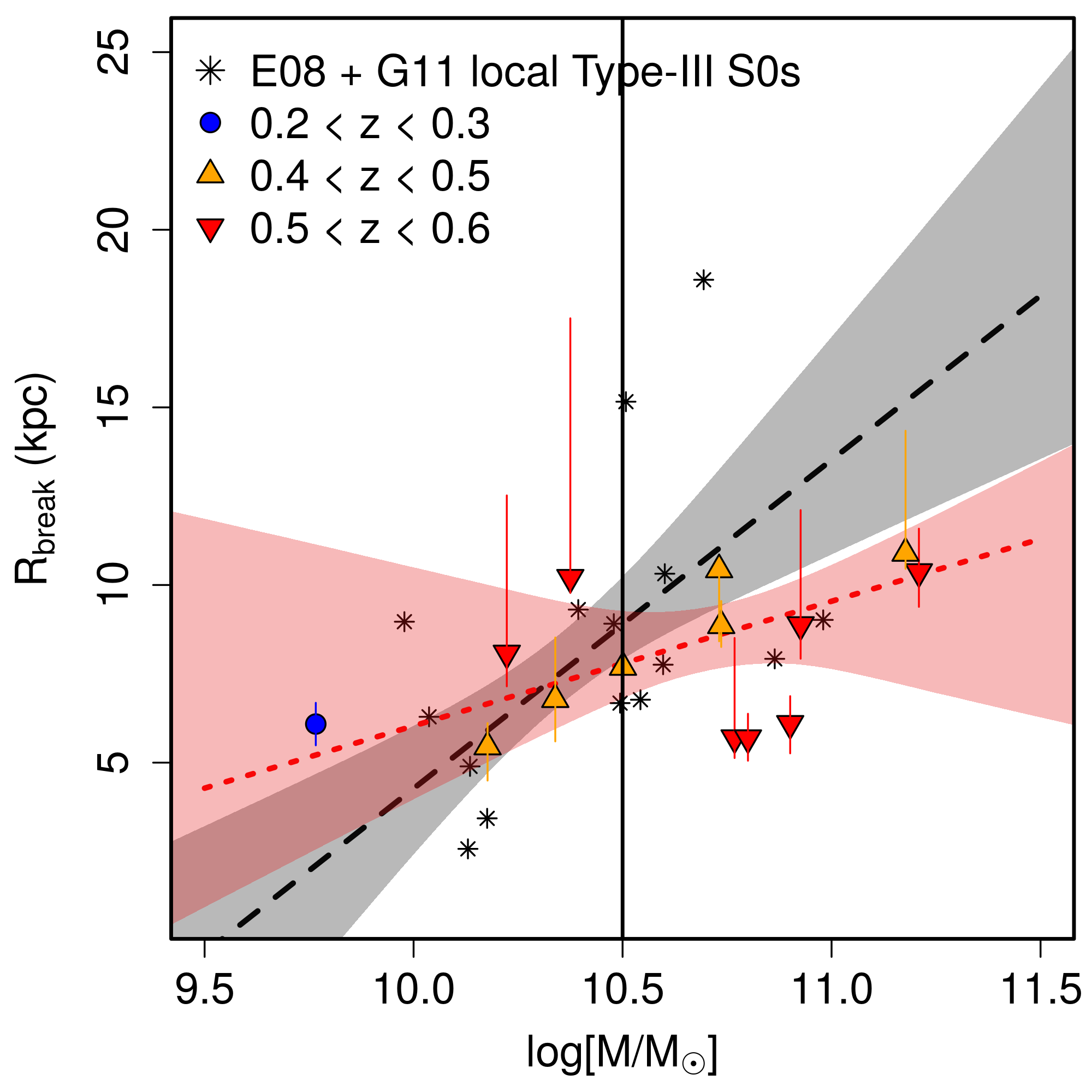}
\includegraphics[width=0.48\textwidth]{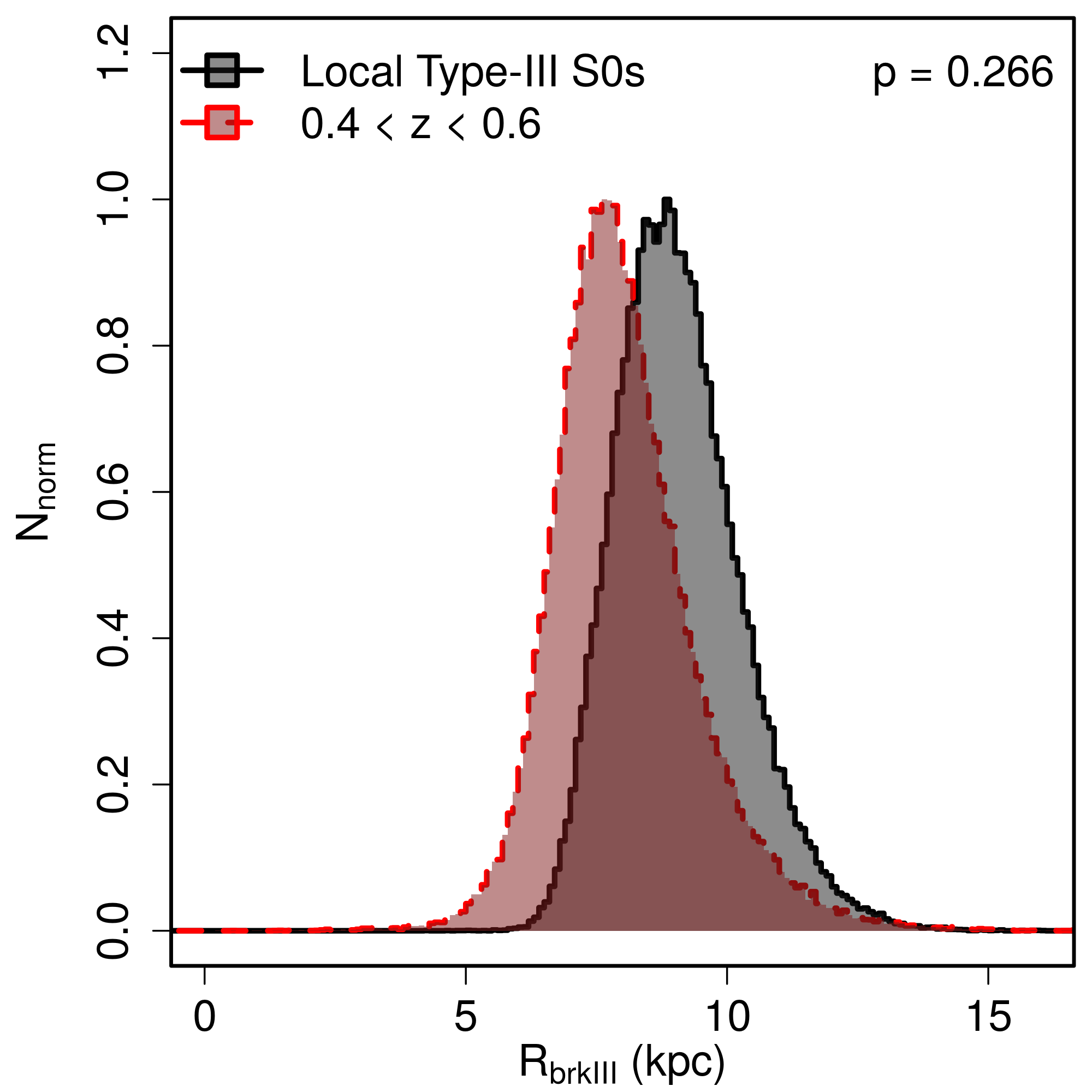}
\vspace{0.35cm}
\caption[]{Position of \rbreak\ as a function of the stellar mass. \emph{Left panel:} Distribution of the break radius \rbreak\ of Type-III S0 galaxies at $0.4<z<0.6$ and at $z\sim0$ with the stellar mass. The solid vertical black line in the left panel represents the fixed value at which we compare the linear fits in the right one. \emph{Right panel:} Comparison of the PDDs of the break radius \rbreak\ at fixed stellar mass $\log_{10}(M/M_{\odot})= 10.5$. \emph{Grey histogram with black solid line:} PDDs for the Type-III S0 galaxies in the local Universe selected from E08 and G11 samples. \emph{Red histogram with dashed line}: PDDs for the Type-III S0 galaxies at $0.4<z<0.6$ from the SHARDS sample. The probability ($p-$value) for both samples to have compatible values of \rbreak\ at fixed stellar mass of $\log_{10}(M/M_{\odot})= 10.5$ is $p=0.266$ (we refer to the legend in the panels for a key of the different lines and symbols).}
\vspace{0.35cm}
\label{fig:Rbrk_z}
\end{figure*}

\subsection{Evolution of \rbreak\ with $z$ at a fixed stellar mass}
\label{Subsec:rbreak_evol}

With the aim of investigating if the break radius \rbreak\ presents any signs of evolution with redshift, we follow the method described in \citet{2008ApJ...684.1026A}. In this paper, the authors analysed a sample of Type-II galaxies at several redshift intervals, from the local Universe, up to $z=1$. In order to compare the break position taking into account the size of each galaxy, the authors estimated the typical value of \rbreak\ as a function of 1) the Johnson $B$-band absolute magnitude and 2) the total stellar mass. After this, they compare the expected values for \rbreak\ at the reference values of M$_\mathrm{B} = -20$ and $\log_{10}(M/M_{\odot}) = 10$ at different redshifts to check whether there is evolution or not. We follow the same procedure to test if there is any noticeable difference in the position of the break (\rbreak) of Type-III S0s between the local universe and our sample at $0.4<z<0.6$. 

In the left panel of Fig.\,\ref{fig:Rbrk_z}, we show the distribution of \rbreak\ as a function of the decimal logarithm of the stellar mass $\log_{10}(M/M_{\odot})$. We compare the expected values for \rbreak\ at the reference value of $\log_{10}(M/M_{\odot}) = 10.50$ at different redshifts to check whether there is evolution or not. This reference value corresponds to the median value for $\log_{10}(M/M_{\odot})$ including the Type-III S0 galaxies from E08 and G11 and our sample at $0.2<z<0.6$.

In order to do that, we fit a linear model to each sample in the diagram of the left panel of Fig.\,\ref{fig:Rbrk_z}. The uncertainty intervals for each linear fit (i.e. the probability distributions of the intercept and slope) were obtained by performing 100,000 Bootstrapping + Monte Carlo simulations (see Table \ref{tab:fits_results} in Appendix \ref{Appendix:fits}). The right column of Fig.\,\ref{fig:Rbrk_z} presents the normalised probability distributions of \rbreak\ at the characteristic value of $\log_{10}(M/M_{\odot}) = 10.50$ for the local Universe sample and the objects at $0.4<z<0.6$. These histograms were calculated using the probability distributions of the intercept and slope of each fit for both the local Universe sample and our sample at $0.4<z<0.6$ and interpolating each case to the fixed reference value of $\log_{10}(M/M_{\odot}) = 10.50$. 

The main result of this analysis is that we do not find any significant differences between the probability distributions of the break radius \rbreak\ for Type-III S0 galaxies in the local Universe and their analogues at $0.4<z<0.6$ when we take into account the mass of the galaxy. We estimate that the probability of being compatible in \rbreak\ is $p=0.260$ when using the stellar mass for the comparison. We have tested whether this result could be dependent on the reference value of the mass that we have chosen. We performed the same test using $\log_{10}(M/M_{\odot}) = 10.20$ ($p=0.44$) and $10.80$ ($p=0.08$), finding a similar result: the distributions of \rbreak\ of S0 galaxies at $0.4<z<0.6$ and $z\sim0$ were statistically compatible even accounting for the mass of the galaxy. Nevertheless, as a consequence of the dispersion in the \rbreak\ versus $\log_{10}(M/M_{\odot})$ diagram and the limited size of both local and $0.4<z<0.6$ samples, we cannot rule out variations on the \rbreak\ lower than $\Delta \rbreak \sim 3$ kpc between $z\sim0.5$ and $z\sim0$ for the reference value, which corresponds to the mean range between the 0.025 and 0.975 quantiles of the distributions of \rbreak\ at the characteristic value of $\log_{10}(M/M_{\odot}) = 10.50$ (see the left panel of Fig.\,\ref{fig:Rbrk_z}).

Finally, we conclude that we do not find any significant difference between the position of \rbreak\ of local Universe Type-III S0 galaxies and those at $0.4<z<0.6$ for similar masses. This result poses strong constraints on the evolution mechanisms that rule the formation of these structures.

\begin{figure*}[tb]
\centering
\vspace{-0.3cm}
\includegraphics[width=0.40\textwidth]{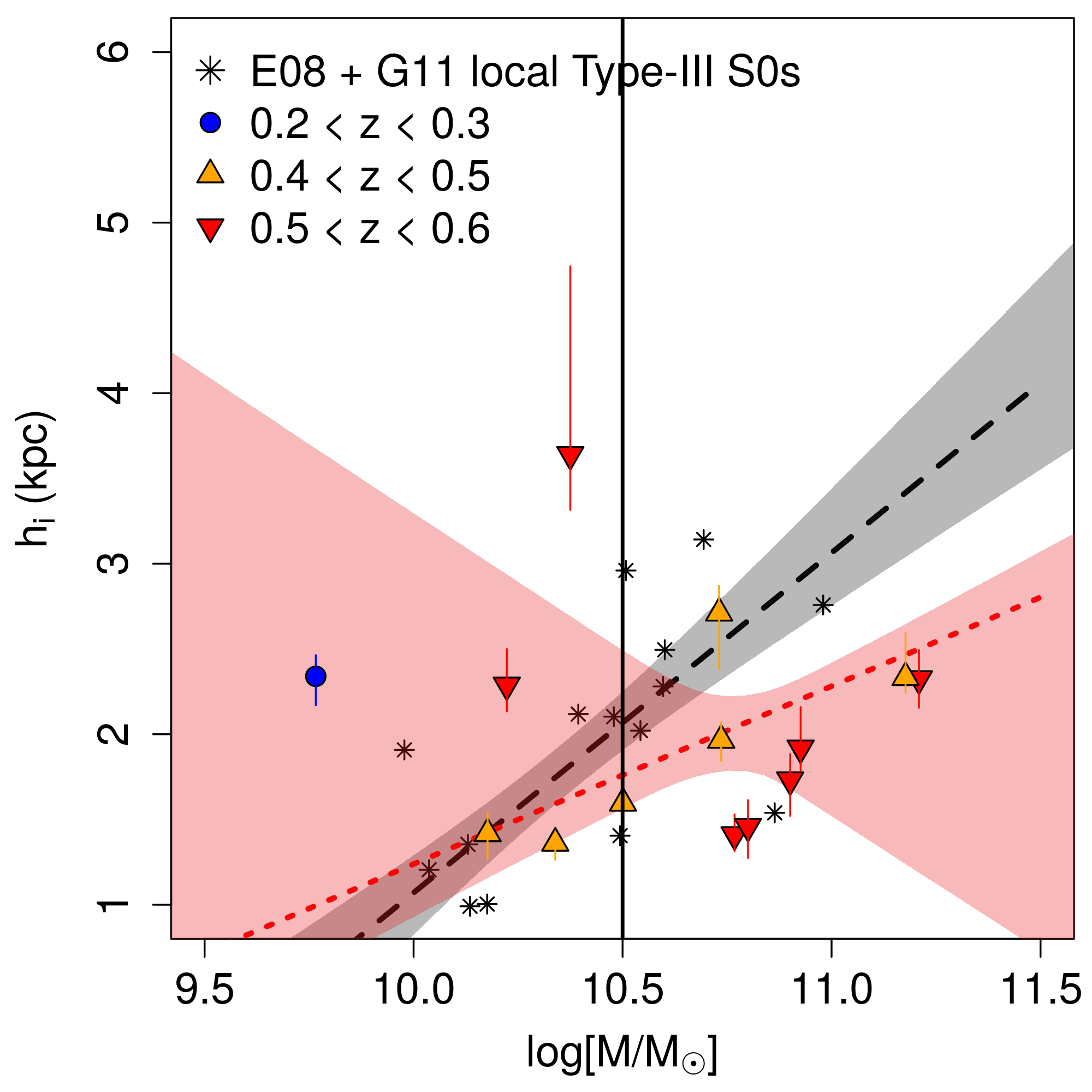}
\includegraphics[width=0.40\textwidth]{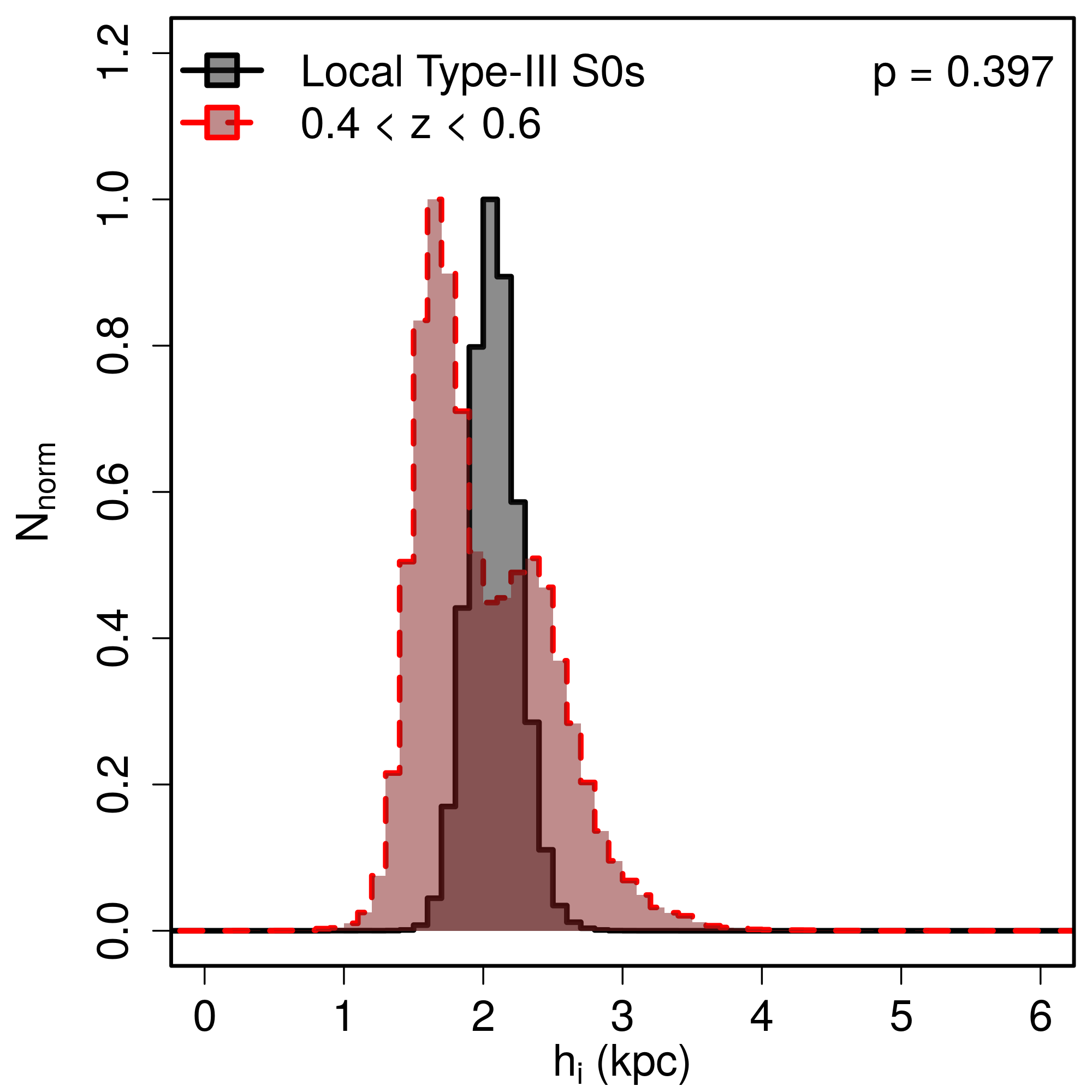}
\includegraphics[width=0.40\textwidth]{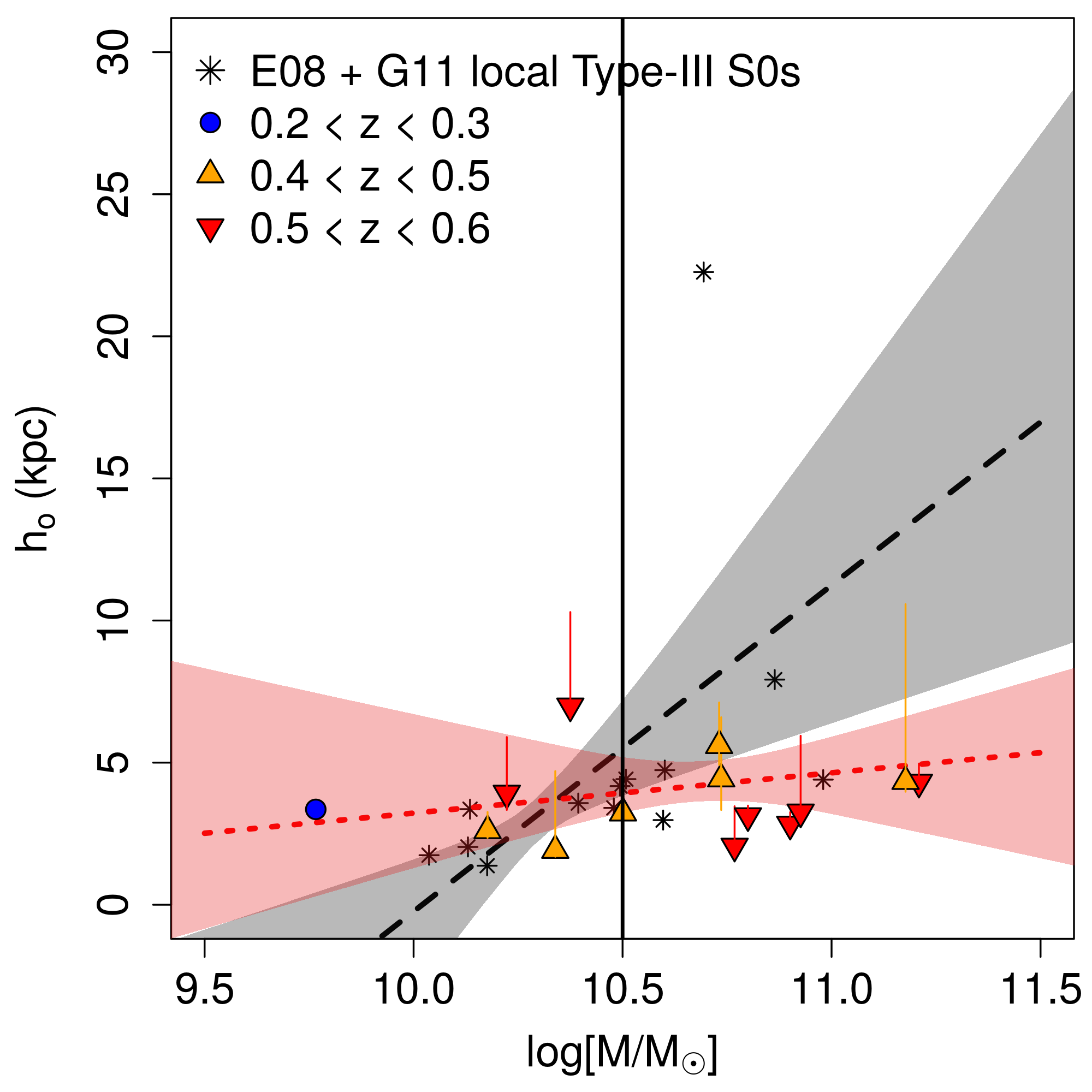}
\includegraphics[width=0.40\textwidth]{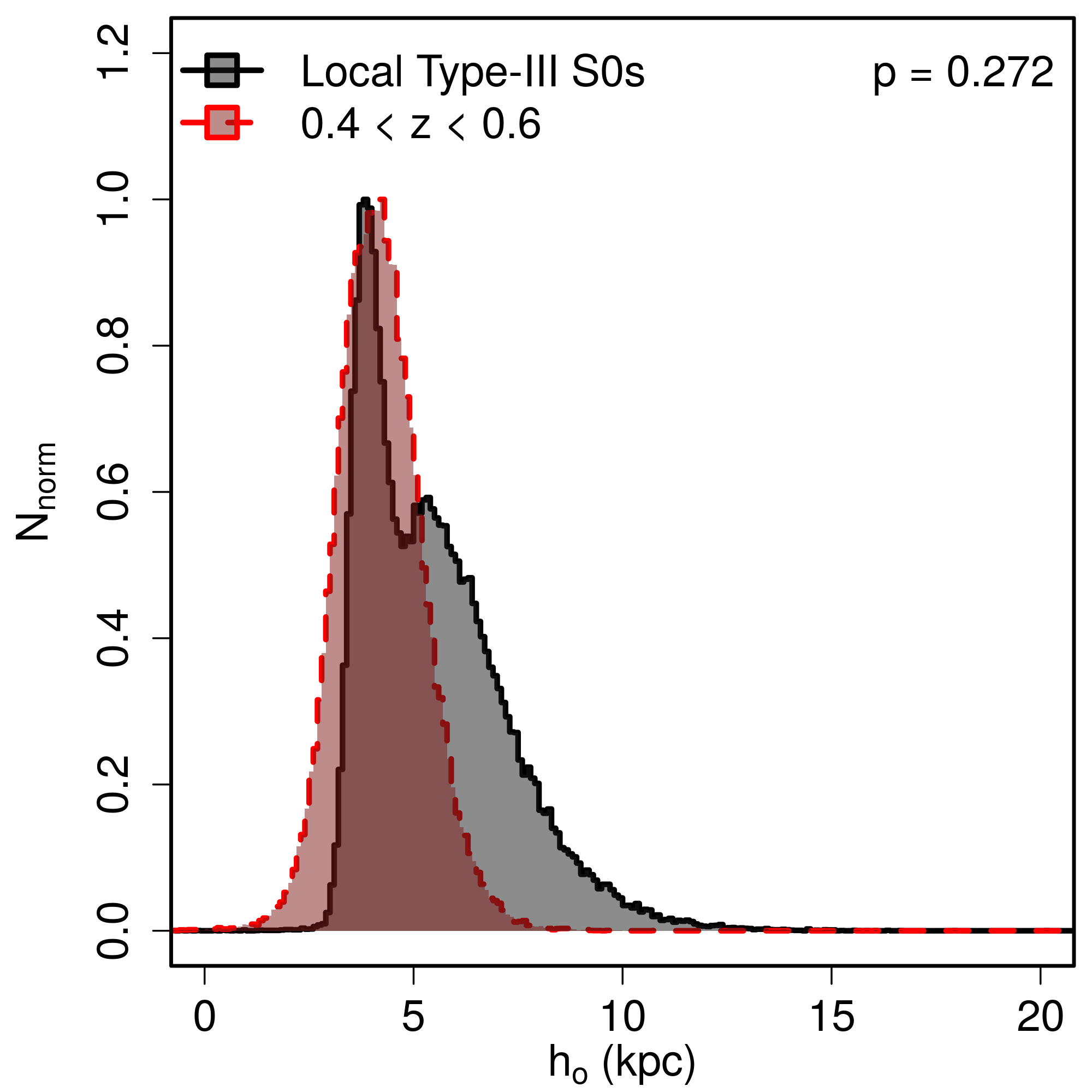}
\includegraphics[width=0.40\textwidth]{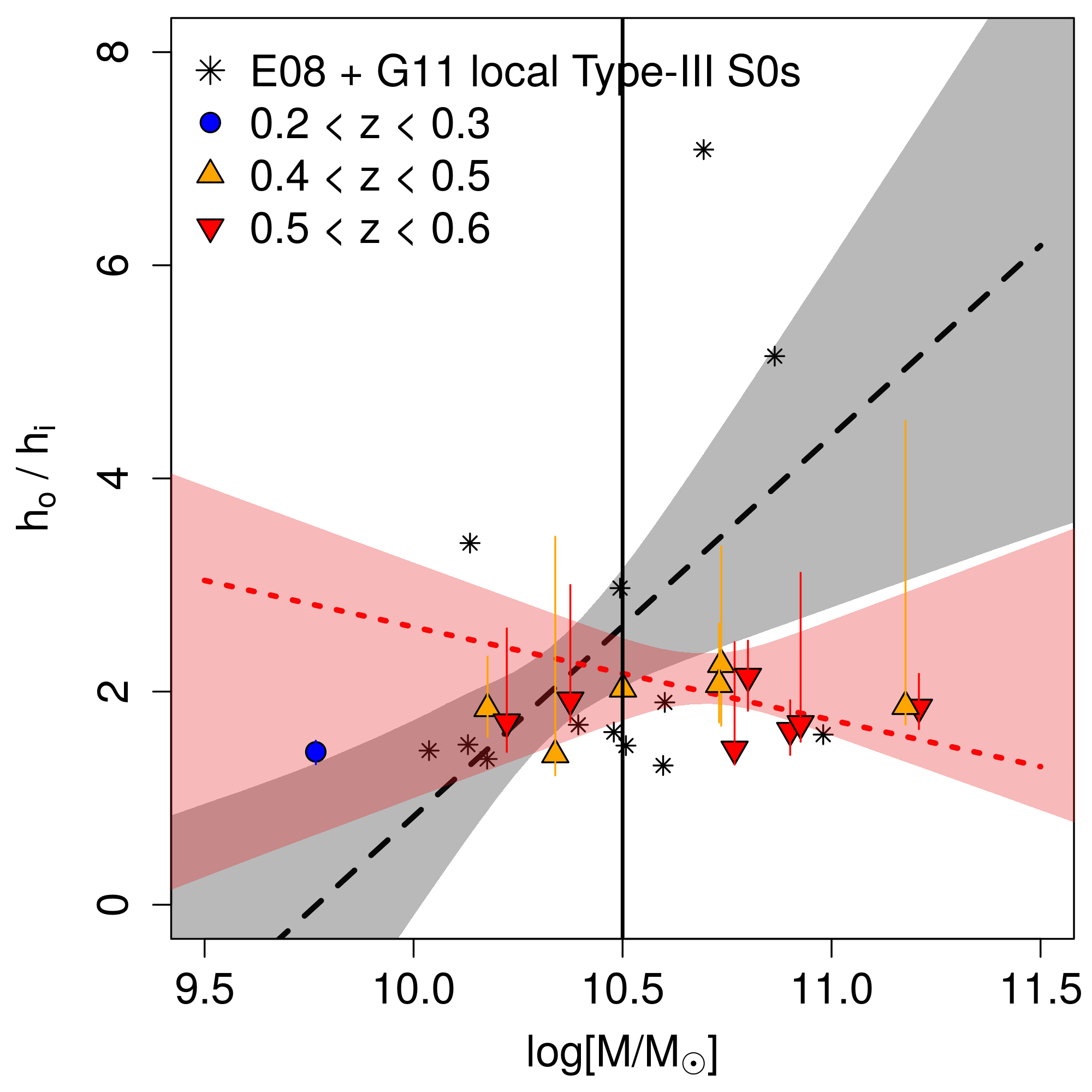}
\includegraphics[width=0.40\textwidth]{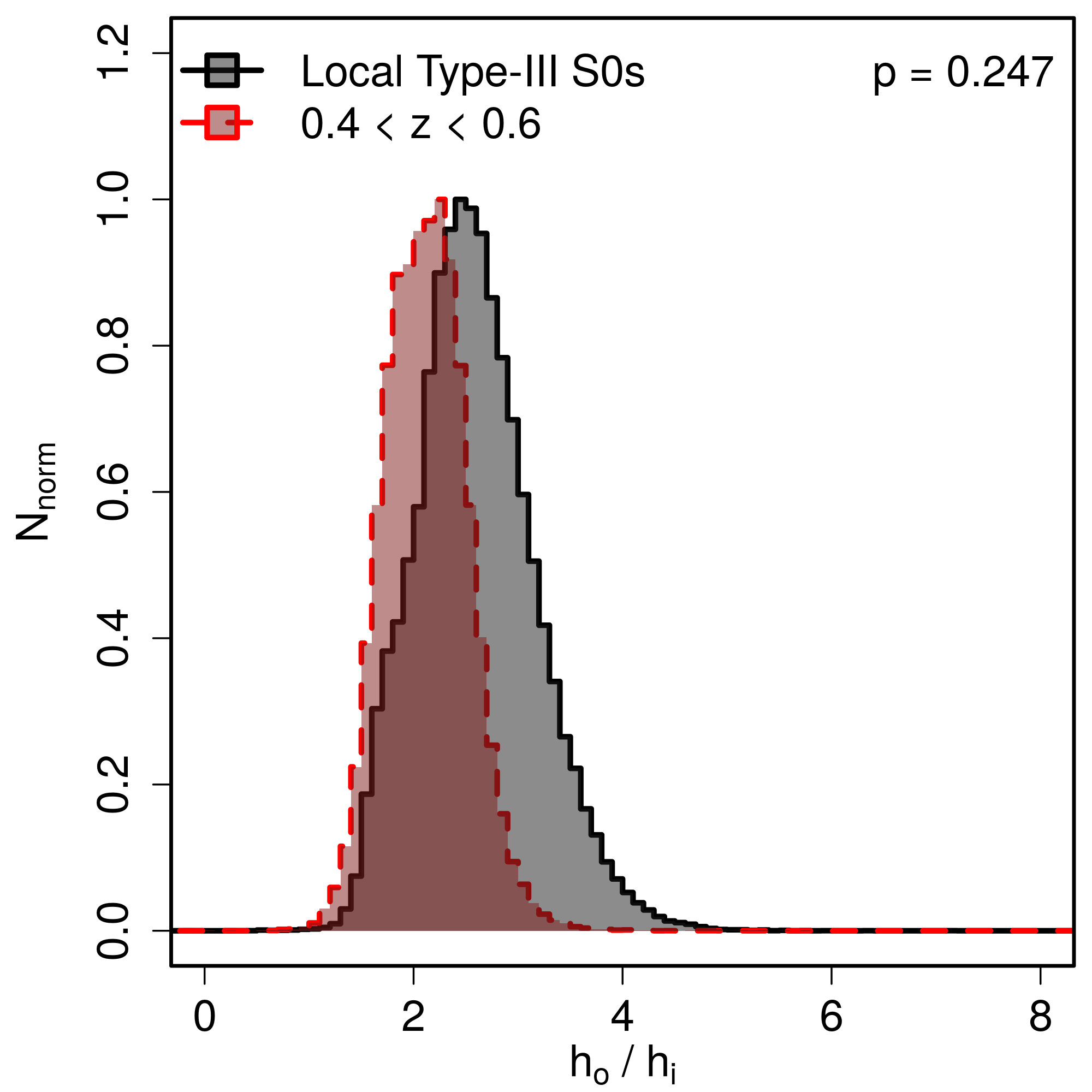}
\caption[]{Scale-lengths of the inner (\hi, top row) and outer profiles (\ho, middle row) and the ratio \ho/\hi\ (bottom row) as a function of the stellar mass. \emph{Left column:} distribution of \hi, \ho\ and \ho/\hi\ of Type-III S0 galaxies at $0.2<z<0.6$ and at $z\sim0$ with the stellar mass. The solid vertical black lines in the left panel represent the fixed value at which we compare the linear fits in the right one. \emph{Right column:} comparison of the PDDs of the scale lengths \hi, \ho\ and the ratio \ho/\hi\ at fixed stellar mass $\log_{10}(M/M_{\odot})= 10.5$. \emph{Grey histogram with black solid line:} PDDs for the Type-III S0 galaxies in the local Universe selected from E08 and G11 samples. \emph{Red histogram with dashed line}: PDDs for the Type-III S0 galaxies at $0.4<z<0.6$ from the SHARDS sample. The probabilities ($p-$value) for both samples to have compatible values at fixed stellar mass of $\log_{10}(M/M_{\odot})= 10.5$ are $p=0.397$ (\hi), $p=0.272$ (\ho) and $p=0.247$ (\ho/\hi). We refer to the legend in the panels for a key of the different lines and symbols.}
\label{fig:hohi_z}
\end{figure*}
\clearpage 

\begin{figure*}[tb]
\centering
\vspace{-0.3cm}
\includegraphics[width=0.40\textwidth]{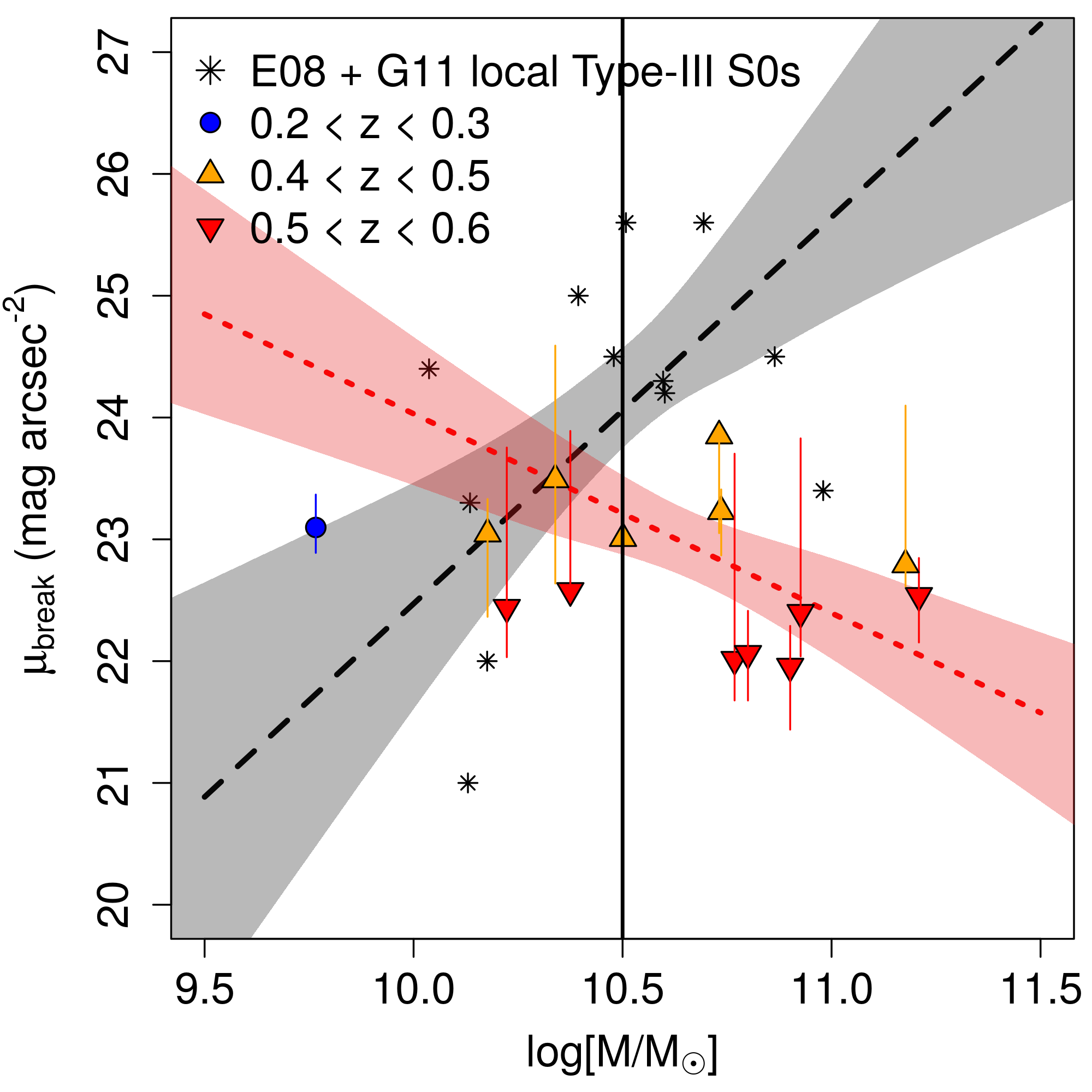}
\includegraphics[width=0.40\textwidth]{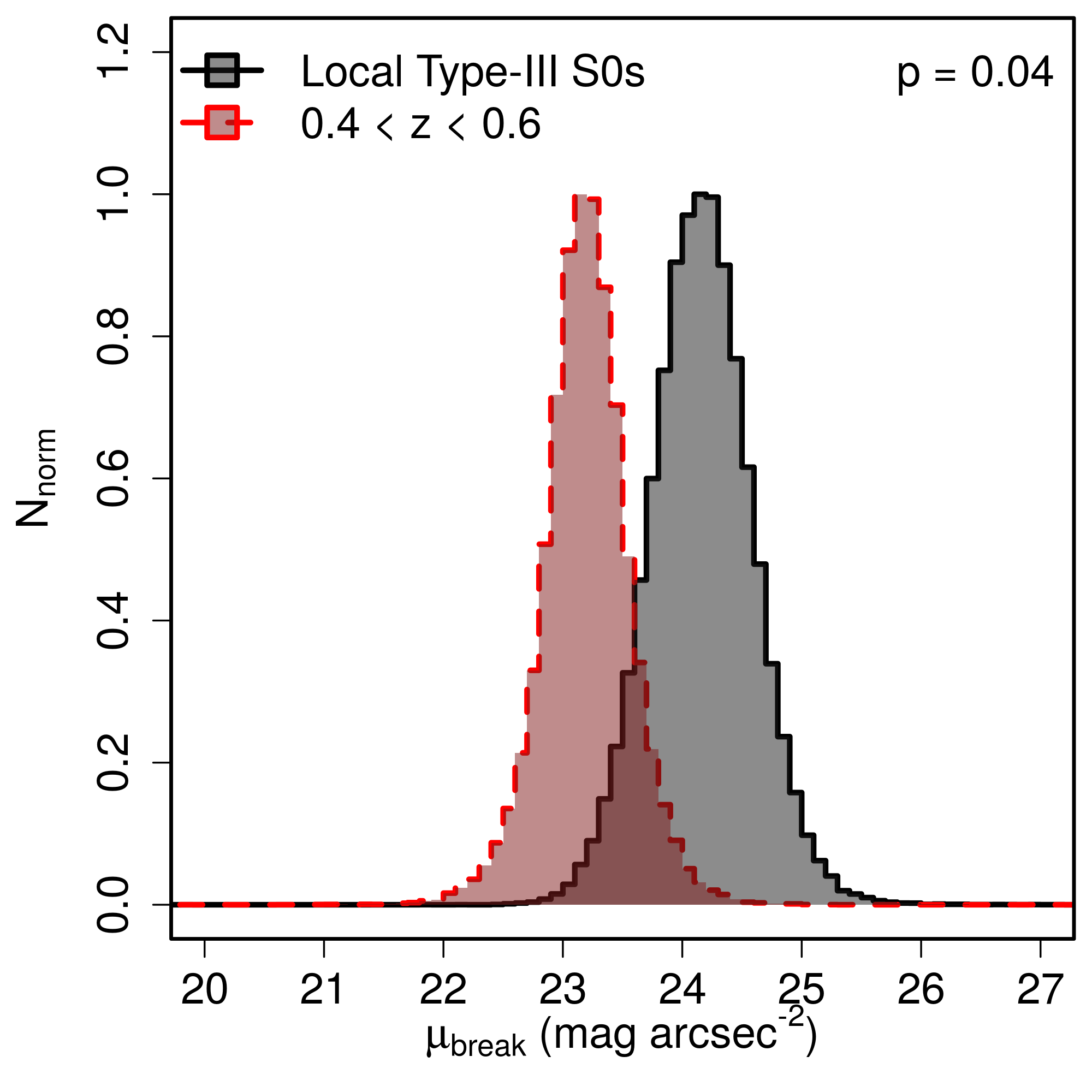}
\includegraphics[width=0.40\textwidth]{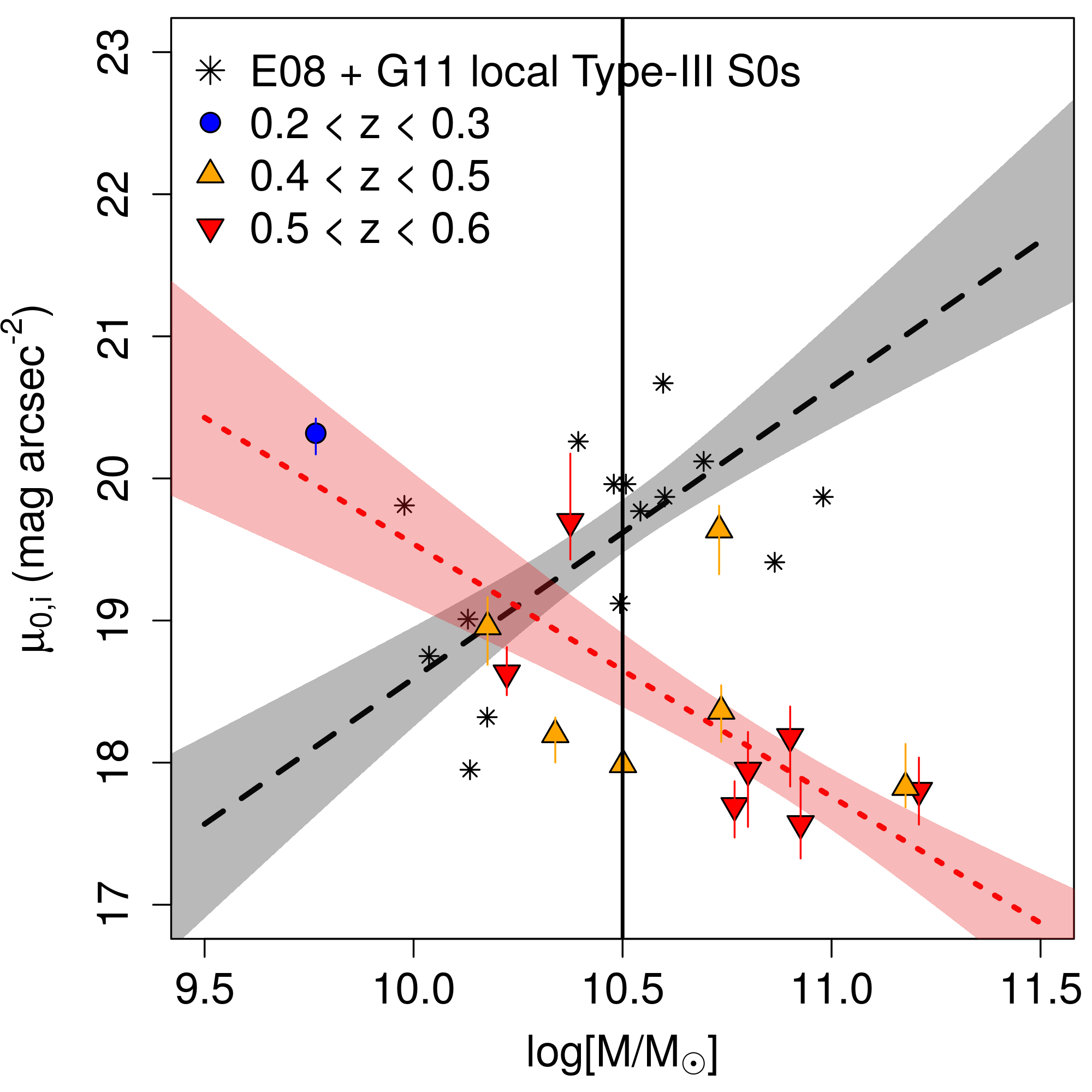}
\includegraphics[width=0.40\textwidth]{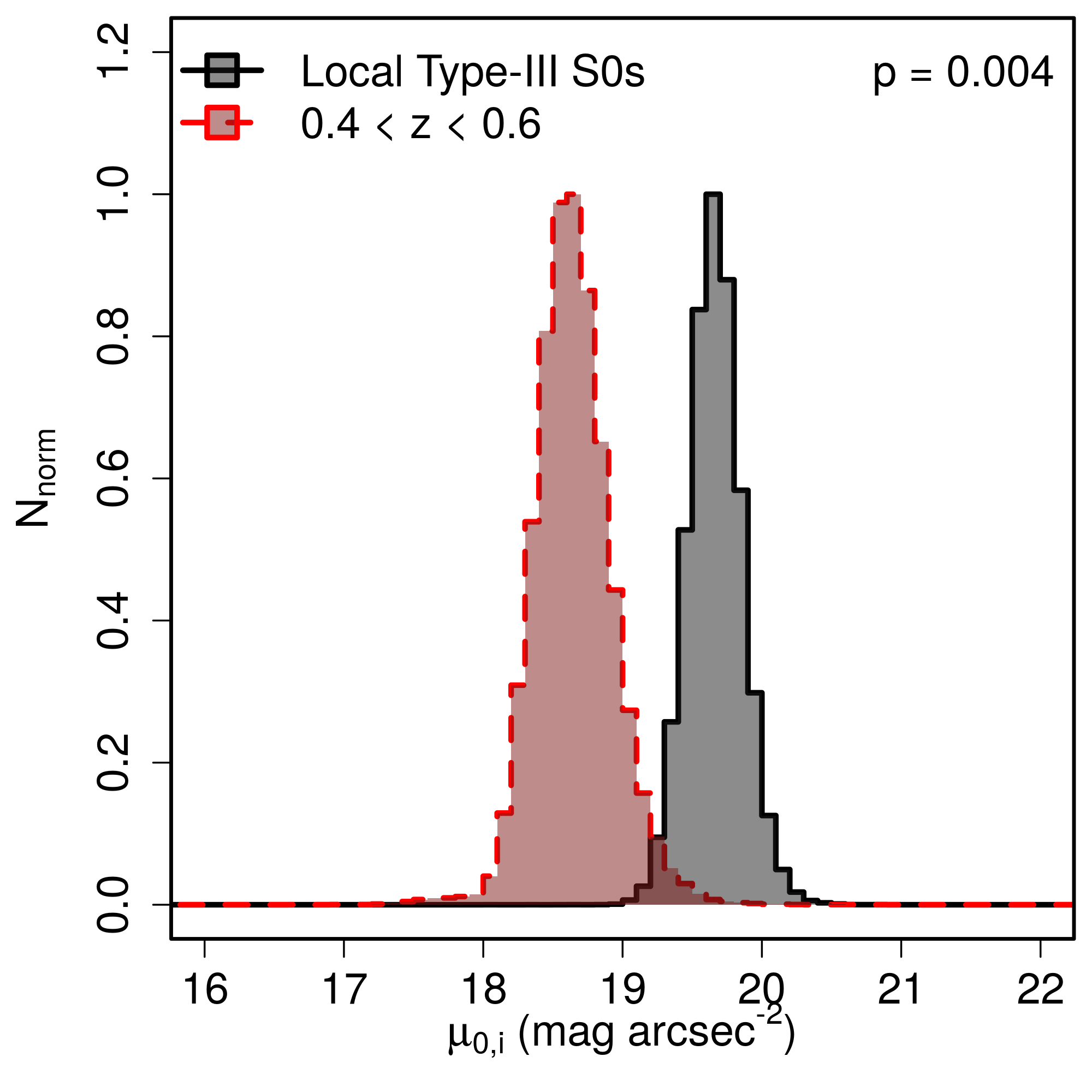}
\includegraphics[width=0.40\textwidth]{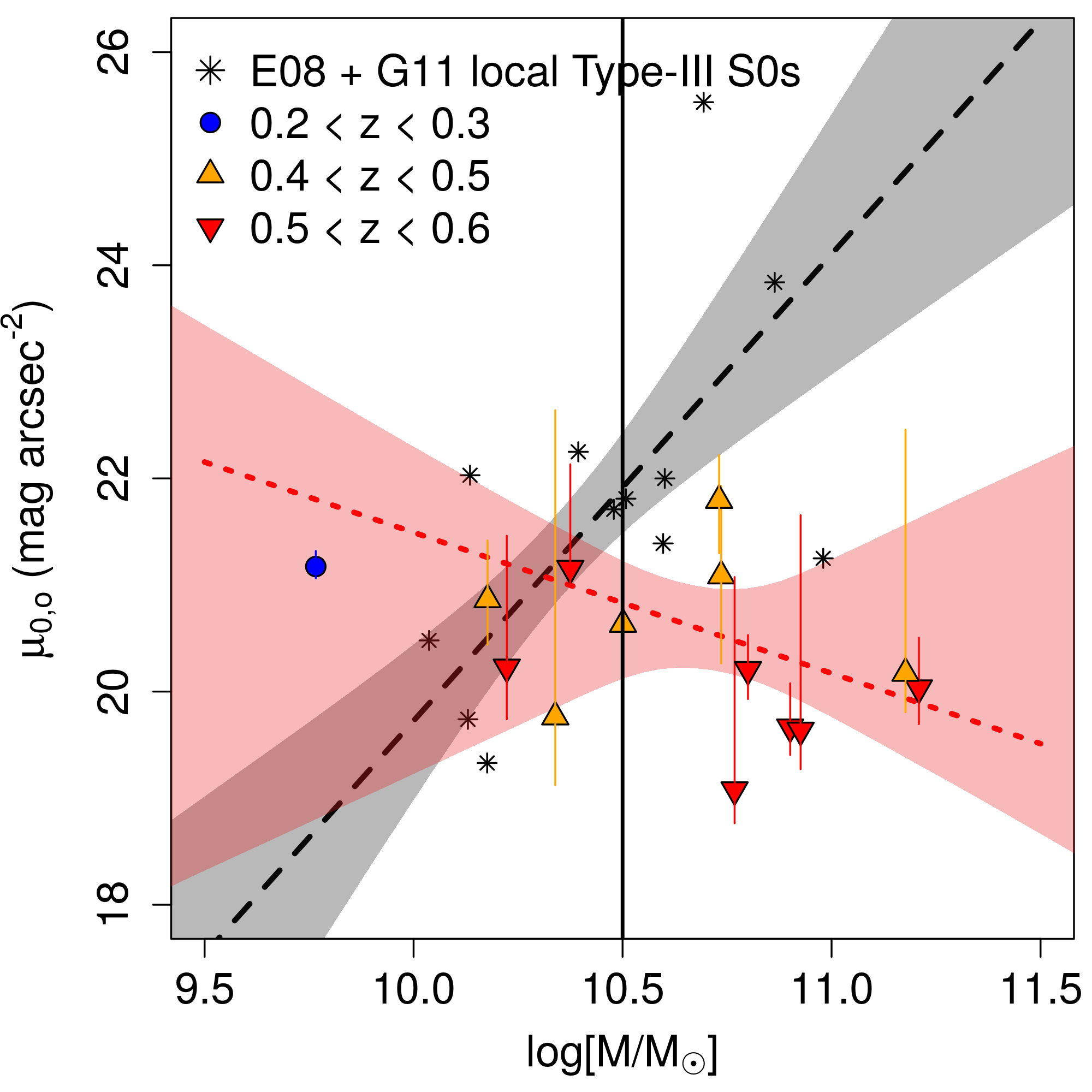}
\includegraphics[width=0.40\textwidth]{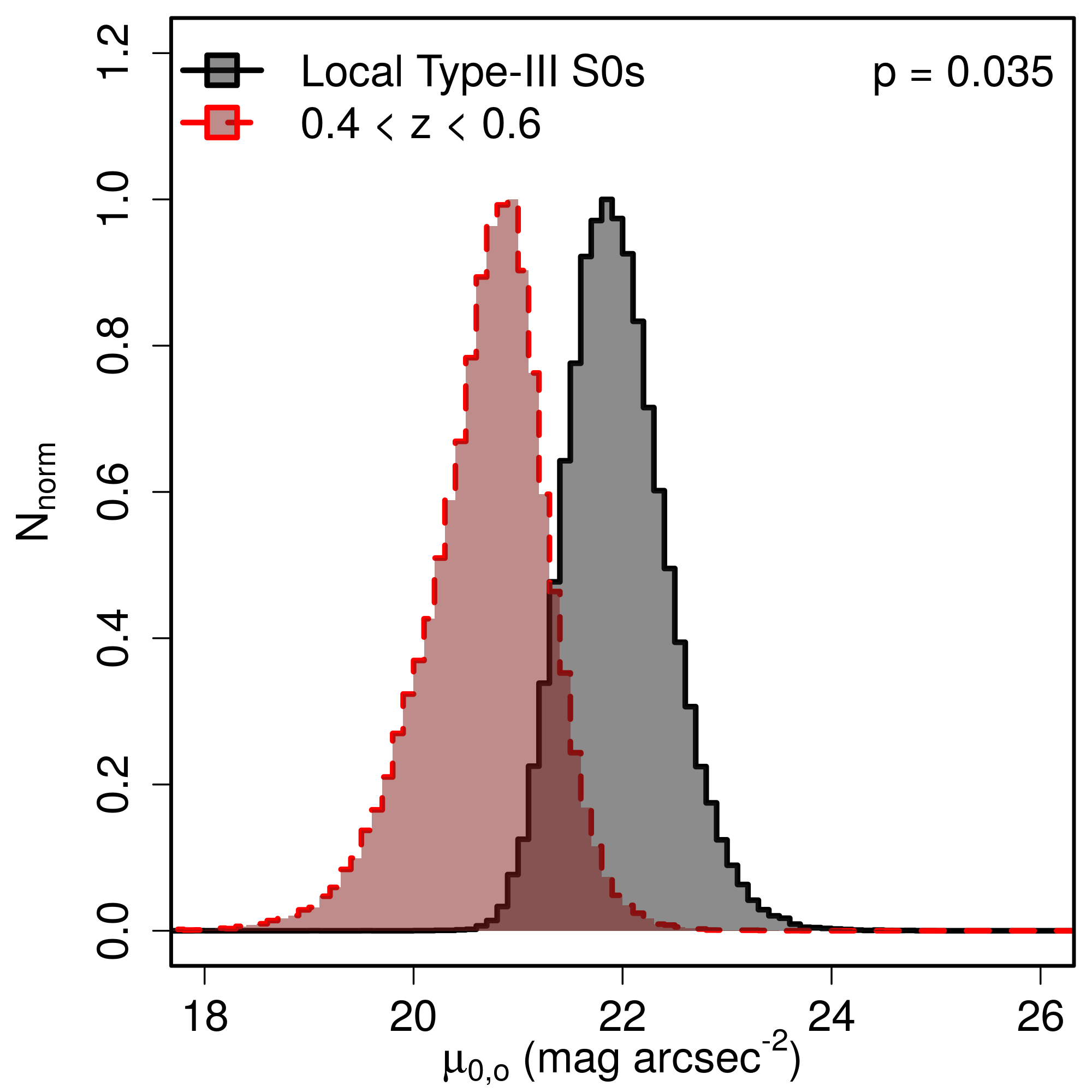}

\caption[]{Surface brightness of the break radius (\mubreak, top row) and central surface brightness of the inner (\mui, central row) and outer profiles (\muo, bottom row) as a function of the stellar mass. \emph{Left column:} distribution of \mubreak, \mui\ and \muo\ of Type-III S0 galaxies at $0.2<z<0.6$ and at $z\sim0$ with the stellar mass. The solid vertical black lines in the left panel represent the fixed value at which we compare the linear fits in the right panel. \emph{Right column:} comparison of the PDDs of \mubreak, \mui\ and \muo\ at fixed stellar mass $\log_{10}(M/M_{\odot})= 10.5$. \emph{Grey histogram with black solid line:} PDDs for the Type-III S0 galaxies in the local Universe selected from E08 and G11 samples. \emph{Red histogram with dashed line}: PDDs for the Type-III S0 galaxies at $0.4<z<0.6$ from the SHARDS sample. The probabilities ($p-$value) that both samples have compatible values at fixed stellar mass of $\log_{10}(M/M_{\odot})= 10.5$ are $p=0.04$ (\mubreak), $p=0.004$ (\mui) and $p=0.035$ (\muo). We refer to the legend in the panels for a key of the different lines and symbols.} 
\label{fig:mubrk_z}
\end{figure*}
\clearpage

\subsection{Evolution of \hi\ and \ho\ with $z$ at a fixed stellar mass}
\label{Subsec:hiho_evol}

We repeat the previous analysis in order to investigate whether or not there is any significant difference between the scale lengths of the inner and outer profiles (\hi\ and \ho) between the samples at $z\sim0.5$ and $z\sim0$ at a fixed stellar mass. In the left panels of Fig.\,\ref{fig:hohi_z}, we show the distributions of \hi, \ho\ and the ratio \ho/\hi\ as a function of the decimal logarithm of the stellar mass $\log_{10}(M/M_{\odot})$. Again, we fit a linear model to each sample in the diagrams of the left column of Fig.\,\ref{fig:hohi_z}. The results are shown in Table \ref{tab:fits_results} of Appendix \ref{Appendix:fits}. The panels of the right column present the normalised probability distributions of \hi, \ho\ and \ho/\hi\ at the characteristic value of $\log_{10}(M/M_{\odot}) = 10.50$ for the local Universe sample and the objects at $0.4<z<0.6$ obtained from the fits performed in the corresponding left panels.

As in the case of \rbreak, we do not find any significant differences between the expected values of \hi, \ho\ or the ratio \ho/\hi\ at fixed stellar mass $\log_{10}(M/M_{\odot}) = 10.50$ between the Type-III S0 galaxies at $z\sim0$ and our sample at $0.4<z<0.6$. We find that the probability that the samples are in \hi\ and \ho\ is $p=0.397$ and $p=0.272$, respectively, when using the stellar mass for the comparison. Additionally, we do not find any significant difference between the ratio \ho/\hi\ of the samples at $z\sim0$ and $0.4<z<0.6$ either ($p=0.247$), despite the large \ho/\hi\ ratios ($\ho/\hi >5$) shown by two of the Type-III S0 galaxies of the local sample (NGC3900 and NGC3998).

We conclude that we do not find any significant differences between the general structure of local Type-III S0 galaxies and those of our sample at $0.4<z<0.6$ at a fixed stellar mass of  $\log_{10}(M/M_{\odot}) = 10.50$. This result agrees well with the observed similarity of the structural scaling relations of \hi\ and \ho\ with \rbreak\ and \rbreak/\risoph\ (see Fig.\,\ref{fig:hiho_rbrkr23} on Sect.\,\ref{Subsec:Scaling_relations}).

\subsection{Evolution of \mubreak, \mui\ and \muo\ with $z$ at a fixed stellar mass}
\label{Subsec:mu_evol}

Finally, we extend the previous analysis to analyse whether or not there is any significant differences between the \mubreak, \mui\ and \muo\ values of the surface brightness profiles of the Type-III S0 galaxies from the local sample and our sample at $0.4<z<0.6$, taking into account the stellar mass. In Fig.\,\ref{fig:mubrk_z} we present the distributions of \mubreak, \mui\ and \muo\ as functions of the decimal logarithm of the stellar mass $\log_{10}(M/M_{\odot})$. We fit a linear model to each sample in the diagrams of the left column of the figure (see Table \ref{tab:fits_results} in Appendix \ref{Appendix:fits}). Again, the panels of the right column present the normalised probability distributions of \mubreak, \mui\ and \muo\ at the characteristic value of $\log_{10}(M/M_{\odot}) = 10.50$ for the local Universe sample and the objects at $0.4<z<0.6$ according to the fits performed in the left panels.

We find that the Type-III S0 galaxies from the local Universe and our sample at $0.4<z<0.6$ do not present compatible distributions in the diagrams of \mubreak, \mui\ and \muo\ versus $\log_{10}(M/M_{\odot})$ for a fixed stellar mass. Interestingly, the linear fits to the local Universe sample present opposite trends to those drawn by our sample at $0.4<z<0.6$. The distributions of the surface brightness parameters of the two samples are approximately compatible for the lower stellar mass range ($\log_{10}(M/M_{\odot}) \sim 10$) but not at higher values of the stellar mass. Nevertheless, this result has to be treated carefully, because we do not have enough objects with ($\log_{10}(M/M_{\odot}) \leq 10$) to perform any conclusive analysis. In contrast to this, we find that there is a clear separation between the distribution of both samples at the intermediate to high mass range. We quantitatively analyse this in the panels of the left column of Fig.\,\ref{fig:mubrk_z}. The Type-III S0 galaxies at $0.4<z<0.6$ present brighter values for \mubreak\ ($p=0.04$), \mui\ ($p=0.004$) and \muo\ ($p=0.035$) by $\sim1.5$ \magarc\ for a fixed value of $\log_{10}(M/M_{\odot}) = 10.5$ than their local analogues. This result is compatible with the lack of agreement between the trends followed by each sample in the diagrams of Fig.\,\ref{fig:muimuo_rbrkr23}, and suggests that the dimming observed in Fig.\,\ref{fig:MbMkMass_hist} affects similarly to the inner and outer profiles of the disc. Therefore, these galaxies have probably experienced a global dimming in the whole disc structure. 

We conclude that the surface brightness profiles from Type-III S0 galaxies at $0.4<z<0.6$ are significantly brighter than similar objects in the local Universe at a fixed stellar mass of $\log_{10}(M/M_{\odot}) = 10.5$.

\subsection{Evolution of the \mubreak\ vs. \rbreak\ relation with $z$}
\label{Subsec:mubreak_evol}

\begin{figure}[]
\vspace{-0.5cm}
\includegraphics[width=0.49\textwidth]{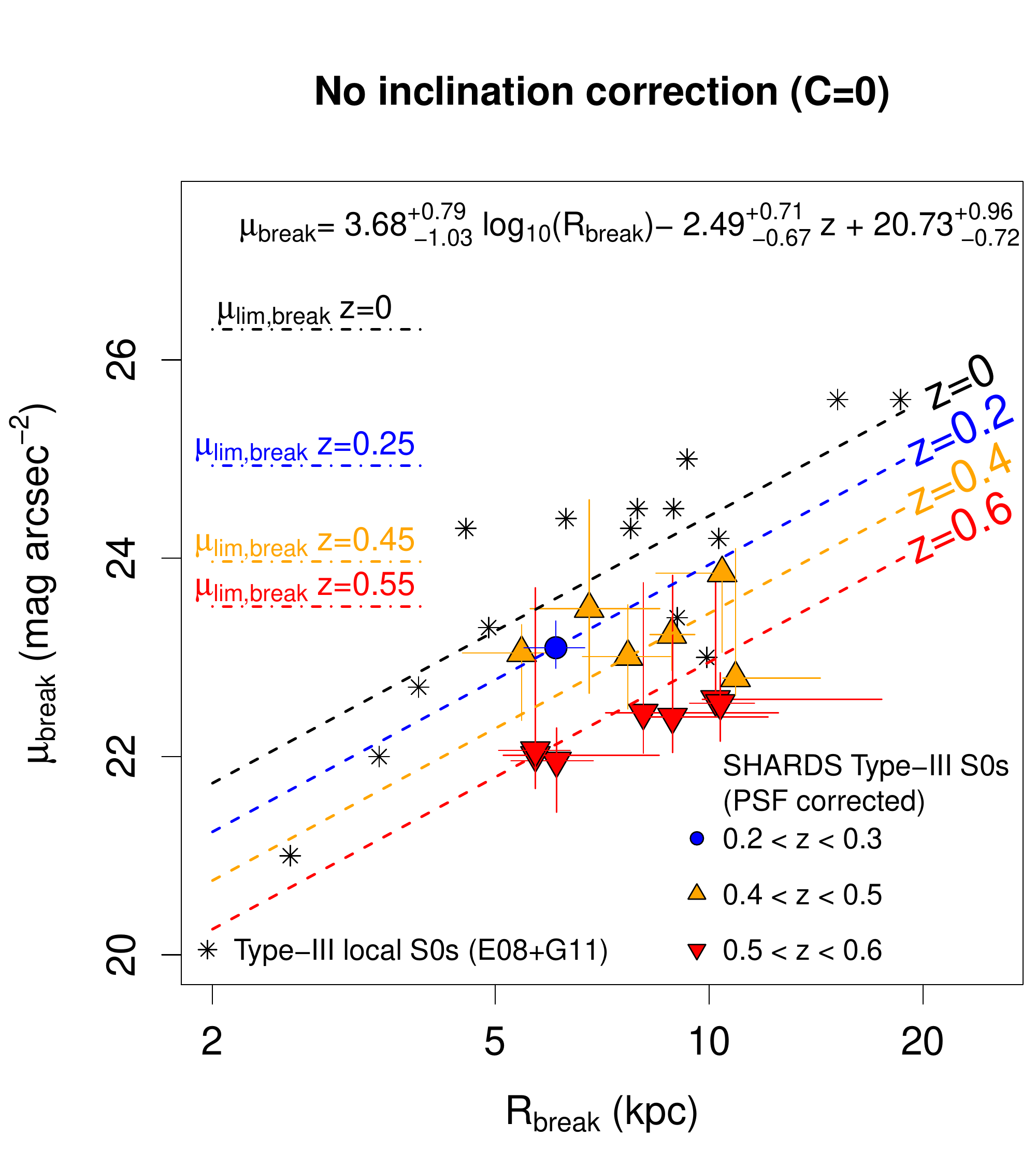}
\caption[]{Distribution of the Type-III S0 galaxies at $0.2<z<0.6$ vs. the local sample from E08 and G11 in the  \mubreak\ -- \rbreak\ diagram. The relation fitted was in logarithmic scale for the x-axis. The 3D model analysis for the \mubreak\ values without inclination correction is also shown. \emph{Dashed lines:} Contours of equal redshift for the surface brightness at the break radius (\mubreak) as a function of \rbreak\ and redshift (from Eq.\,\ref{eq:mubreak_rbreak_z}). \emph{Dashed-dotted lines:} Detection limit for Type-III profiles as a function of redshift. Consult the legend in the figure for the redshift colour classification and 3D fit results.}
\label{fig:mubrk_rbrk_z}
\end{figure}

In Fig.\,\ref{fig:mubrk_rbrk_z} we present \mubreak\ as a function \rbreak\ for the local sample and our sample at $0.2<z<0.6$. Their distributions show a clear separation between the samples at different redshifts. This goes in the sense that objects at higher $z$ tend to present brighter \mubreak\ values than those at lower $z$. In B14, the authors reported a strong correlation between \mubreak\ and \rbreak\ for local Type-III S0 galaxies. In this study, we analyse if there is a significant trend with $z$ embedded into the \mubreak\ vs. \rbreak\ relation. 

In the left panel of Fig.\,\ref{fig:mubrk_rbrk_z} we compare the trend followed by Type-III S0s at $z\sim0$ and $0.4<z<0.6$ by fitting the following 3D model to the data:
\begin{equation} \label{eq:mubreak_rbreak_z}
\mubreak = 3.68^{+0.79}_{-1.03}\cdot \log_{10}(\rbreak) - 2.49^{+0.71}_{-0.67} \cdot\ z + 20.73^{+0.96}_{-0.72},
\end{equation}

\begin{figure*}[]
\vspace{-0.5cm}
\includegraphics[width=0.49\textwidth]{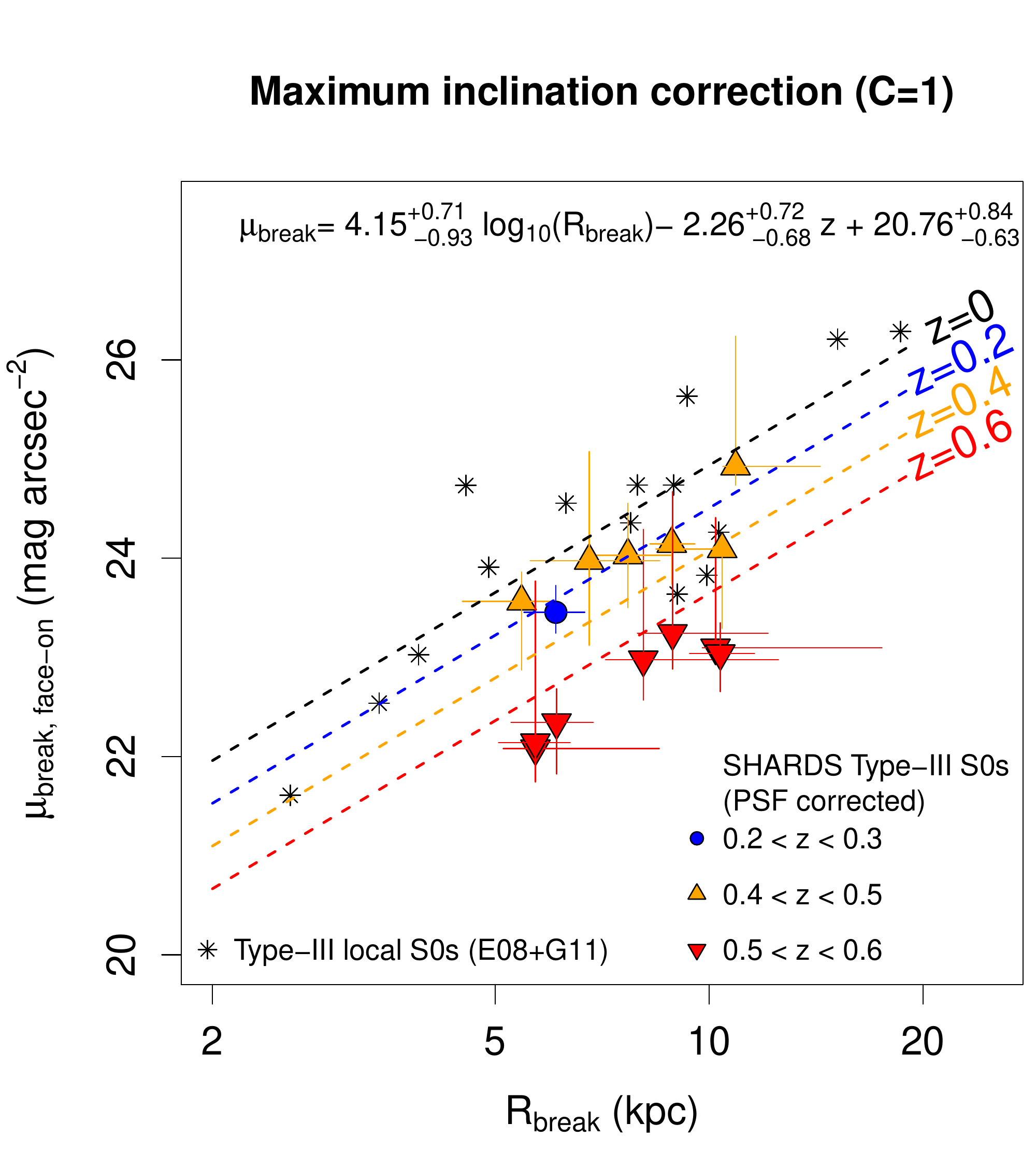}
\includegraphics[width=0.49\textwidth]{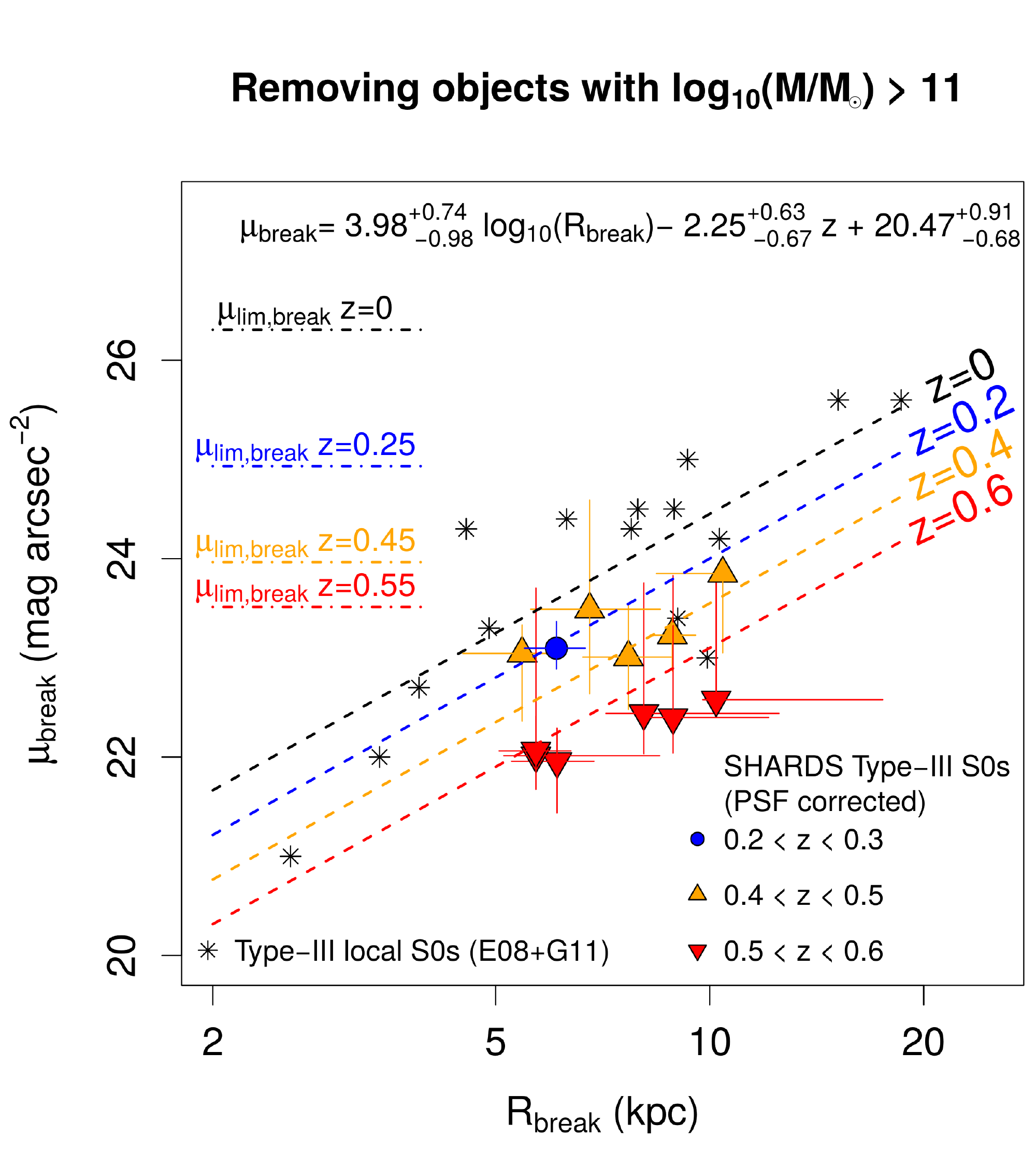}

\caption[]{Distribution of the Type-III S0 galaxies at $0.2<z<0.6$ vs. the local sample from E08 and G11 in the  \mubreak\ -- \rbreak\ diagram. The relation fitted was in logarithmic scale for the x-axis. \emph{Left panel:} Diagram and 2D model analysis for the \mubreak\ values with maximum inclination correction ($C=1$, optically transparent). \emph{Right panel:} Diagram and 2D model analysis for the \mubreak\ values without inclination correction, removing the objects with stellar masses larger than the local sample (SHARDS10009610 and SHARDS20000827). \emph{Dashed lines:} Contours of equal redshift for the surface brightness at the break radius (\mubreak) as a function of \rbreak\ and redshift (from Eq.\,\ref{eq:mubreak_rbreak_z}). \emph{Dashed-dotted lines:} Detection limit for Type-III profiles as a function of redshift. Consult the legend in the figure for the redshift colour classification and 3D fit results.}
\label{fig:mubrk_rbrk_z_effects}
\end{figure*}

\begin{figure*}[]
 \begin{center}
 \vspace{-0.25cm}
\includegraphics[width=0.45\textwidth]{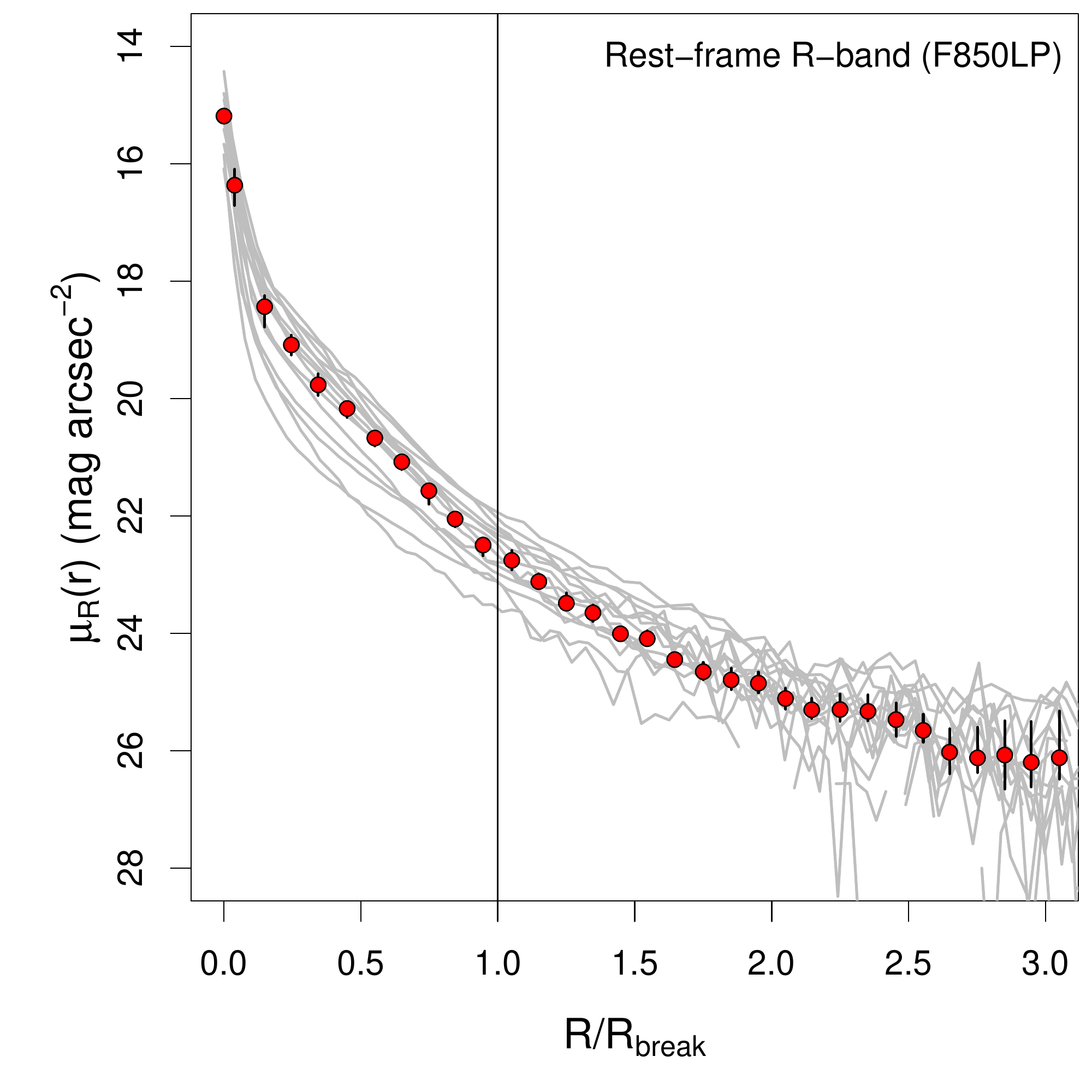}
\includegraphics[width=0.45\textwidth]{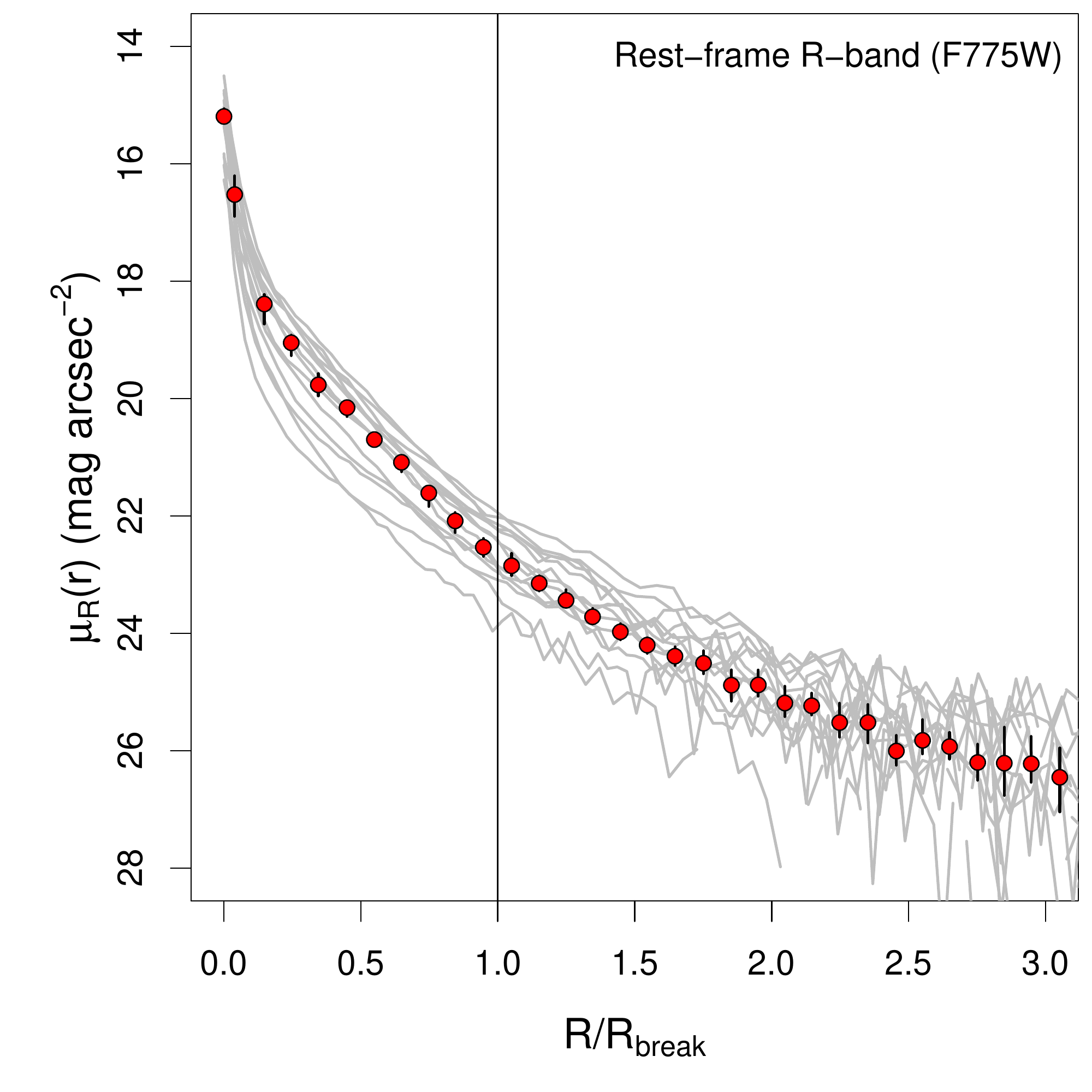}
\includegraphics[width=0.45\textwidth]{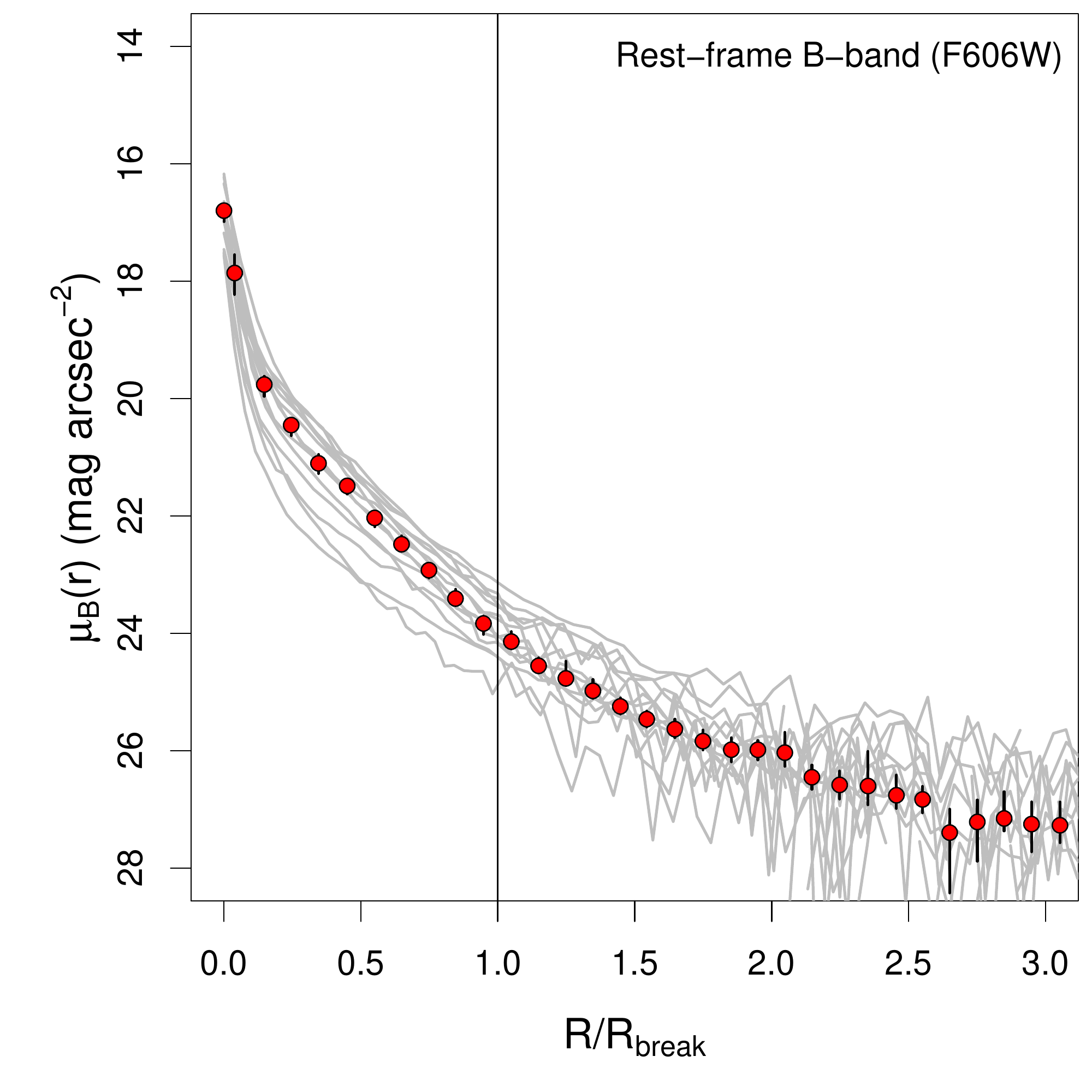}
\includegraphics[width=0.45\textwidth]{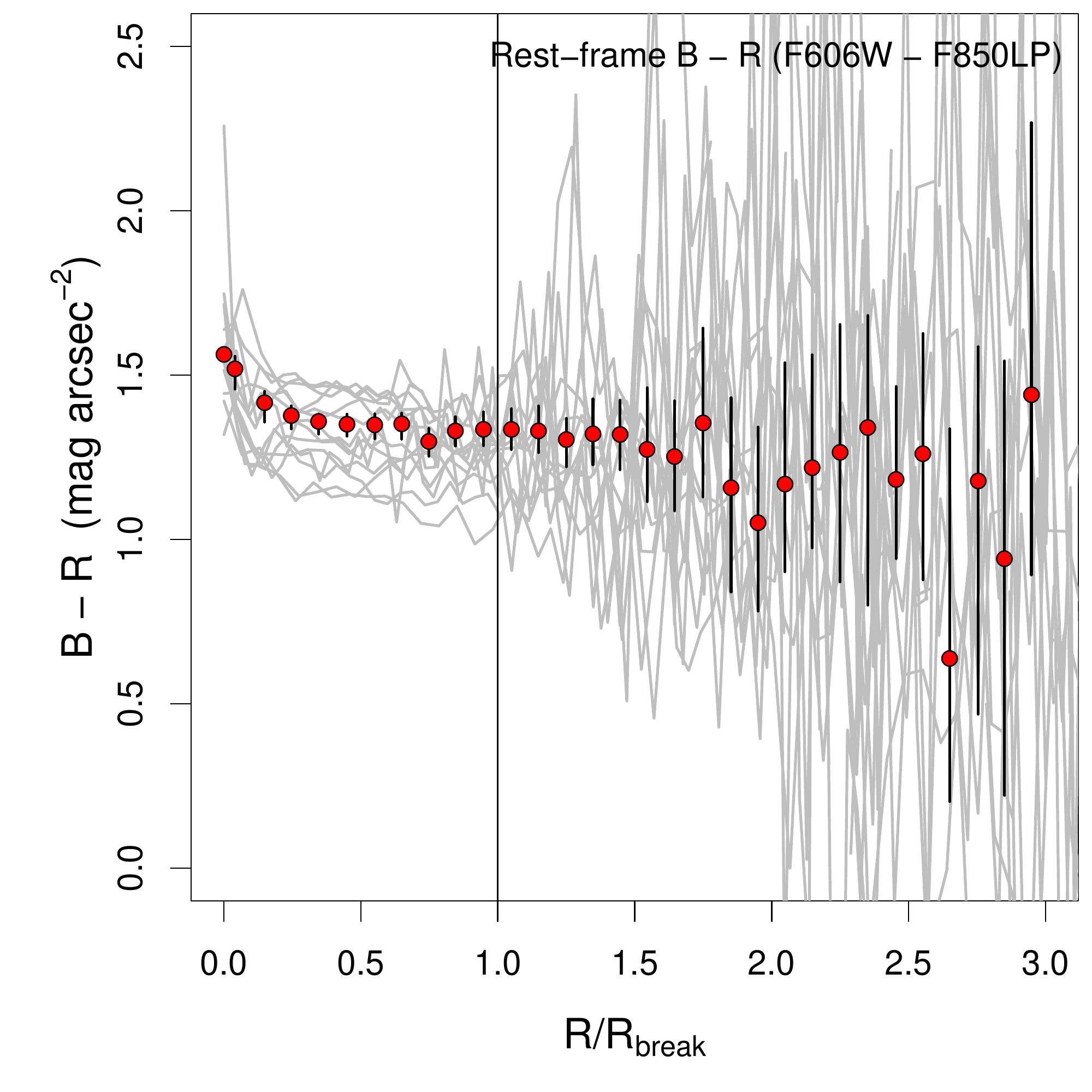}
 \vspace{-0.25cm}

\caption[]{Median profiles for the sample of Type-III S0--E/S0 galaxies at $0.2<z<0.6$, as a function of the normalised radius ($R/\rbreak$). \emph{Top left panel:} median surface brightness profile in the $R$ band, obtained from the F850LP band of HST/ACS. \emph{Top right panel:} median surface brightness profile in the $R$ band, obtained from the F775W band of HST/ACS. \emph{Bottom left panel:} median surface brightness profile on the $B$ band, obtained from the F606W band of HST/ACS. \emph{Bottom right panel:} median $(B-R)$ colour profile. The vertical solid lines represent the break radius for all the profiles ($R/\rbreak=1$). Red dots represent the median profiles in each case. The grey lines represent the individual profiles for each object.}
\label{fig:median_profiles}
 \vspace{-0.5cm}

\end{center}
\end{figure*}

We have used the least-squares method to minimise the 3D model, using Bootstrapping and Monte Carlo simulations to estimate the uncertainties of the final fit. Additionally, we use these simulations to estimate the probability for each coefficient of being equal to zero. The coefficients for \rbreak\ and $z$ have significance levels of $p=9.9\cdot10^{-4}$ and $p=4.4\cdot10^{-4}$, respectively. Therefore, there is a statistically significant evolution in surface brightness between the local and $0.2<z<0.6$ samples, confirming the results in Fig.\,\ref{fig:mubrk_z}. This fact would explain why the photometric relations shown in Fig.\,\ref{fig:muimuo_rbrkr23} present a displacement towards brighter magnitudes with respect to those of the local Universe sample. Additionally, note that a global dimming of the Type-III S0 galaxies since $z\sim0.6$ does not affect the structural relations at different redshifts (see Sect.\,\ref{Subsec:Scaling_relations} and Fig.\,\ref{fig:hiho_rbrkr23}).

\subsubsection{Effect of the limiting magnitude on the measured global dimming}
\label{subsubsec:limitingmagnitude}

We studied the possibility that the variation in the limiting magnitude as a function of $z$ may systematically bias the \mubreak\ versus $z$ distribution towards brighter values with increasing redshift, creating an apparent correlation. The median limiting magnitude for the F775W profiles is $\mu_{\mathrm{F775W,lim}} = 27.092^{+0.024}_{-0.032}$ \magarc\ (see Sect.\,2.4 on Paper I). Taking into account the K-correction, the Milky Way dust extinction and the cosmological dimming (Sect.\,2.5, Paper I), the limiting magnitude in the rest-frame $R-$band can be approximated as:

\begin{equation}
\label{eq:limit_magnitude_rest_frame}
\mu_{\mathrm{R,lim}} = 27.432^{+0.041}_{-0.053} - 1.619^{+0.065}_{-0.072} \cdot\ z - 10\cdot\log_{10}(1+z).
\end{equation}

In addition to this, we have to take into account that in order to properly detect a break we need to adequately detect the outer profile. We estimated how many magnitudes brighter a typical Type-III profile must be than the limiting magnitude in order to be detected. We created a simulated surface brightness profile using the median values from the sample of local Type-III S0 galaxies from E08 and G11 (see Table 2 in Paper I). We modified the limiting magnitude from 27 to 24 \magarc, analysing with {\tt{Elbow}} the probability in each profile of detecting the Type-III break at a certain limiting magnitude. We find that in order to detect a Type-III break at a 99\% confidence level, the surface brightness at the break radius (\mubreak) must be at least $1.1\pm0.5$ \magarc\ brighter than the limiting magnitude of the surface brightness profile. In Fig.\,\ref{fig:mubrk_rbrk_z} we represent with dashed-dotted horizontal segments the limiting magnitude at which we would detect a Type-III profile according to the limiting magnitude of our images and the results from the simulations with {\tt{Elbow}} ($\mu_{\mathrm{lim,break}} = \mu_{\mathrm{lim}} - 1.1$). The results show that the Type-III breaks appear to be at least one magnitude brighter than the theoretical limit where we would be able to detect them. This difference is more noticeable for the objects at higher redshift ($0.5<z<0.6$). Therefore, our results concerning the global dimming by $\sim1.5$ \magarc\ of the whole disc structure of Type-III S0s since $z\sim0.6$ are not affected by the limiting magnitude of the data. 

\subsubsection{Effect of the galaxy inclination on the measured global dimming}
\label{subsubsec:galaxyinclination}

We tested whether or not the differences between the surface brightness at the break radius \mubreak\ for the sample of Type-III S0 galaxies at $z=0$ and our sample at $0.2<z<0.6$ can be explained as an effect of inclination. \citet{2001MNRAS.326..543G} demonstrated that high-inclination galaxies can present central surface brightness values that are significantly greater in the $K-$band than those of low-inclination galaxies. If the galaxy system is optically thin (transparent) the total magnitude should be the same, regardless of the apparent inclination. Nevertheless, for a disc-like morphology, the projected area decreases with the inclination angle to the observer. Therefore, the surface brightness should increase with the inclination angle. The author presented the following inclination correction model:

\begin{equation} \label{eq:inclination_correction}
\mu_{face-on} = \mu_{incl} - 2.5\cdot C\cdot\log_{10}(b/a),
\end{equation}

\noindent where $\mu_{face-on}$ is the inclination-corrected surface brightness, $\mu_{incl}$ is the measured surface brightness, and $C$ is the optical depth of the system ($C=1$ for an optically thin galaxy, $C=0$ for an optically thick galaxy). Lenticular galaxies can be approximated as optically thin systems, because of their lower levels of gas and intergalactic dust compared to those of spiral galaxies. However, dust lanes have been frequently observed in nearby S0 galaxies \citep{2010A&A...519A..40A,2010MNRAS.407.2475F}. As a consequence of this, assuming a uniform and thick optical depth would be an oversimplification of reality, and a blind application of this type of correction may introduce additional uncertainties in the final results. Thus, in order to estimate to what point our results can be affected by the inclination effects, we assume the worst scenario by applying the maximum inclination correction ($C=1$, see left panel of Fig.\,\ref{fig:mubrk_rbrk_z_effects}) and compare the results with the non-inclination-corrected \mubreak\ versus \rbreak\ diagram in Fig.\,\ref{fig:mubrk_rbrk_z}. 

We measured the inclination from our Type-III S0 galaxies using {\tt{GALFIT3.0}}, during the 2D modelling step needed for the PSF-correction. In their papers, E08 and G11 provide the inclination angles for their sample. We transform from inclination angles to axis ratios using the following expression: 

\begin{equation} \label{eq:inclination_to_q}
i = \cos ^{ - 1} \Bigg(\sqrt{\cfrac{q^2 - q_0^2}{1-q_0^2}}\Bigg),
\end{equation}

\noindent where $q$ is the minor-to-major axis ratio (b/a) and $q_{0}$ is the intrinsic axis ratio. We assume that $q_{0} = 0.25$, which is the median value expected for S0 galaxies following the results from \citet{2014MNRAS.444.3340W}. In the left panel of Fig.\,\ref{fig:mubrk_rbrk_z_effects} we analyse the variation of \mubreak\ after inclination correction with \rbreak\ and $z$. Again, we fit the distribution to a three-dimensional (3D) model, obtaining:
\begin{equation} \label{eq:mubreak_rbreak_z_incl_corr}
\mubreak = 4.15^{+0.71}_{-0.93}\cdot \log_{10}(\rbreak) - 2.26^{+0.72}_{-0.68} \cdot\ z + 20.76^{+0.84}_{-0.63}.
\end{equation}

We find that the coefficients for \rbreak\ and $z$ have a significance level of $p=6.4\cdot10^{-4}$ and $p=4.1\cdot10^{-3}$ respectively. The $p$-value of the dependence with $z$ is higher than in the case of no inclination correction, therefore, this means that the differences between both samples are partially reduced if we assume the maximum inclination correction, but it is not enough to fully explain the observed trend of \mubreak\ and \rbreak\ with $z$, as observed in the right panel of Fig.\,\ref{fig:mubrk_rbrk_z}. The 3D model predicts a dimming in brightness of
$\Delta\mubreak \sim-1.32^{-0.42}_{+0.45}$ \magarc\ between $z=0.6$ and $z=0$ if we take into account the maximum inclination correction (left panel, Fig.\,\ref{fig:mubrk_rbrk_z_effects}) against the $\Delta\mubreak\sim-1.49^{-0.37}_{+0.39}$ \magarc\ measured dimming of the \mubreak\ without inclination corrections (Fig.\,\ref{fig:mubrk_rbrk_z}).

\subsubsection{Effect of the objects with higher stellar masses on the measured global dimming}
\label{subsubsec:stellarmassesondimming}

Finally, we tested if the observed variation of \mubreak\ with $z$ could be a bias effect created by the objects from the $0.2<z<0.6$ sample with higher masses. We removed SHARDS10009610 ($\log_{10}(M/M_{\odot}) = 11.209$, $z = 0.557$) and
SHARDS20000827 ($\log_{10}(M/M_{\odot}) = 11.177$, $z = 0.409$) from the $0.2<z<0.6$ sample. These are the only two objects that present larger stellar masses than the most massive object from the local sample (NGC4459, $\log_{10}(M/M_{\odot}) = 10.98$). We then repeated the 3D fit with this sub-sample. The new \mubreak\ versus \rbreak\ diagram is presented in the right panel of Fig.\,\ref{fig:mubrk_rbrk_z_effects}. We obtained the following results from the 3D fit analysis:

\begin{equation} \label{eq:mubreak_rbreak_z_mass_cut}
\mubreak = 3.98^{+0.74}_{-0.98}\cdot \log_{10}(\rbreak) - 2.25^{+0.63}_{-0.67} \cdot\ z + 20.47^{+0.91}_{-0.68}.
\end{equation}

We find that the coefficients for \rbreak\ and $z$ have a significance level of $p=1.2\cdot10^{-3}$ and $p=1.8\cdot10^{-3}$, respectively. In a similar way as done with the inclination correction test, both $p$-values of the dependence with $z$ and \rbreak\ are higher than in the full sample case without inclination correction, but still significant. The 3D model predicts a dimming in brightness of
$\Delta\mubreak \sim-1.35^{-0.37}_{+0.40}$ \magarc\ between $z=0.6$ and $z=0$, which is lower than in the case with the full sample but compatible within errors. Therefore, the presence of objects with higher stellar masses is not enough to fully explain the observed trend of \mubreak\ and \rbreak\ with $z$. We conclude that there is a significative dimming trend in the \mubreak\ versus \rbreak\ diagram with $z$, independently of inclination effects and the limiting magnitude of the data. Considering that there is no significant evolution in \rbreak\ between $z=0.6$ and $z=0$ (see Sect.\,\ref{Subsec:rbreak_evol}), Type-III S0 galaxies in the local Universe are $\sim1.5$ \magarc\ dimmer than analogous objects at $z=0.6$.

In addition to this, the central surface brightnesses \mui\ and \muo\ of Type-III S0s at $z\sim0.5$ are also brighter by $\sim 1.5$ \magarc\ than their local analogues (see Figs.\,\ref{fig:muimuo_rbrkr23} and \ref{fig:mubrk_z}), while they present similar \hi\ and \ho\ values (see Figs.\,\ref{fig:muimuo_rbrkr23} and \ref{fig:mubrk_z}). Therefore, the dimming seems to affect the whole disc structure of these Type-III S0s (i.e., both the inner and outer profiles). Nevertheless, we note that our sample size is small, and therefore these results must be interpreted with care. Deeper and more complete observations are necessary to confirm this evolution in brightness for Type-III S0 galaxies since $z=0.6$. We analyse the possible consequences of a general dimming of 1.5 \magarc\ of the surface brightness profiles of these galaxies between $z=0.6$ and $z=0$ in Sect.\,\ref{Subsec:stellar_pop_analysis}. 

\subsection{Colour profiles}
\label{Subsec:colour_profiles}
In Fig.\,\ref{fig:median_profiles} we show the median profiles in the corrected (i.e. corrected for galactic dust, cosmological dimming and K-correction) rest-frame $B$ and $R$ bands and the median ($B-R$) colour profiles of the Type-III S0 galaxies at $0.2 < z < 0.6$. We present the surface brightness profiles as a function of the galactocentric radius normalised by the break radius ($R/\rbreak$). The upper row of Fig.\,\ref{fig:median_profiles} presents the median surface brightness profile in the rest-frame $R$ band obtained by two different methods: one using the F775W data and another using F850LP data. The agreement between both profiles is excellent, and they confirm two major assumptions in Paper I and the present work: 1) we can apply a K-correction across $0.2<z<0.6$ as a function of redshift for any E/S0 and S0 object as computed in Paper I, and 2) the structure of the surface brightness profiles of the galaxies does not change drastically within these filters. In order to ensure the validity of our results, we also analysed the surface brightness profiles using the F850LP corrected band and compared the results to the values of the F775W band. We present the results in Fig.\,\ref{fig:F775WvsF850LP} (see Appendix \ref{Appendix:F775WvsF850LP}). We conclude that the results are compatible between both bands with a median dispersion of $\Delta\rbreak \sim 25\%$ and $\Delta\mubreak \sim 0.525$ \magarc.

In the lower row of Fig.\,\ref{fig:median_profiles}, we present the $B-$band and the $(B-R)$ median profiles. For all bands, the break is clearly visible in the median surface brightness profiles. This ensures that we have enough signal in the outer parts of the objects to analyse their colour profile. In addition to this result, we found that there is not a clear difference between the inner and outer colour profile in the mean $(B-R)$ profile. The median values for the colour $(B-R)$ are $1.310^{+0.077}_{-0.040}$ \magarc\ in the inner region and $1.208^{+0.029}_{-0.017}$ \magarc\ in the outer region. This difference of $\Delta (B-R) \sim 0.1$ between the colours of the inner and outer regions of Type-III S0 galaxies in our sample is significant at a level of $p\sim0.039$. Thus, we do not find any strong differences between the inner and outer median profiles of Type-III S0--E/S0 galaxies.

The results from \citet{2008ApJ...683L.103B} in Type-III spiral galaxies show that these initially appear slightly blue at the inner regions followed by a reddening toward the break radius. Nevertheless, the amplitude of this colour change is $\Delta(g'-r') \simeq 0.1$. This blue-to-red trend is not observed in our median $(B-R)$ profile, which appears to be mostly flat (bottom-right panel in Fig.\,\ref{fig:median_profiles}). This result can be interpreted as a consequence of the low SFR values of S0 galaxies along their whole disc compared to the spirals. These latter cases are expected to produce more complex colour profiles due to the different star formation levels in their discs.

Nevertheless, we find some objects with significant differences between the inner and outer parts of their individual colour profiles. The surface brightness profiles in the $B$ and $R$ bands, as well as the rest-frame colour profiles are available in Appendix \ref{Appendix:profiles}. Some of them present bluer colours in their outskirts (SHARDS10000840, SHARDS10002942, SHARDS10003647 and SHARDS20000827), while others appear to present a reddening of their outer parts when compared to the inner regions (SHARDS10000327 and SHARDS10002730). The rest of the cases present mostly flat colour profiles. In no cases are the differences between the inner and outer profiles of the individual galaxies larger than 0.35 \magarc. We conclude that although there are significant differences between some regions of individual galaxies in their colour profiles, in general, there is no difference between the inner and outer median profiles of Type-III S0--E/S0 galaxies at $0.2 < z < 0.6$. 

\subsection{Stellar population evolution analysis}
\label{Subsec:stellar_pop_analysis}

Finally we analyse if a decay of $\Delta \mu = -1.5$ \magarc\ from $z\sim0.6$ to $z=0$ is compatible with the passive evolution expected for several SSP models. In order to do so, we take advantage of the E-MILES\footnote{E-MILES is available at the MILES/IAC project website http://www.iac.es/proyecto/miles/} stellar population synthesis models \citep{2016MNRAS.463.3409V}. We simulate the magnitude in the $R$ band of an SSP burst with different ages (from 0 to 14 Gyr) and metallicities ($[M/H] = -0.96, -0.35, -0.25, +0.06$ and +0.15). We assume a Kroupa universal initial mass function \citep[IMF, ][]{2001MNRAS.322..231K} with a slope of 1.3 and a BaSTI isocrone \citep{2004ApJ...612..168P} transformed to the observational plane, following empirical relations \citep{1996A&A...313..873A, 1999A&AS..140..261A}. 

\begin{figure}[t]
 \begin{center}
\includegraphics[width=0.5\textwidth]{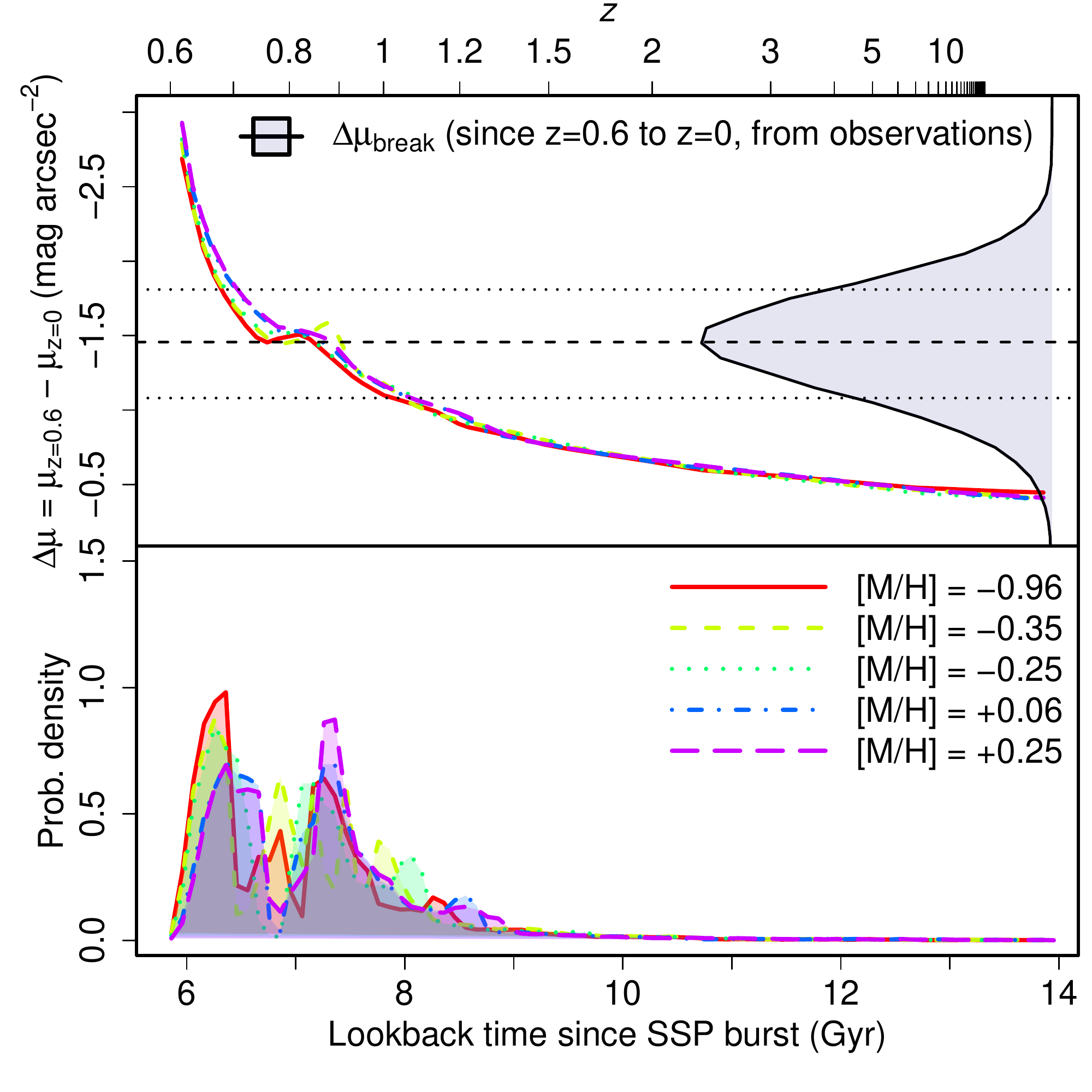}
\caption[]{Analysis of the surface brightness dimming with SSP EMILES models. \emph{Top:} Surface brightness magnitude dimming between $z=0.6$ and $z=0$ as a function of the age of the SSP model. The five different lines cover a range in metallicities from -0.96 to +0.25. The black vertical histogram represents the probability density of the observed dimming in the surface brightness at the break radius ($\Delta\mubreak$) since $z=0.6$ according to our observational data. \emph{Bottom:} Projected probability distributions for the lookback time to the SSP burst, assuming the different SSP models and the observed dimming in \mubreak. We refer to the legend in the panels for a key of the different lines.}

\label{fig:EMILES_test}
\end{center}
\end{figure}

\begin{figure*}[t]
 \begin{center}
\includegraphics[width=0.49\textwidth, page=1]{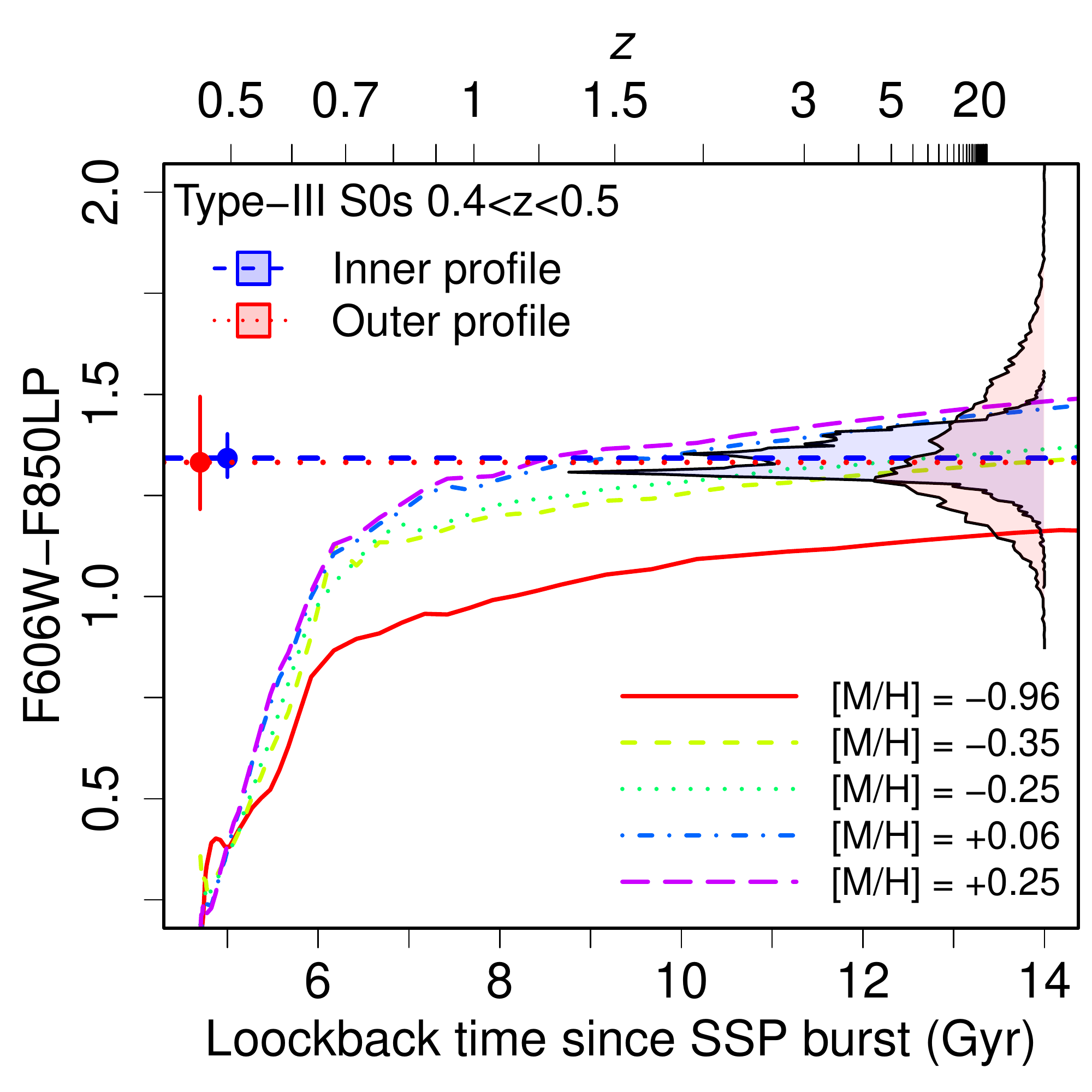}
\includegraphics[width=0.49\textwidth, page=2]{IMAGES/EMILES/SPA_2.pdf}
\caption[]{Analysis of the (F606W - F850LP) colours with SSP EMILES models. \emph{Left panel:} Observed (F606W-F850LP) colour at $z = 0.45$ for an SSP burst as a function of the age of the SSP model. \emph{Right panel:} Observed (F606W-F850LP) colour at $z = 0.55$ for an SSP burst as a function of the age of the SSP model. The five different lines cover a range in metallicities from -0.96 to +0.25. The blue and red vertical histograms represent the probability density of the observed F606W-F850LP colour in the real Type-III S0 galaxies at $0.4<z<0.5$ (left) and $0.5<z<0.6$ (right), for the inner and the outer regions with respect to the \rbreak. The horizontal dashed and dotted lines represent the median (F606W-F850LP) values of the inner and outer profiles of our data, respectively. We refer to the legend in the panels for a key of the different lines.}
\vspace{-0.5cm}
\label{fig:EMILES_colours}
\end{center}
\end{figure*}

The results are shown in Fig.\,\ref{fig:EMILES_test}. The top panel shows the estimated dimming in surface brightness of an SSP burst between $z=0$ and $z=0.6$ as a function of the lookback time of the redshift at which the burst takes place for several metallicities. Given that we want to compare with the observed dimming since $z=0.6$ in \mubreak, we set the minimum limit for the start of the SSP burst in lookback time to $\sim5.7$ Gyr. The results show that a SSP burst would present a surface brightness magnitude $\sim1.5$ \magarc\ brighter at $z=0.6$ than at $z=0$ if the burst took place at $z\sim0.8-0.9$. 

Additionally in the same panel, we represent in a vertical black histogram the PDD of the observed \mubreak\ difference between $z=0$ and $z=0.6$ according to Eq.\,\ref{eq:mubreak_rbreak_z}. We find that the central regions of the PDD for $\Delta\mubreak$ correspond to an age for the SSP burst of $\sim7$ Gyr. In the lower panel of Fig.\,\ref{fig:EMILES_test} we present the projected probability for the age of an SSP burst assuming different metallicities and the observed distribution of $\Delta\mubreak$ between $z=0.6$ and $z=0$. We find that the results for the different metallicities are very similar. The PDDs for the lookback time since an SSP burst present two peaks for most metallicities, one at $z \sim 0.65-0.7$ and a second one at $z\sim0.9$. This is caused by the low variation of $\Delta \mu$ between $z=0$ and $z=0.6$ expected for the SSP models at a lookback time since the SSP burst of $\sim7$ Gyr.

The upper uncertainty interval of the observed $\Delta \mubreak$ is high enough to impede a reasonable determination of the lower limit age of the SSP burst. Nevertheless, we find that the SSP models can not explain the observed change in surface brightness if the SSP started before $z\sim1.2$, that is, if the dominating component of the stellar emission is older than $\sim8$ Gyr. Although this result points to a strong constraint in the stellar formation models, we must remark that this is an over-simplified interpretation of the real evolution and formation mechanisms of S0 galaxies. In a forthcoming paper we will perform a deep analysis of the SEDs of these objects with more complex star formation history (SFH) models such as several SSPs and exponentially-declining SFHs.

In Fig.\,\ref{fig:EMILES_colours} we study whether or not the predicted colours from the SSP models are compatible with those obtained from observations in Fig.\,\ref{fig:median_profiles}). Again, we represent with different lines the colours of several SSP bursts as a function of their age, as they would be observed at $z=0.45$ (left panel) and $z=0.55$ (right panel). The figure shows that the dependence of the colour with the SSP burst metallicity is stronger than in the case of the analysis of the surface brightness dimming in Fig.\,\ref{fig:EMILES_test}. Additionally, we observe that the uncertainties in the observed colour distributions (0.7 - 1 mag) for the inner and outer profiles are compatible with the predicted colour at each redshift for a wide range of metallicities and ages, impeding a similar analysis for the age of the SSP burst as done with the surface brightness dimming. For most metallicities, we find SSP models that can reproduce the colour of the profiles of our Type-III S0s at $z\sim0.5$ if the burst took place at $z>0.6$. Therefore, we cannot restrict an upper limit to the age of the SSP model simply by analysing the median colours of the discs of Type-III S0s at $z\sim0.5$. We conclude that the observed distributions of (F606W-F850LP) in Sect.\,\ref{Subsec:colour_profiles} of the inner and outer profiles are compatible with the predicted (F606W-F850LP) colours from SSP models with ages $> 6-7$ Gyr, especially with those with intermediate metallicities.


\section{Discussion}
\label{Sec:Discussion}

Any hypothetical formation scenario of Type-III S0 galaxies has to be able to explain several observations presented in this study:
\begin{enumerate}
    \item The presence of Type-III S0 galaxies since $z\sim0.6$
    \item The small variation (compatible with no significant variation) of \rbreak\ during the past $\sim 6$ Gyr and the structural stability of its correlation with the inner and outer scale lengths (\hi\ and \ho).
    \item The observed dimming of $\sim1.5$ \magarc\ in surface brightness of Type-III S0 galaxies since $z\sim0.6$ when comparing to the objects of the same type at $z=0$ at the same break radius. 
\end{enumerate}

Although there is no current agreement with the formation of Type-III galaxies (see Sect. \ref{Sec:Intro}), we can summarise all the possible scenarios that have been proposed in three different classes:
\begin{enumerate}
    \item Spiral galaxies form Type-III profiles in their disc structure and they fade afterwards by secular evolution into S0 galaxies.
    \item Type-III profiles form in S0 galaxies after being transformed from spiral galaxies by secular evolution. 
    \item Type-III profiles form in S0 galaxies that did not evolve passively from spirals, but through gravitational interactions (i.e., mergers). 
\end{enumerate}

The latter scenario has previously been studied by \citet{2007ApJ...670..269Y} and \citet{2014A&A...570A.103B}. In these papers the authors demonstrated that Type-III galaxies can be formed through minor and major mergers. Nevertheless, the fact that there are no significant differences between the structural and
photometric parameters of spiral and S0 Type-III galaxies at $z=0$ \citep{2015A&A...580A..33E} suggests that the Type-III profiles may be highly stable structures that can survive passive evolution from spirals into S0 galaxies, preserving their properties. This indicates that Type-III profiles may not be due to transitions between different stellar formation rates or populations, but created by the stellar mass profile of the galaxy, as observed in \citet{2008ApJ...683L.103B}. In addition to this, the observed dimming of $\sim1.5$ \magarc\ in the surface brightness profiles of Type-III S0 galaxies between $z=0.6$ and $z=0$ without any noticeable change in \rbreak\ is compatible with a passive evolution of these objects for the last $\sim6$ Gyr. However, the comparison of this result with SSP models from E-MILES allows us to reject the hypothesis that Type-III S0 galaxies formed preferentially through monolithic collapse before $z=1.2$, evolving passively since then. If Type-III S0 galaxies are older, they should have more complex SFHs than an SSP (e.g. extended SFHs, gas infall, minor/major mergers, etc). Finally, radial migration mechanisms  \citep[see][]{2015MNRAS.448L..99H, 2017MNRAS.470.4941H, 2017A&A...608A.126R} should be able to explain why the general structure of the Type-III profiles remain stable after $\sim6$ Gyr. Nevertheless, there has not yet been a test of whether a Type-III profile caused by radial migration in a barred spiral galaxy would remain stable long enough to survive a secular process of transformation from spirals to S0 galaxies. We conclude that the observed stability of the break radius and the inner and outer scale-lengths since $z\sim0.6$, as well as the observed dimming of 1.5 \magarc\ in the surface brightness profiles since $z\sim0.6$ to $z=0$ pose important constraints on the proposed formation mechanisms of Type-III S0 galaxies. 

\section{Conclusions}
\label{Sec:Conclusions}
We have studied the structural and photometric properties of a sample of Type-III S0--E/S0 galaxies at $0.2<z<0.6$ from the GOODS-N field, and compared them to a sample of galaxies of the same type in the local Universe. In order to do that, we have used a sample of Type-III S0-E/S0 galaxies identified by using the F775W band of HST/ACS in a previous paper (Paper I, see Sect.\,\ref{Sec:Methods}). The images of these objects were morphologically classified and corrected for PSF effects. Additionally, their surface brightness profiles were analysed and classified as a function of their shape. In this paper, we also analysed the corresponding images in the adjacent bands F606W and F850LP, in order to study the differences of their colour profiles in the inner and outer regions of the break. Furthermore, we analysed if the surface brightness profile parameters of the Type-III S0 galaxies at $0.4<z<0.6$ (\rbreak, \mubreak, \hi, \ho, \mui\ and \muo) present a different distribution to the Type-III S0 galaxies in the local Universe, taking into account the stellar mass of the galaxy. Finally, we studied whether or not there is a significant variation of the surface brightness at the break radius (\mubreak) with $z$ and compared it with the expected evolution in brightness from a small grid of SSP models.    

We found several interesting results. First, we report that several structural scaling relations from those reported in B14 in local S0 Type-III galaxies are also found in Type-III S0--E/S0 galaxies at $0.4<z<0.6$ (Sect.\,\ref{Subsec:Scaling_relations}). The structural relations of \hi\ and \ho\ with \rbreak\ and the normalised break radius (\rbreak/\risoph) of both samples are also compatible with the simulations of Type-III S0 galaxies from B14, and present similar correlations, despite the observational limitations. We have found that the photometric scaling relations (those involving \mui\ and \muo\ as well as \mubreak) do not present distributions that are compatible with those from the local sample or the simulations. This is due to a general dimming in surface brightness from $z\sim0.6$ to $z=0$, which is also visible in the normalised structural diagrams. We found no significant differences between the rest-frame $(B-R)$ colours of the inner and outer parts with respect to the break radius \rbreak.

Furthermore, we found that there are no significant differences between the positions of the break (\rbreak) or the scale-lengths (\hi\ and \ho) in local Universe Type-III S0 galaxies and our sample at $0.4<z<0.6$, taking into account the stellar mass of the galaxy. This strongly contrasts with the results found by \citet{2008ApJ...684.1026A} in Type-II spiral galaxies, where the authors report that the break radius has increased with time by a factor of $1.3\pm0.1$ since $z\sim1$ to $z=0$, in agreement with the inside-out formation scenario of disc galaxies. For Type-III profiles, the stability of the structural scaling relations and the break radius as well as the lack of differences in the rest-frame colours contrast with the evolution of Type-II spiral galaxies, and suggests gravitational and dynamical formation scenarios rather than the evolution of stellar populations. Nevertheless, our sample is limited to S0 galaxies and a similar study would be necessary to analyse a possible evolution of the break radius in Type-III spiral galaxies.


We find that the distributions of \mubreak, \mui\ and \muo\ from Type-III S0 galaxies at $0.4<z<0.6$ are not compatible with those of the local Universe sample. Furthermore, there is a significant dependency of \mubreak\ not only on \rbreak\ but also on $z$ (see Sect.\,\ref{Subsec:mubreak_evol}). The \mubreak\ of Type-III S0 galaxies at $z\sim0.6$ appears to be $\sim1.5$ \magarc\ brighter than the objects of the same type at $z=0$ at the same break radius. The cause of this seems to be a general dimming of the whole structure of Type-III S0 galaxies from $z\sim0.6$ to $z=0$. In this paper we investigate whether or not a single stellar population (SSP) model can predict a dimming of $\sim1.5$ \magarc\ from $z=0.6$ to $z=0$ ($\sim6$ Gyr). An instantaneous stellar formation burst should not be older than $\sim8$ Gyr in order to fit the observed change in surface brightness; that is, if the dominant population of Type-III S0 galaxies were formed in a short star formation period, this should have occurred after $z\sim1.2$. This poses a strong constraint on the proposed formation scenarios of Type-III S0 galaxies. If confirmed, the presence of relatively young stellar populations ($z\lesssim1.2$) or more extended SFHs during the formation of Type-III S0 galaxies would be favoured over monolithic collapse at high redshift ($z > 2-3$). Nevertheless, this study is strongly limited by the size of the sample. Deeper and more complete observations are required to confirm or reject these results. In a forthcoming paper we intend to present a detailed analysis of the SED and SFH of each object with more complex models and multiple populations to provide more clues to the origin of these structures.



\begin{acknowledgements}
We thank the anonymous referee for their detailed comments and suggestions to the original version of this manuscript, which helped to improve it considerably. Supported by the Ministerio de Econom\'{\i}a y Competitividad del Gobierno de España (MINECO) under project AYA2012-31277 and project P3/86 of the Instituto de Astrofisica de Canarias. PGP-G acknowledges support from MINECO grants AYA2015-70815-ERC and AYA2015-63650-P. NCL acknowledges support from MINECO grants AYA2013-46724-P and AYA2016-75808-R. This work has made use of the Rainbow Cosmological Surveys Database, which is operated by the Universidad Complutense de Madrid (UCM) partnered with the University of California Observatories at Santa Cruz (UCO/Lick,UCSC). We are deeply grateful to the SHARDS team, since this work would not have been possible without their efforts. Based on observations made with the Gran Telescopio Canarias (GTC) installed at the Spanish Observatorio del Roque de los Muchachos of the Instituto de Astrof\'{\i}sica de Canarias, in the island of La Palma. This paper made use of R: A language and environment for statistical computing \citep[][https://www.R-project.org/]{Rcore} and Astropy, a community-developed core Python package
for Astronomy \citep{Astropy2013, Astropy2018}. We thank all the GTC Staff for their support and enthusiasm with the SHARDS project. This work is based on observations taken by the 3D-HST Treasury Program (GO 12177 and 12328) with the NASA/ESA HST, which is operated by the Association of Universities for Research in Astronomy, Inc., under NASA contract NAS5-26555. 
\end{acknowledgements}

\normalsize

\begin{appendix}
\onecolumn

\section{K-correction}
\label{Appendix:kcorr}
In this appendix we represent the K-correction diagrams used to transform between the HST/ACS F606W and F850LP bands and the $B$ and $R$ rest-frame bands. The K-correction fit results are detailed in Table \ref{tab:fits_results}. 

\begin{figure*}[hp]
 \begin{center}
  \vspace{-0.3cm}
\includegraphics[width=0.43\textwidth]{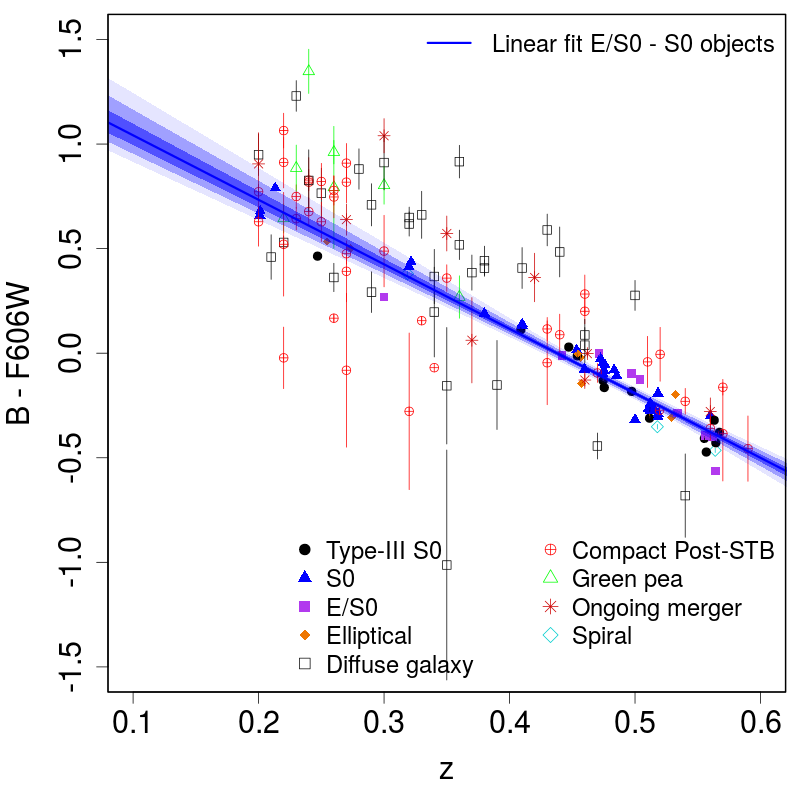}
\includegraphics[width=0.43\textwidth]{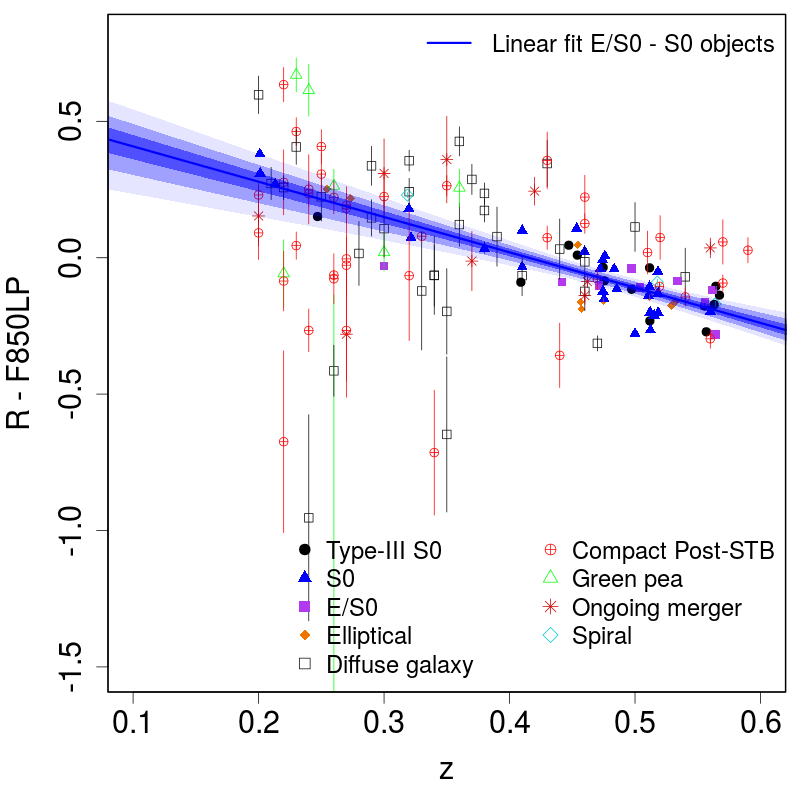}
  \vspace{-0.35cm}
\caption[]{\emph{Left panel:} Estimated K-correction for the F606W filters to the $B$ band as a function of $z$. \emph{Right panel:} Estimated K-correction for the F850LP filters to the $R$ band as a function of $z$. The linear fit applies to the observed correlation of the S0 and E/S0 galaxies of the sample. The intensity levels of the linear fit represent the 1, 2 and 3$\sigma$ confidence regions of the fit. We refer to Paper I for more information about the morphological classification of the objects. See the legend in the figure for identification of symbols.}
\vspace{-1.0cm}

\label{fig:Kcorr_zcomp}
\end{center}
\end{figure*}

\section{F775W vs. F850LP surface brightness profile parameters}
\label{Appendix:F775WvsF850LP}
\begin{figure*}[hp]
 \begin{center}
 \vspace{-0.25cm}

\includegraphics[width=0.43\textwidth]{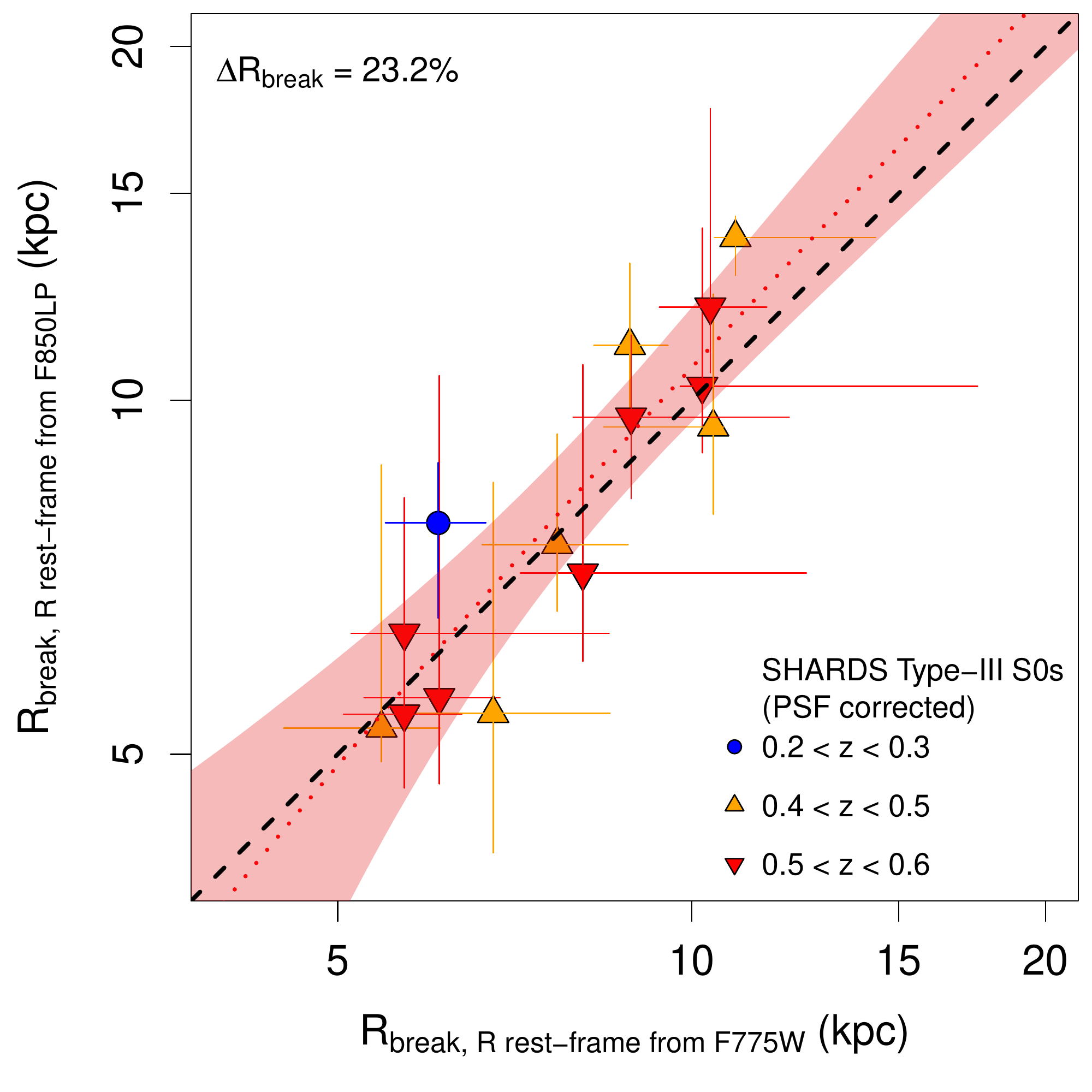}
\includegraphics[width=0.43\textwidth]{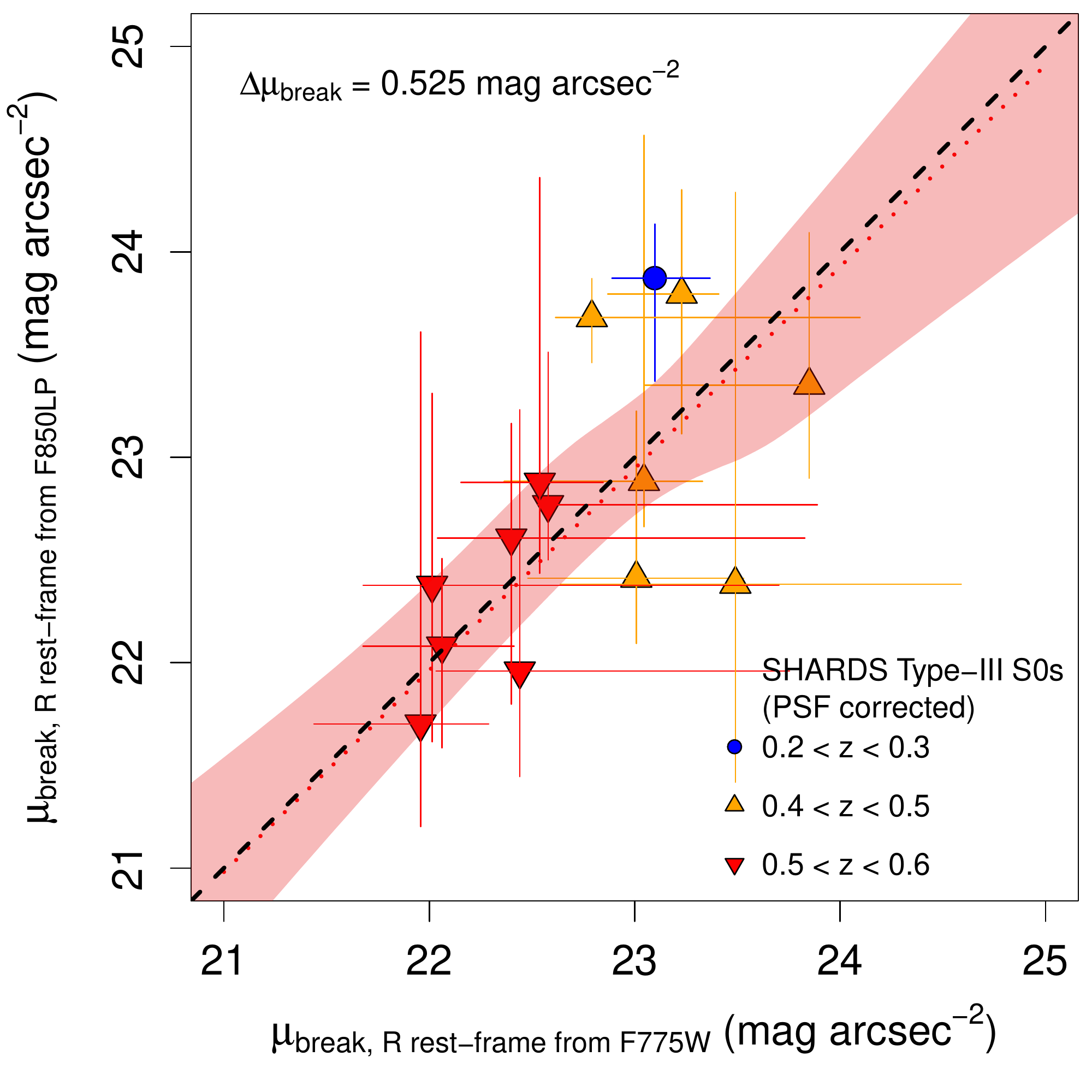}
  \vspace{-0.3cm}
\caption[]{\emph{Left panel:} \rbreak\ measured in the F850LP profile as a function of the \rbreak\ measured in the F775W profile, both after correction to the rest-frame $R$-band. \emph{Right panel:} \mubreak\ measured in the F850LP profile as a function of the \mubreak\ obtained from the F775W profile, both after correction to the rest-frame $R$-band. The dashed black line represents the values that the parameters should have if they were identical. The red dotted line is the linear fit performed to the data in each panel. The colour contour of the linear fit represents the $1\sigma$ confidence region of the fit. We refer to Paper I for more information about the morphological classification of the objects. Consult the legend in the figure.}  
\label{fig:F775WvsF850LP}
\end{center}
\end{figure*}

In this appendix we compare the main parameters of the surface brightness profile analysis in the rest-frame $R$ band obtained from the F775W and F850LP bands. Fig.\,\ref{fig:F775WvsF850LP} shows that the results obtained from the rest-frame $R$-band profiles obtained from both bands concerning \rbreak\ and \mubreak\ are quite compatible, with a median dispersion
of $\Delta\rbreak \sim25\%$ and $\Delta\mubreak \sim 0.525$ \magarc. 

\clearpage

\section{Linear fits to the trends in structural, photometric, and stellar mass planes of Type-III S0--E/S0 galaxies at $z=0$ and $0.4<z<0.6$}
\label{Appendix:fits}

\setlength{\tabcolsep}{4.5pt}
\renewcommand{\arraystretch}{1.2}

\begin{table*}[!hp]
\caption{Linear fits to the trends in several photometric planes of Type-III S0--E/S0 galaxies at $z=0$ and $0.4<z<0.6$, to the K-corrections for the F606W and F850LP filters, and to the trends of the anti-truncation parameters with the logarithm of the stellar mass}
\vspace{-0.7cm}
\label{tab:fits_results}
\begin{center}
{\footnotesize
\begin{tabular}{llccrrlccrr}
\toprule
 &  & \multicolumn{4}{c}{Type-III S0s at $0.4<z<0.6$} & & \multicolumn{4}{c}{Local Universe Type-III S0s}\\
\vspace{-0.3cm}\\\cline{3-6}\cline{8-11}\vspace{-0.3cm}\\
\multicolumn{1}{l}{No.}& \multicolumn{1}{l}{Photometric relation} & $a$ & $b$ & $\rho_{\mathrm{Spearman}}$ & p-value & & $a$ & $b$ & $\rho_{\mathrm{Spearman}}$ & p-value\\
(1) & \multicolumn{1}{c}{(2)} & (3)&(4)&(5)&(6)& & (7)&(8)&(9)&(10)\\
\midrule

1& $\log_{10}(\hi)$ vs. $\log_{10}(\rbreak)$ & $0.990^{+0.108}_{-0.095}$ & $-0.618^{+0.091}_{-0.100}$ & 0.835 & $<10^{-5}$ & & $0.852^{+0.142}_{-0.115}$ & $-0.50^{+0.11}_{-0.13}$ & 0.866 & $<10^{-5}$\\%

2& $\log_{10}(\ho)$ vs. $\log_{10}(\rbreak)$ & $1.56^{+0.71}_{-0.32}$ & $-0.81^{+0.28}_{-0.60}$ & 0.841 & $3\cdot10^{-5}$ & & $1.35^{+0.25}_{-0.25}$ & $-0.57^{+0.21}_{-0.23}$ & 0.809 & $5\cdot10^{-5}$\\%

3& $\mui$ vs. $\log_{10}(\rbreak)$ & $3.5^{+1.0}_{-1.3}$ & $15.19^{+1.21}_{-0.99}$ & 0.269 & 0.085 & & $3.51^{+0.82}_{-0.61}$ & $16.32^{+0.57}_{-0.73}$ & 0.676 & 0.001\\ %

4& $\muo$ vs. $\log_{10}(\rbreak)$ & $9.6^{+6.6}_{-3.3}$ & $12.3^{+2.8}_{-5.6}$ & 0.582 & $3.7 \cdot 10^{-3}$ & & $6.1^{+1.1}_{-1.5}$ & $16.56^{+1.10}_{-0.96}$ & 0.723 & $2.1\cdot 10^{-3}$ \\ %

5& $\mubreak$ vs. $\log_{10}(\rbreak)$ & $5.13^{+0.75}_{-0.70}$ & $18.19^{+0.62}_{-0.67}$ & 0.626 & $8.9\cdot10^{-4}$ & & $5.27^{+0.48}_{-0.64}$ & $19.33^{+0.65}_{-0.42}$ & 0.702 &  $8.8\cdot10^{-3}$\\ %

\vspace{-0.2cm}\\\hline\vspace{-0.2cm}\\

6& $\log_{10}(\hi/\risoph)$ vs. $\log_{10}(\rbreak/\risoph)$ & $0.90^{+0.13}_{-0.12}$ & $-0.629^{+0.014}_{-0.012}$ & 0.670 & $9\cdot10^{-4}$ & & $0.66^{+0.17}_{-0.12}$ & $-0.602^{+0.022}_{-0.025}$ & 0.816 & $4\cdot10^{-5}$\\%

7& $\log_{10}(\ho/\risoph)$ vs. $\log_{10}(\rbreak/\risoph)$ & $1.82^{+1.04}_{-0.48}$ & $-0.276^{+0.080}_{-0.047}$ & 0.769 & $6.7\cdot10^{-4}$ & & $1.29^{+0.35}_{-0.27}$ & $-0.296^{+0.054}_{-0.057}$ & 0.429 & 0.059\\%

8& $\mui$ vs. $\log_{10}(\rbreak/\risoph)$ & $5.5^{+1.5}_{-2.1}$ & $18.484^{+0.108}_{-0.089}$ & 0.456 & 0.03 & & $4.056^{+1.58}_{-0.94}$ & $19.066^{+0.20}_{-0.23}$ & 0.651 & $8\cdot10^{-4}$\\ %

9& $\muo$ vs. $\log_{10}(\rbreak/\risoph)$ & $14.9^{+10.8}_{-4.7}$ & $21.45^{+0.72}_{-0.40}$ & 0.835 & $9\cdot10^{-5}$ & & $7.5^{+2.2}_{-2.2}$ & $21.30^{+0.30}_{-0.33}$ & 0.598 & 0.018\\ %

10& $\mubreak$ vs. $\log_{10}(\rbreak/\risoph)$ & $7.74^{+0.56}_{-0.56}$ & $23.086^{+0.057}_{-0.058}$ & 0.967 & $<10^{-5}$ & & $7.01^{+0.55}_{-0.38}$ & $23.468^{+0.087}_{-0.092}$ & 0.968 & $<10^{-5}$\\ %

\vspace{-0.2cm}\\\hline\vspace{-0.2cm}\\

11& $B$ - F606W vs. $z$ & $-3.08^{+0.14}_{-0.12}$ & $ 1.350^{+0.059}_{-0.069}$ & -0.962 & $<10^{-5}$ & & -- & -- & -- & -- \\
12& $R$ - F850LP vs. $z$ & $- 1.30^{+0.11}_{-0.12}$ & $0.639^{+0.060}_{-0.052}$ & 0.799 & $<10^{-5}$ & & -- & -- & -- & -- \\

\vspace{-0.2cm}\\\hline\vspace{-0.2cm}\\

13& $\rbreak$ vs. $\log_{10}(M/M_{\odot})$ & $3.5^{+2.6}_{-6.2}$ &  $-29^{+66}_{-28}$  & 0.252 & 0.257 & & $9.2^{+4.9}_{-4.0}$ &  $-88^{+41}_{-51}$  & 0.513 & $2.8\cdot10^{-2}$\\%

14& $\hi$ vs. $\log_{10}(M/M_{\odot})$ & $1.04^{+0.33}_{-2.7}$ &  $-9.2^{+29.0}_{-3.4}$  & 0.234 & 0.247 & & $1.99^{+0.53}_{-0.43}$ &  $-18.9^{+4.6}_{-5.5}$  & 0.633 & $8.3\cdot10^{-3}$\\

15& $\ho$ vs. $\log_{10}(M/M_{\odot})$ & $1.4^{+2.9}_{-4.7}$ &  $-10^{+50}_{-31}$  & 0.125 & 0.370 & & $11.5^{+8.6}_{-6.6}$ &  $-115^{+68}_{-89}$  & 0.746 & $3.8\cdot10^{-3}$\\

16& $\ho/\hi$ vs. $\log_{10}(M/M_{\odot})$ & $-0.87^{+2.40}_{-0.59}$ &  $11.34^{+6.5}_{-25.6}$  & -0.073 & 0.433 & & $3.6^{+2.5}_{-2.3}$ &  $-35^{+24}_{-25}$  & 0.347 & 0.148\\

17& $\mubreak$ vs. $\log_{10}(M/M_{\odot})$ & $-1.64^{+0.59}_{-0.83}$ & $40.4^{+8.9}_{-6.2}$  & -0.373 & 0.161 & & $3.17^{+1.75}_{-1.63}$ &  $-9.2^{+17.2}_{-18.3}$  & 0.384 & 0.127\\

18& $\mui$ vs. $\log_{10}(M/M_{\odot})$ & $-1.78^{+0.47}_{-0.58}$ &  $37.3^{+6.3}_{-5.1}$  & -0.546 & $3.6\cdot10^{-2}$ & & $2.05^{+0.70}_{-0.56}$ &  $-1.9^{+5.9}_{-7.2}$  & 0.493 & $3.9\cdot10^{-2}$\\

19& $\muo$ vs. $\log_{10}(M/M_{\odot})$ & $-1.3^{+3.2}_{-1.1}$ &  $35^{+11}_{-34}$  & -0.204 & 0.311 & & $4.4^{+1.9}_{-1.7}$ & $-24^{+17}_{-19}$ & 0.456 & 0.102\\

\bottomrule 
\end{tabular}
\begin{minipage}[t]{\textwidth}{\vspace{0.25cm}
\footnotesize

 \emph{Columns}: (1) ID number of the relation. (2) Fitted relation. (3) Slope of the linear fit to the Type-III S0 galaxies at $0.4<z<0.6$. (4) Y-intercept of the linear fit to the Type-III S0 galaxies at $0.4<z<0.6$. (5) Spearman $\rho$ correlation test statistic for the Type-III S0 galaxies at $0.4<z<0.6$. (6) Spearman correlation test p-value for the Type-III S0 galaxies at $0.4<z<0.6$. (7--10) The same as columns 3--6, but for the linear fits performed to the local samples.}
\end{minipage}
}
\vspace{-0.5cm}
\end{center}
\end{table*}

\clearpage

\section{Surface brightness and colour profiles}
\label{Appendix:profiles}
Here we represent the surface brightness and colour profiles of the Type-III S0 and E/S0 objects within our sample, accounting for their PSF-corrected photometric profiles.\\

\begin{figure}[!h]
\textbf{1) SHARDS10000327:}\\
\begin{minipage}{\textwidth}
\includegraphics[clip,width=0.45\textwidth,page=4]{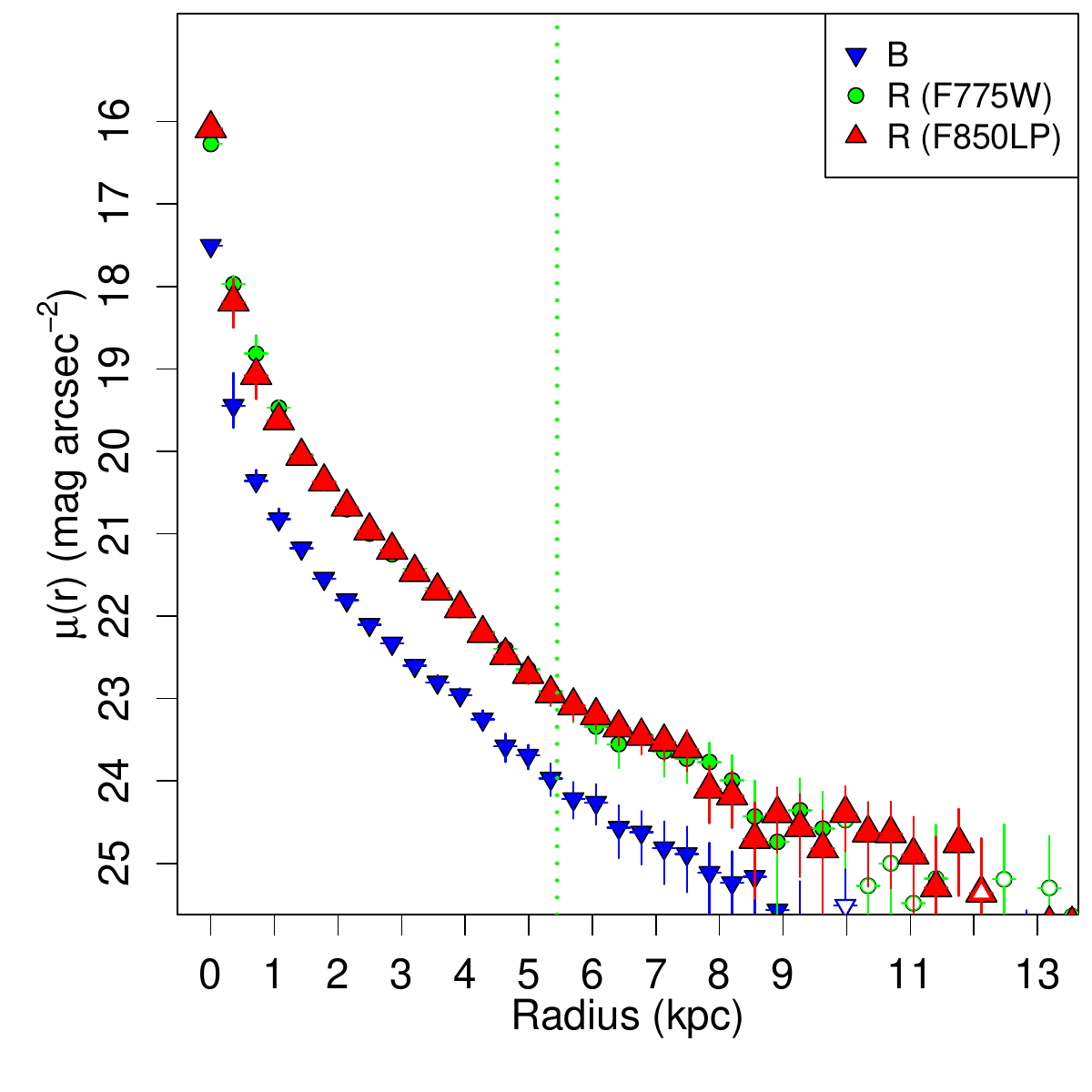}\vspace{-1cm}
\includegraphics[clip,width=0.45\textwidth,page=5]{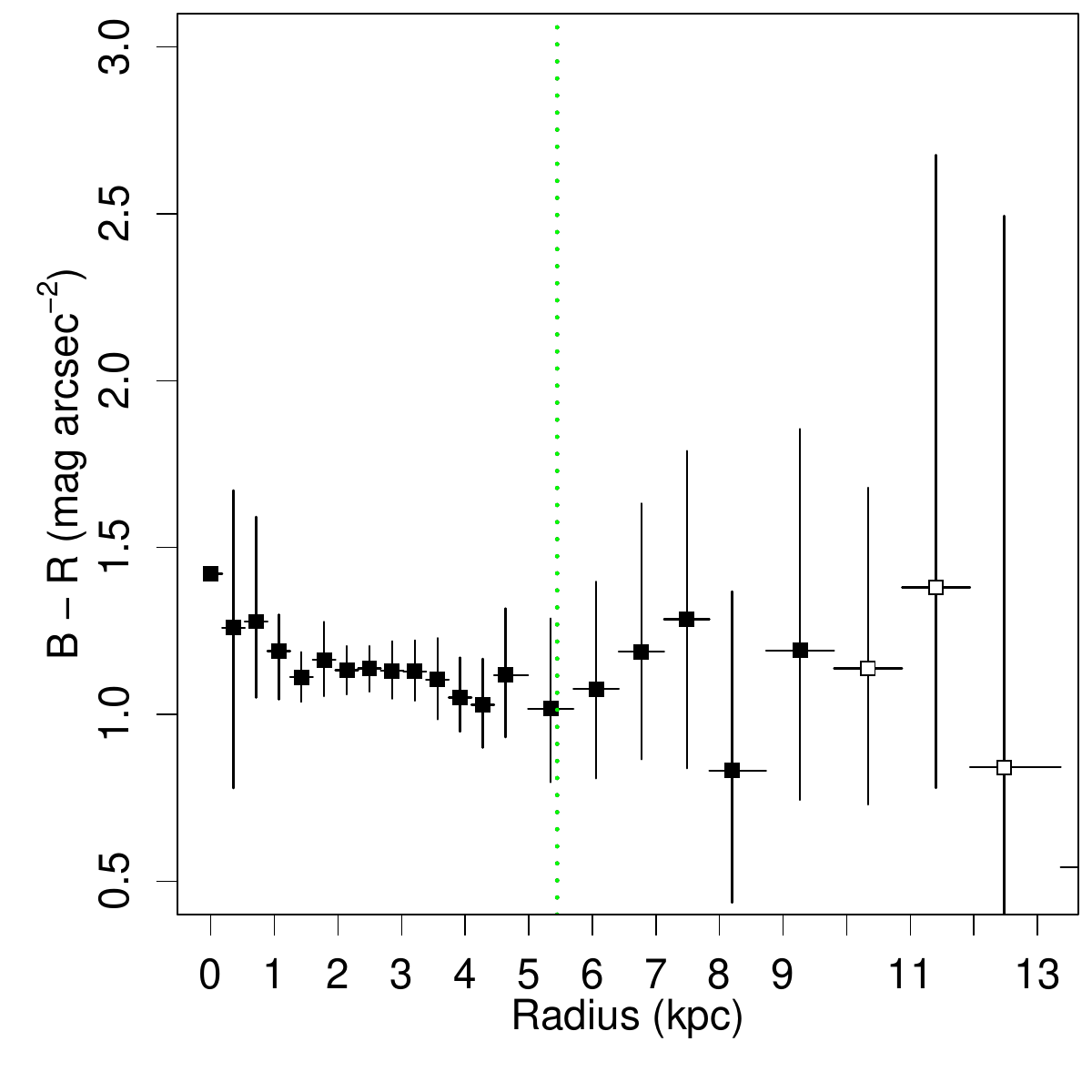}
\end{minipage}%
\vspace{0.5cm}

\caption[]{\emph{Left panel:} Surface brightness profiles corrected for dust extinction, cosmological dimming and \textbf{K-correction to the rest-frame} FORS $B$ band and the Steidel $R$ band. We represent the $R$ band profile calculated with both F775W and the F850LP HST/ACS bands (green and red, see the legend for details). \emph{Right panel:} $(B-R)$ colour profile corrected for dust extinction, cosmological dimming and K-correction to the rest-frame FORS $B$ band and the Steidel $R$ band. Filled symbols present a probability higher than 99.5\% to be over the sky-level. In the case of colour profiles, the limiting radius corresponds to the lower radius of the limiting magnitudes of the $B$ and $R$ bands. The vertical green dotted line represents the break radius \rbreak\ for the F775W band (see Paper I).}   
\label{fig:img_final}
\end{figure}


\begin{figure}[!h]
\textbf{2) SHARDS10000827:}\\
\begin{minipage}{\textwidth}
\includegraphics[clip,width=0.45\textwidth,page=4]{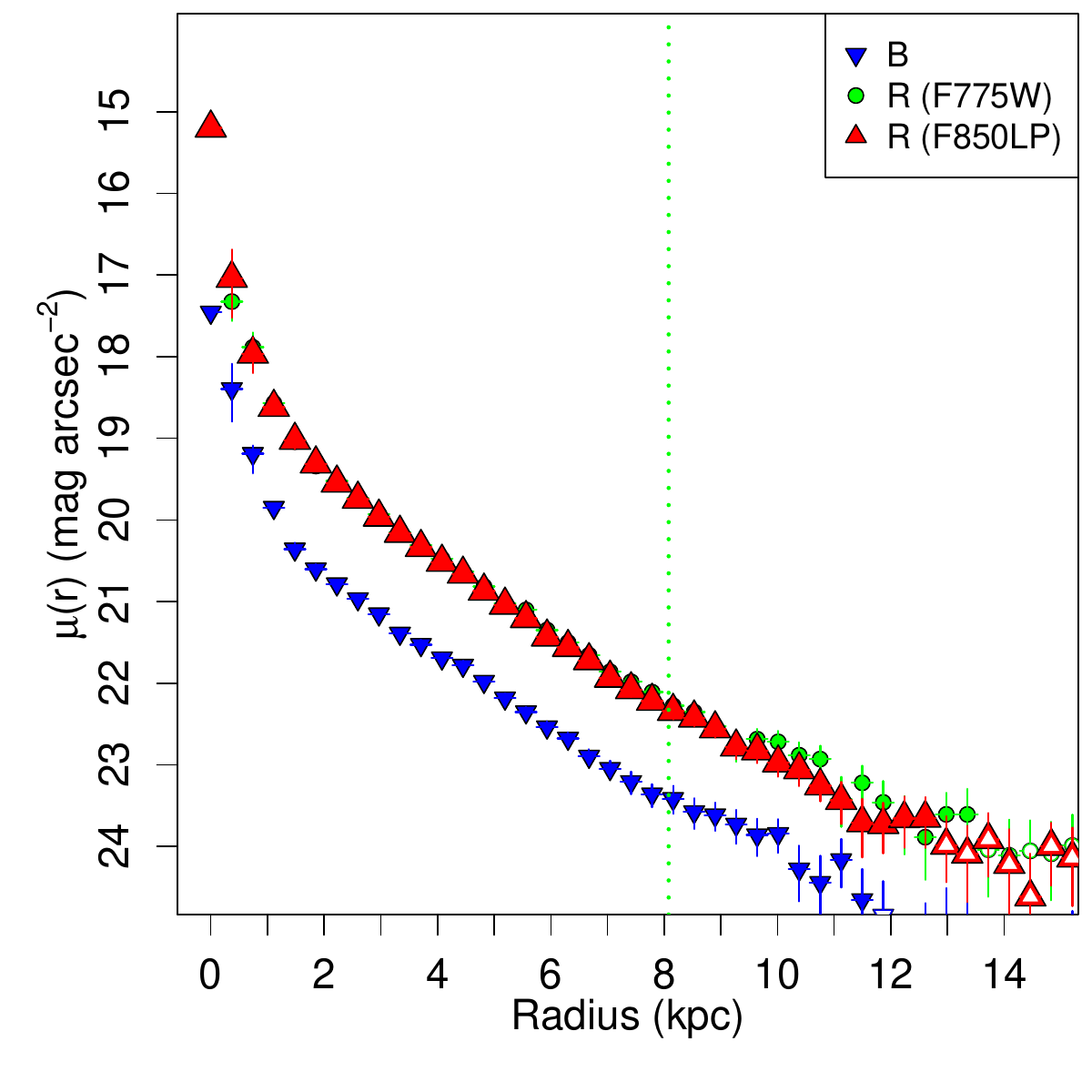}\vspace{-1cm}
\includegraphics[clip,width=0.45\textwidth,page=5]{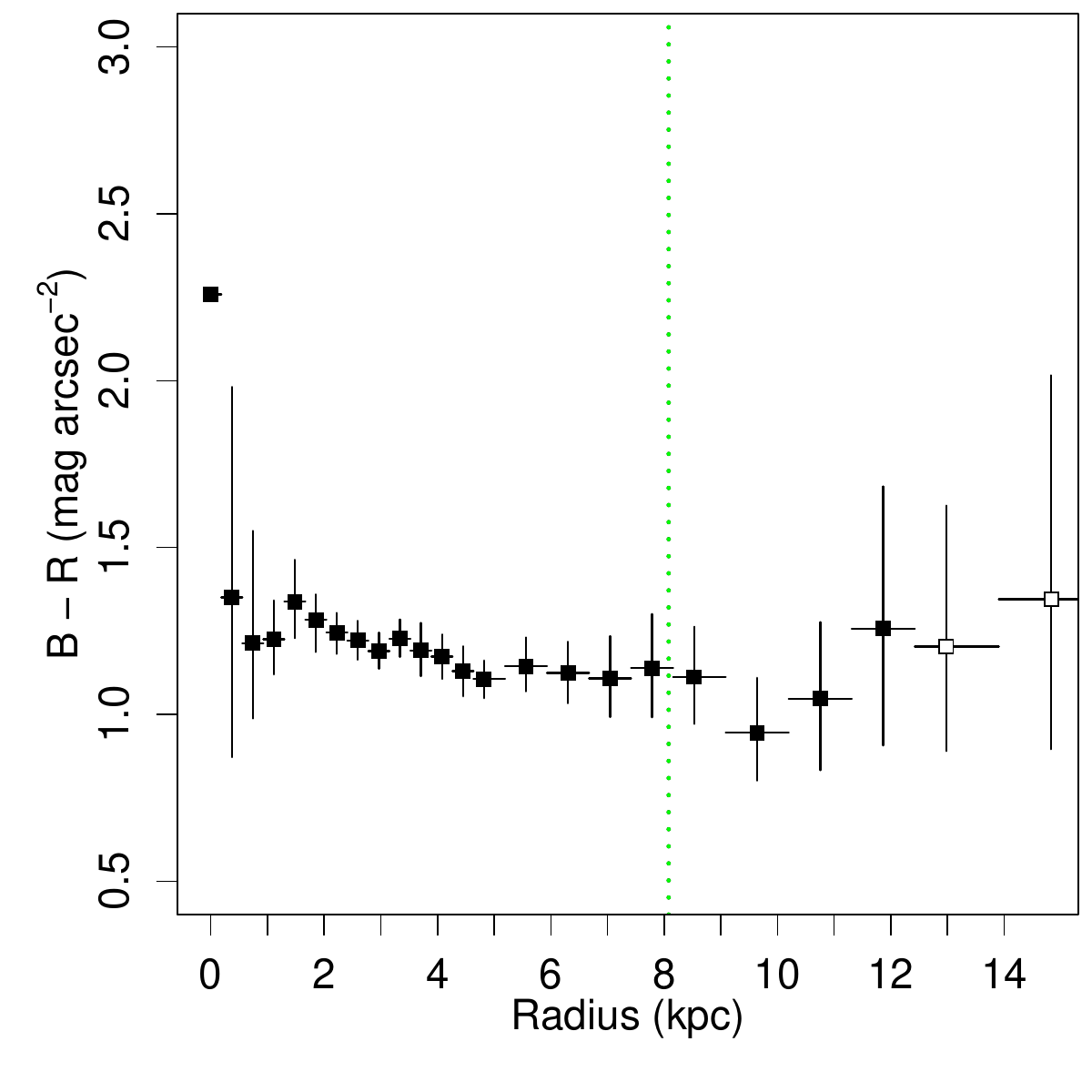}
\end{minipage}%
\vspace{0.5cm}

\caption[]{See caption of Fig.1.}         
\label{fig:img_final}
\end{figure}

\begin{figure}[!h]
\textbf{3) SHARDS10000840:}\\
\begin{minipage}{\textwidth}
\includegraphics[clip,width=0.45\textwidth,page=4]{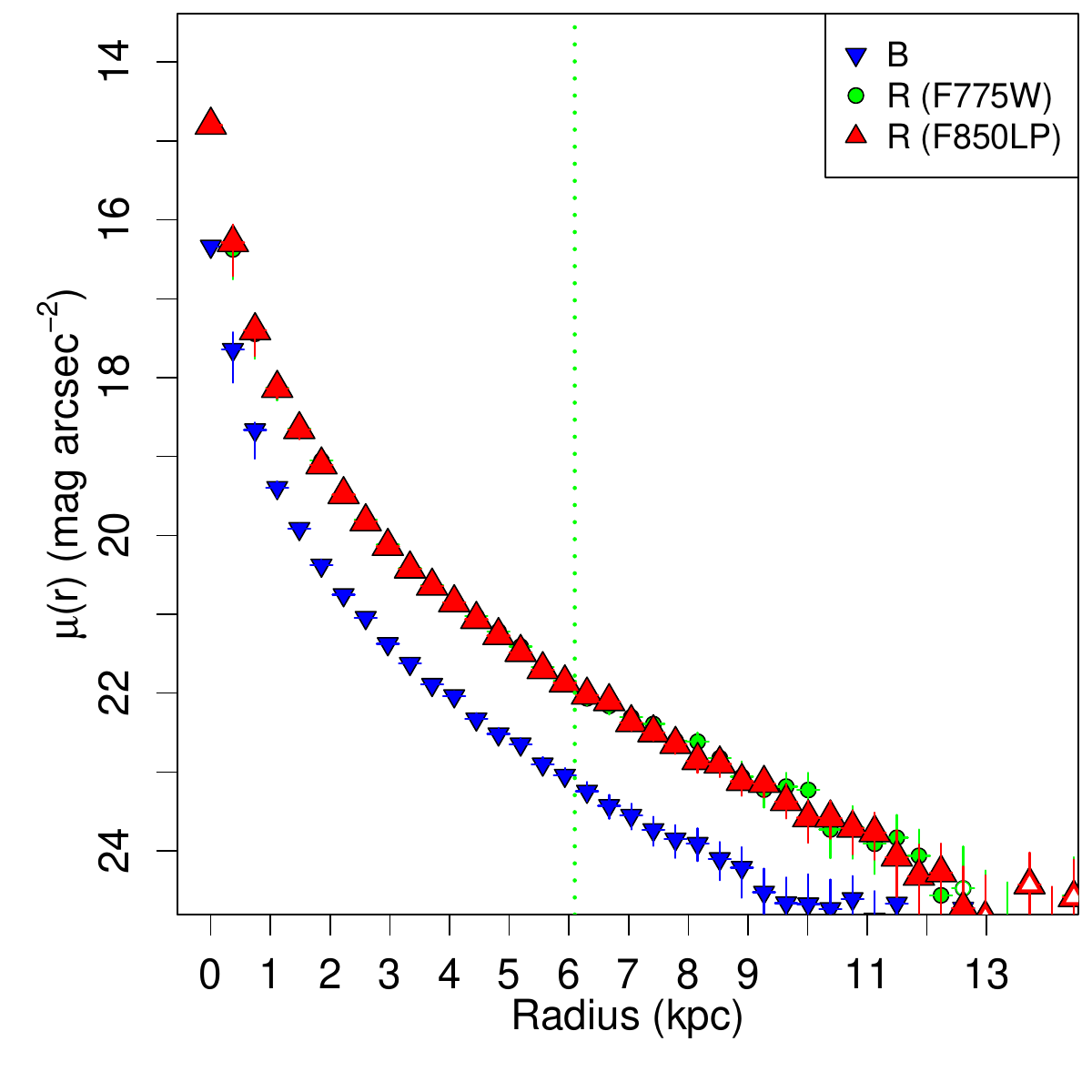}\vspace{-1cm}
\includegraphics[clip,width=0.45\textwidth,page=5]{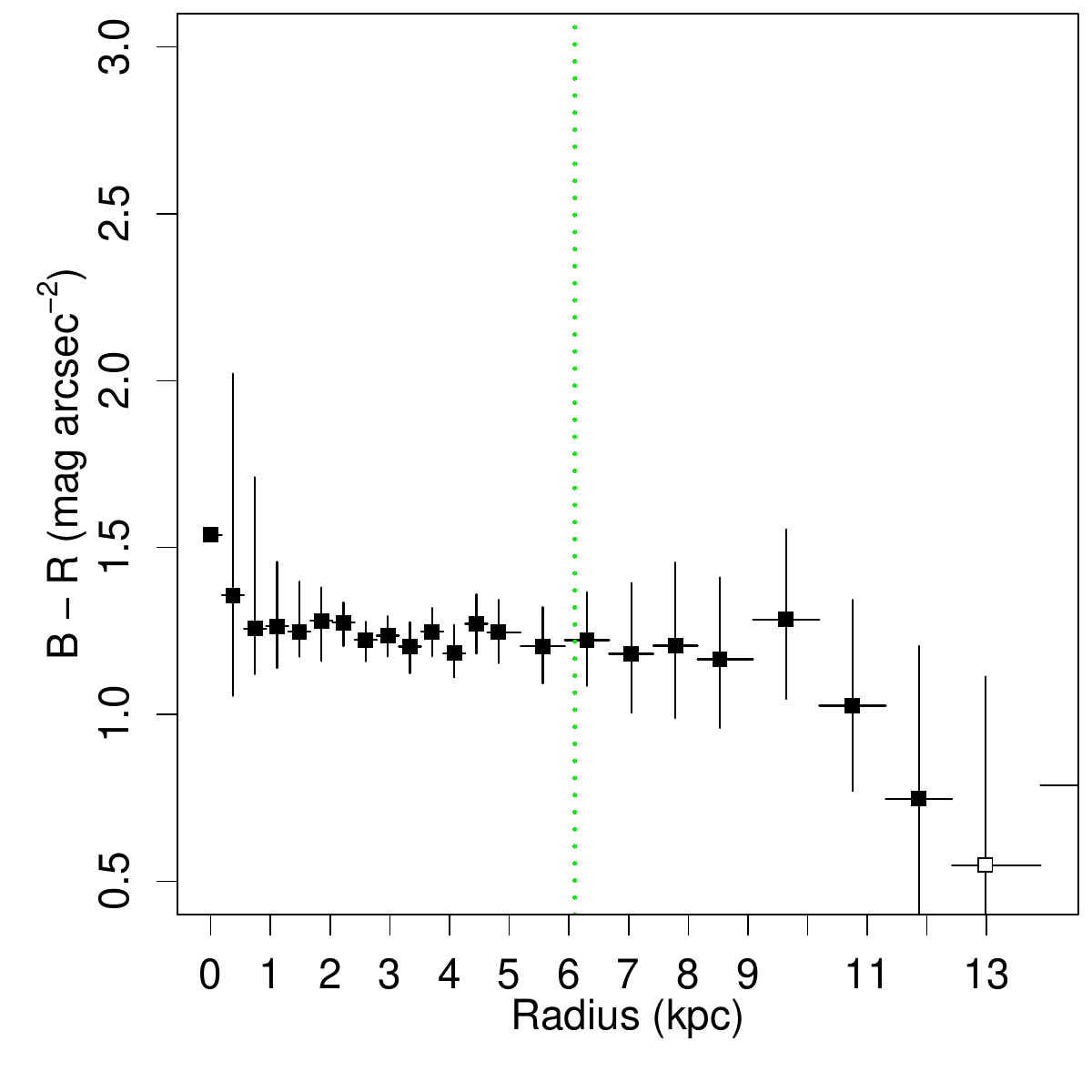}
\end{minipage}%
\vspace{0.5cm}

\caption[]{See caption of Fig.1.}         
\label{fig:img_final}
\end{figure}


\begin{figure}[!h]
\textbf{4) SHARDS10000849:}\\
\begin{minipage}{\textwidth}
\includegraphics[clip,width=0.45\textwidth,page=4]{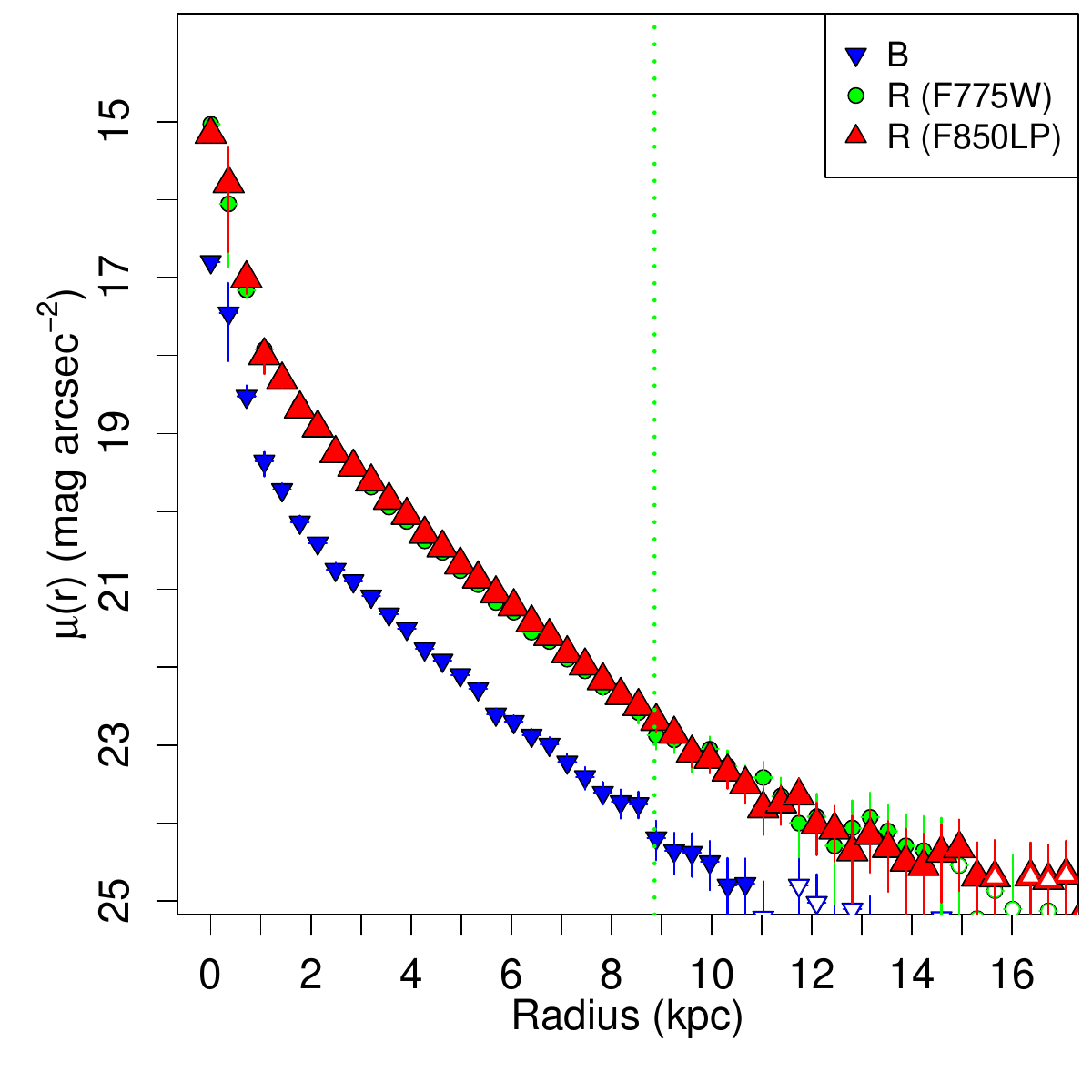}\vspace{-1cm}
\includegraphics[clip,width=0.45\textwidth,page=5]{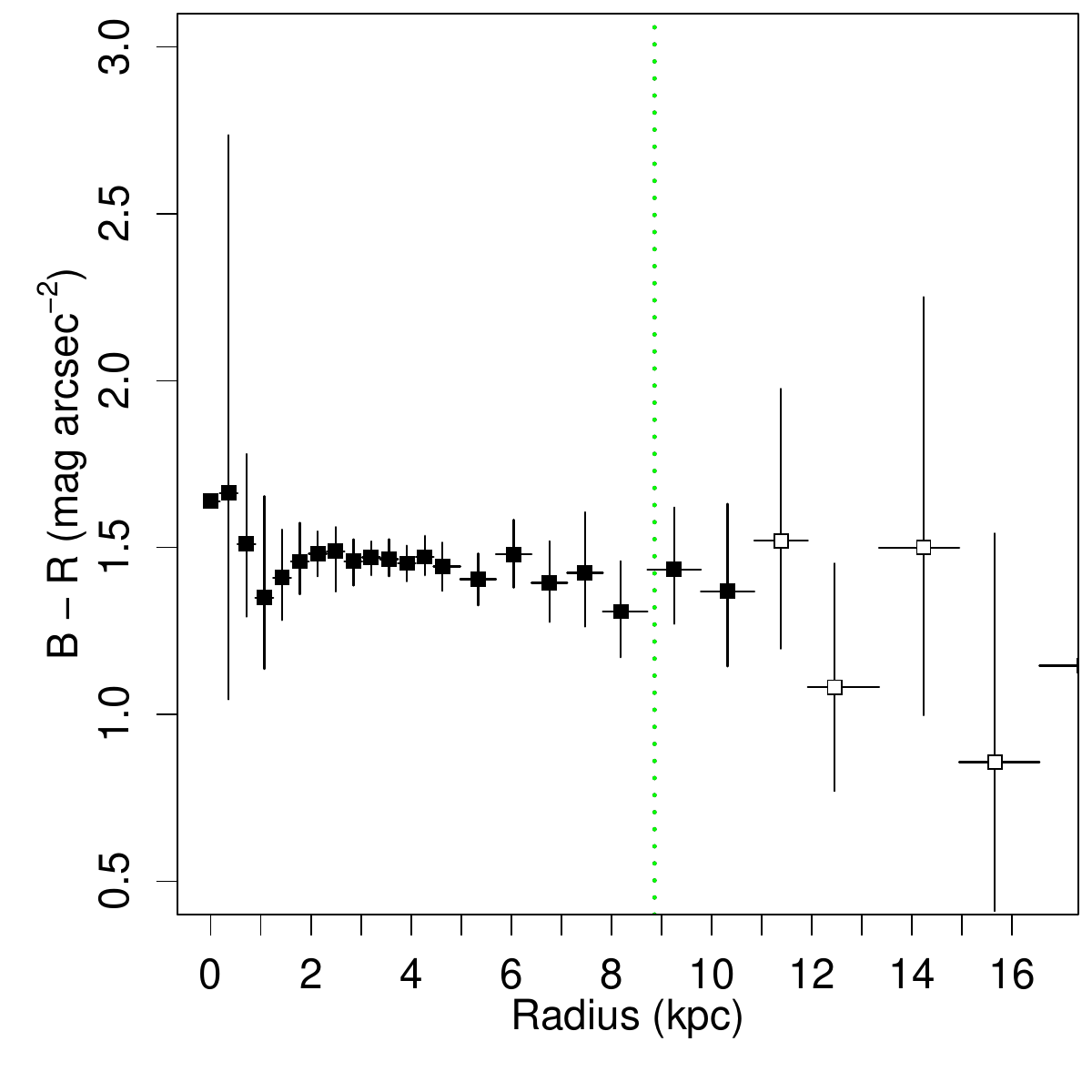}
\end{minipage}%
\vspace{0.5cm}
\caption[]{See caption of Fig.1.}         
\label{fig:img_final}
\end{figure}


\begin{figure}[!h]
\textbf{5) SHARDS10001344:}\\
\begin{minipage}{\textwidth}
\includegraphics[clip,width=0.45\textwidth,page=4]{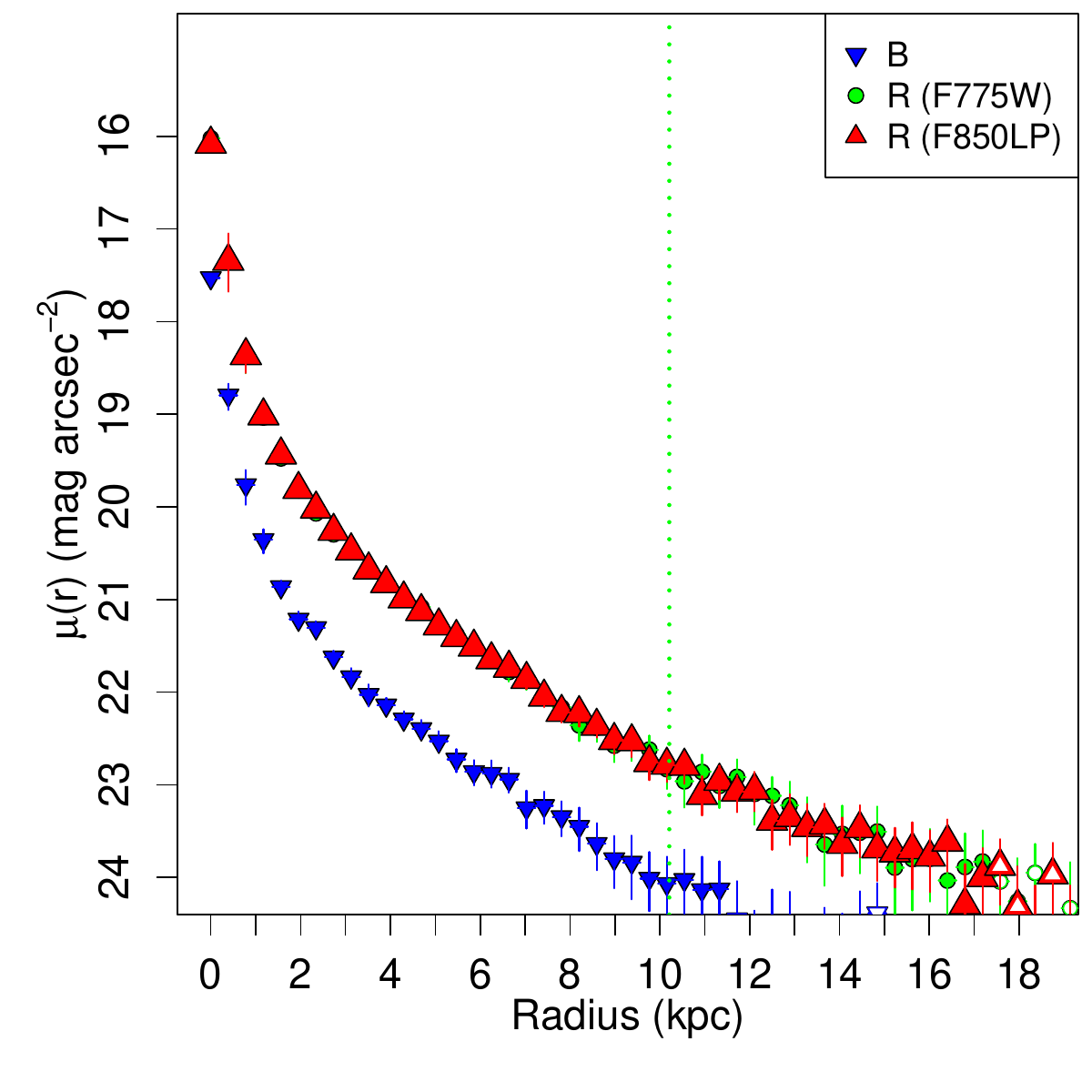}\vspace{-1cm}
\includegraphics[clip,width=0.45\textwidth,page=5]{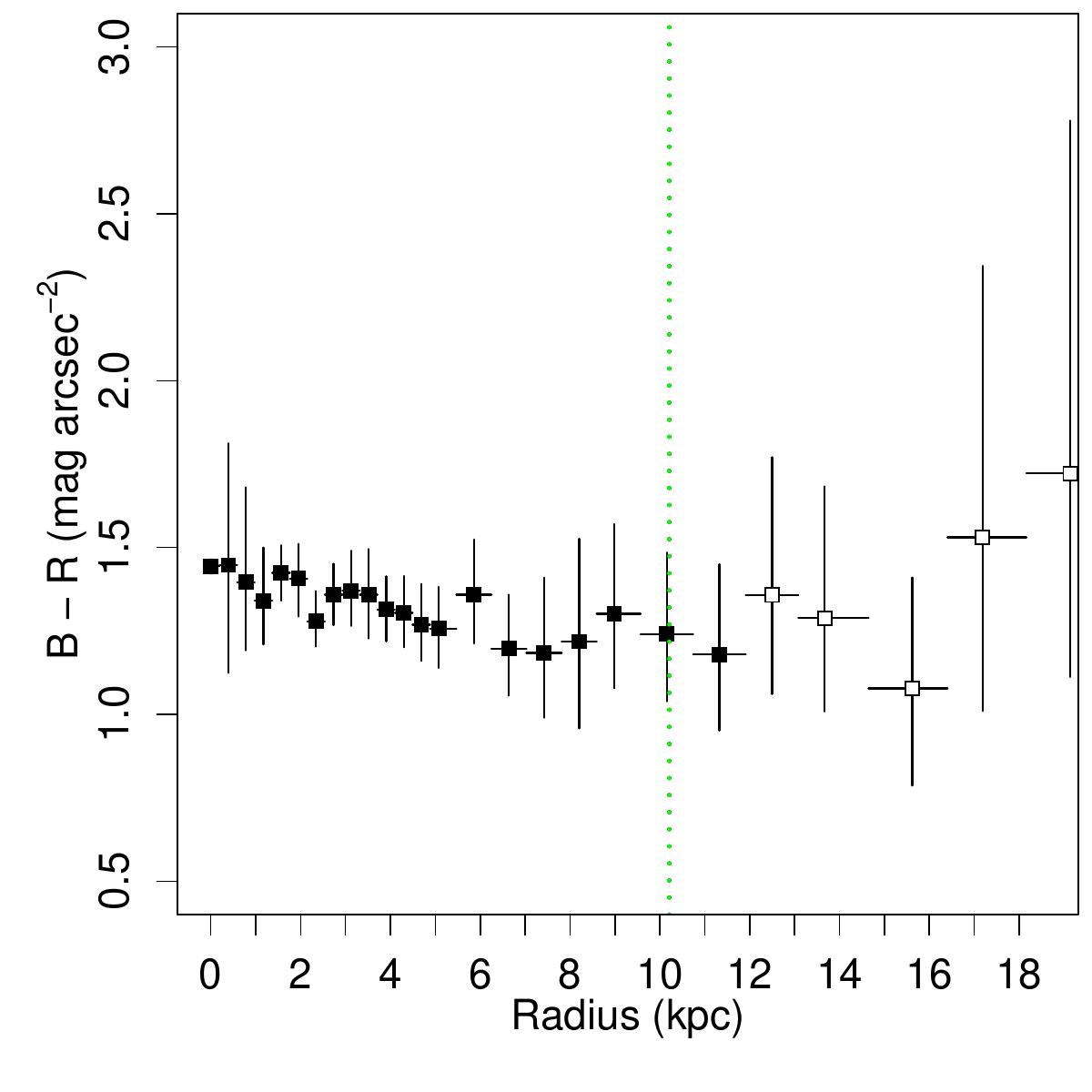}
\end{minipage}%
\vspace{0.5cm}
\caption[]{See caption of Fig.1.}         
\label{fig:img_final}
\end{figure}


\begin{figure}[!h]
\textbf{6) SHARDS10001648:}\\
\begin{minipage}{\textwidth}
\includegraphics[clip,width=0.45\textwidth,page=4]{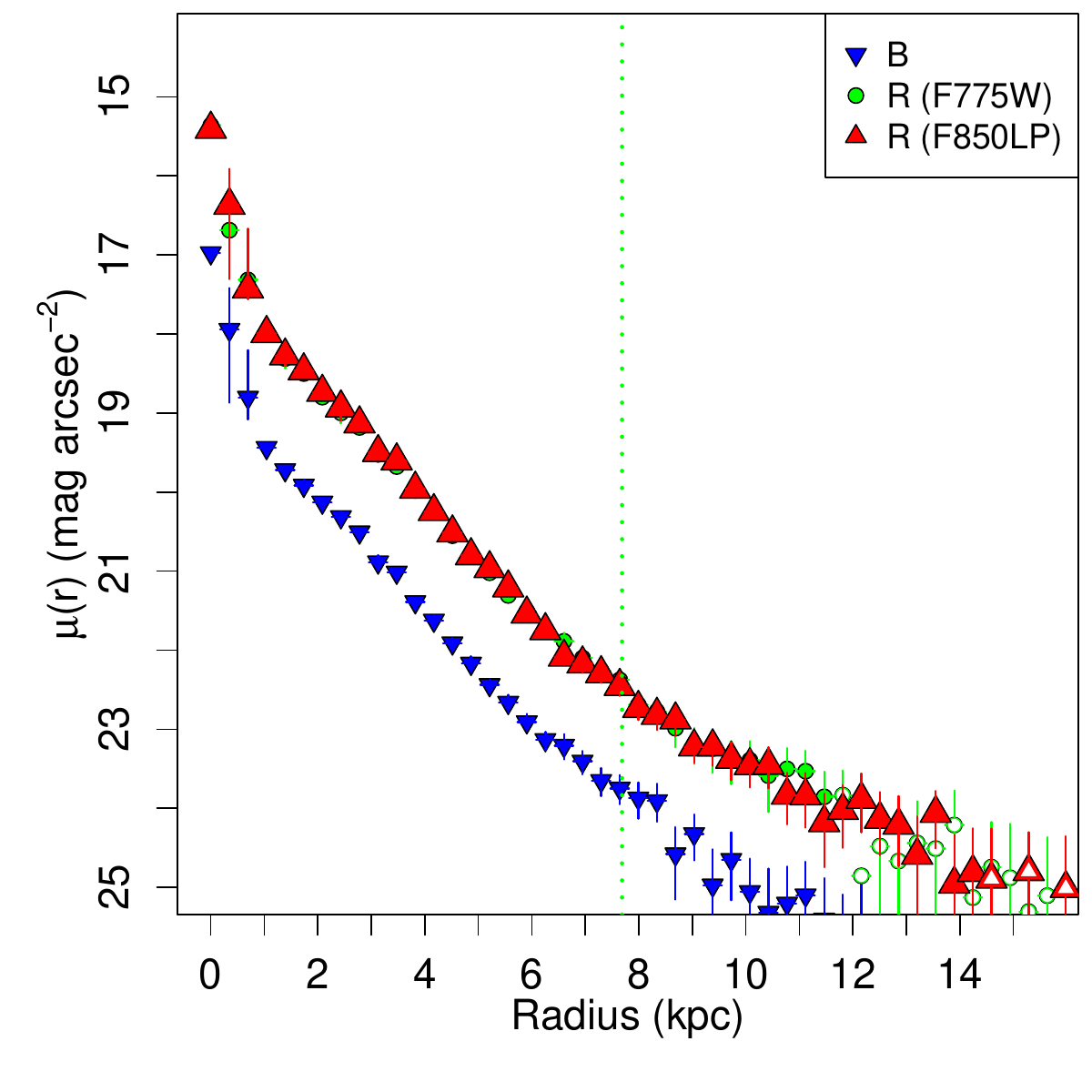}\vspace{-1cm}
\includegraphics[clip,width=0.45\textwidth,page=5]{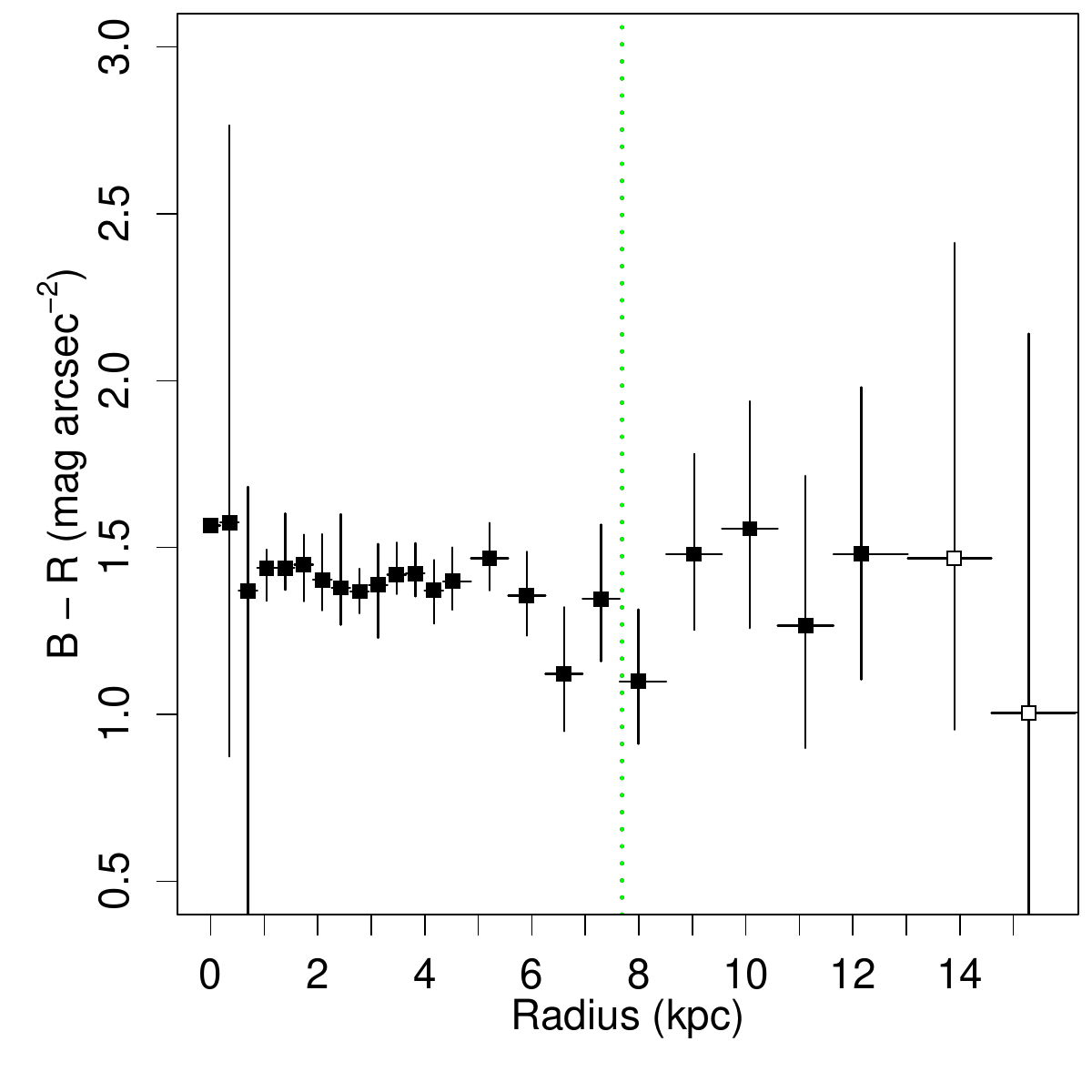}
\end{minipage}%
\vspace{0.5cm}
\caption[]{See caption of Fig.1.}         
\label{fig:img_final}
\end{figure}


\begin{figure}[!h]
\textbf{7) SHARDS10002730:}\\
\begin{minipage}{\textwidth}
\includegraphics[clip,width=0.45\textwidth,page=4]{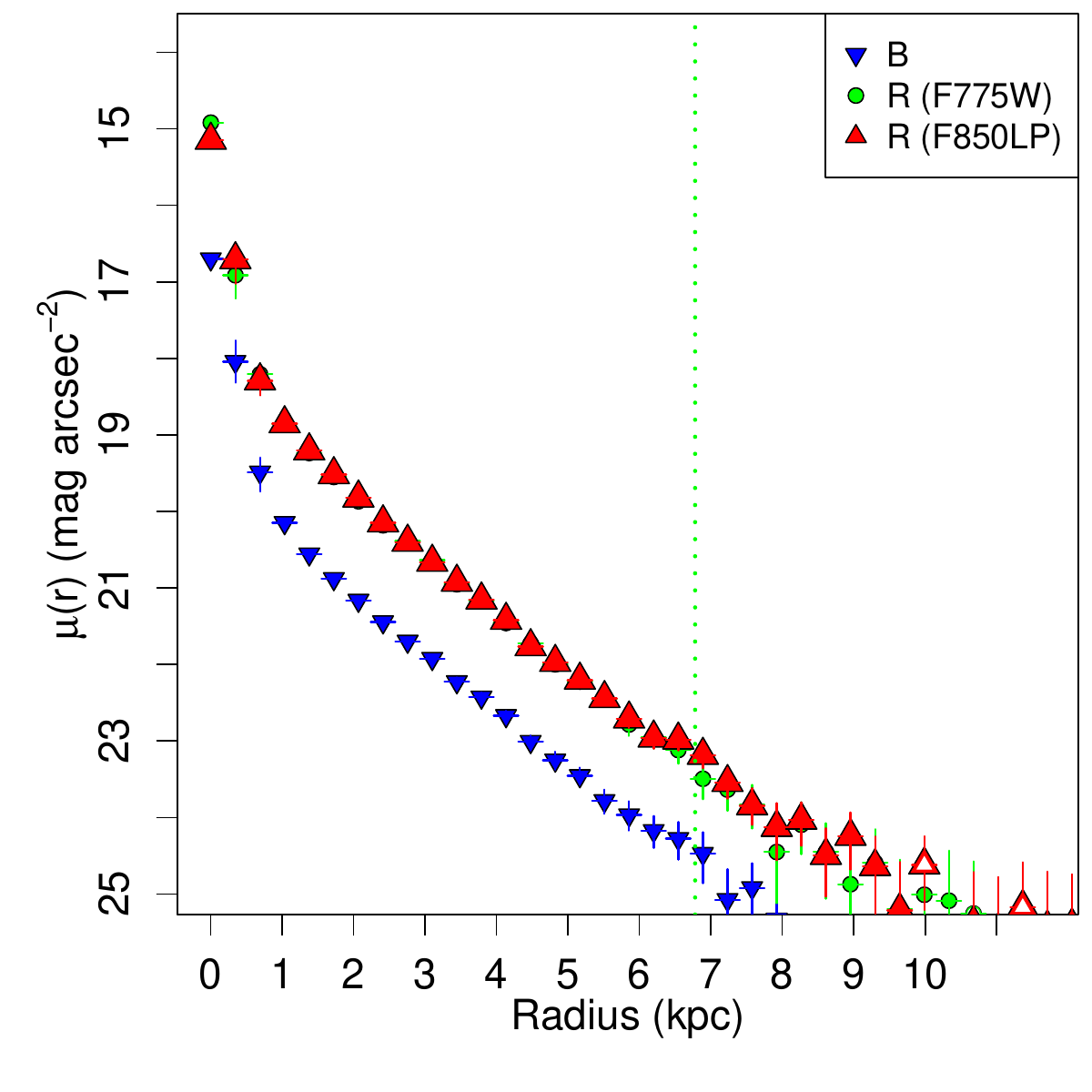}\vspace{-1cm}
\includegraphics[clip,width=0.45\textwidth,page=5]{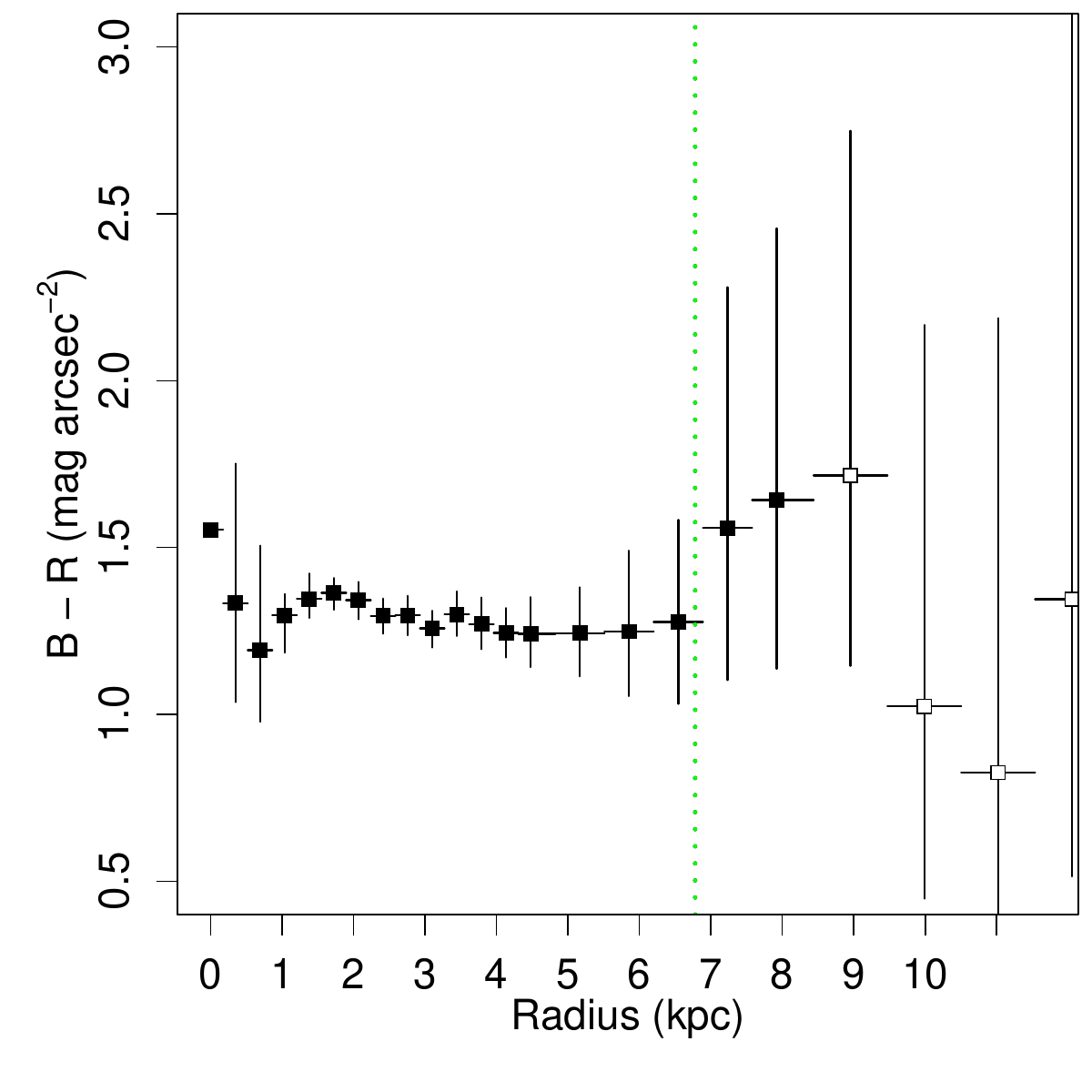}
\end{minipage}%
\vspace{0.5cm}
\caption[]{See caption of Fig.1.}         
\label{fig:img_final}
\end{figure}


\begin{figure}[!h]
\textbf{8) SHARDS10002942:}\\
\begin{minipage}{\textwidth}
\includegraphics[clip,width=0.45\textwidth,page=4]{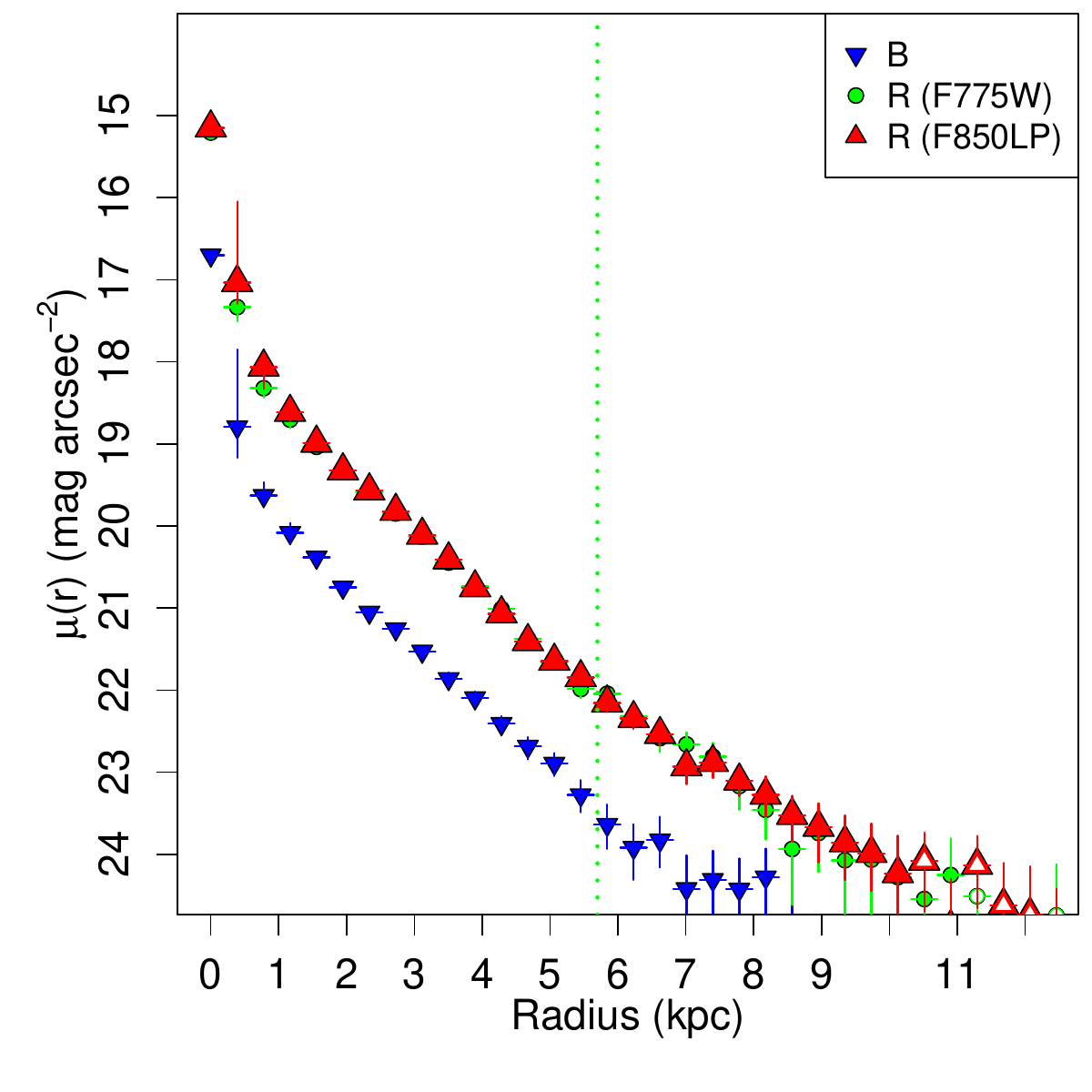}\vspace{-1cm}
\includegraphics[clip,width=0.45\textwidth,page=5]{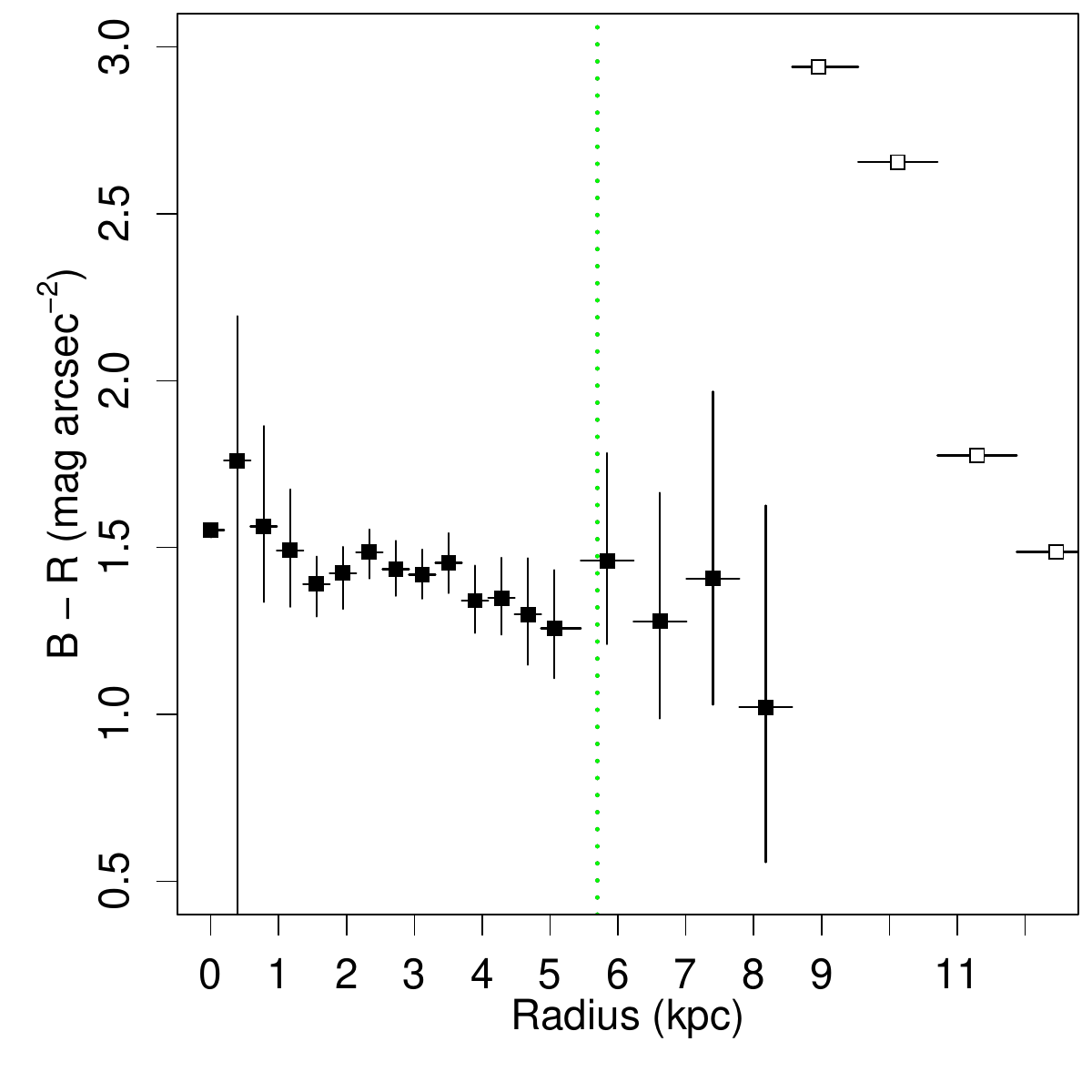}
\end{minipage}%
\vspace{0.5cm}
\caption[]{See caption of Fig.1.}         
\label{fig:img_final}
\end{figure}


\begin{figure}[!h]
\textbf{9) SHARDS10003312:}\\
\begin{minipage}{\textwidth}
\includegraphics[clip,width=0.45\textwidth,page=4]{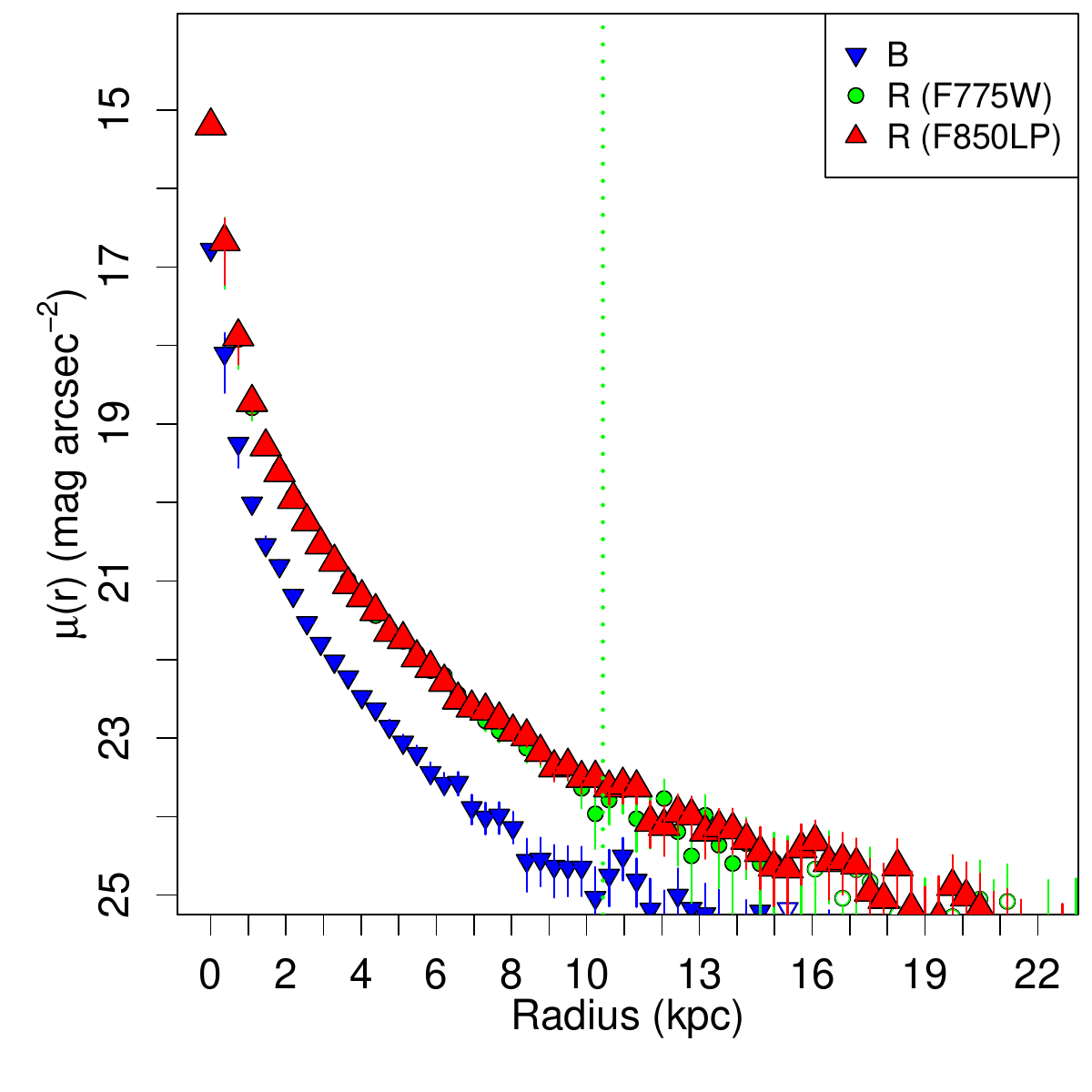}\vspace{-1cm}
\includegraphics[clip,width=0.45\textwidth,page=5]{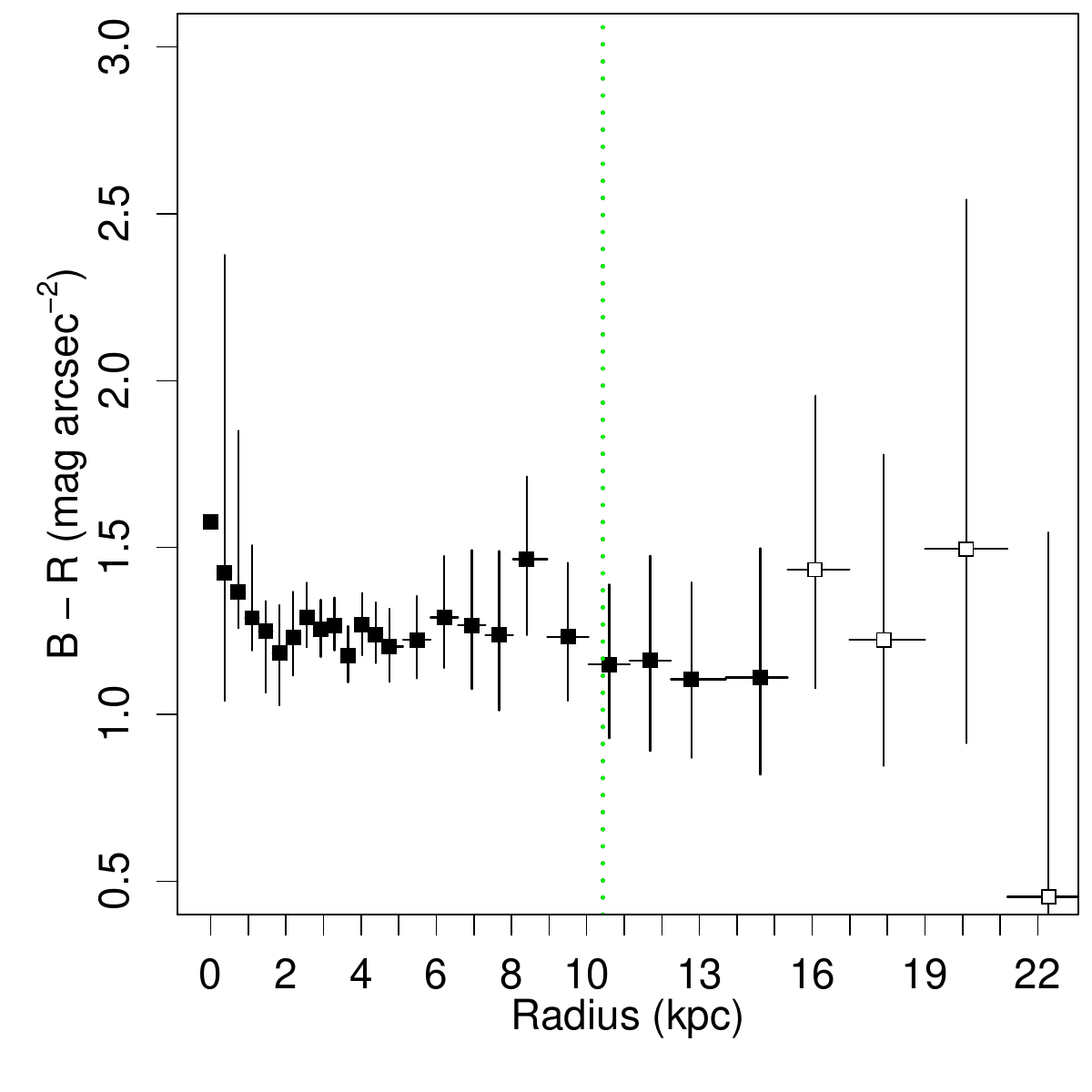}
\end{minipage}%
\vspace{0.5cm}
\caption[]{See caption of Fig.1.}         
\label{fig:img_final}
\end{figure}


\begin{figure}[!h]
\textbf{10) SHARDS10003647:}\\
\begin{minipage}{\textwidth}
\includegraphics[clip,width=0.45\textwidth,page=4]{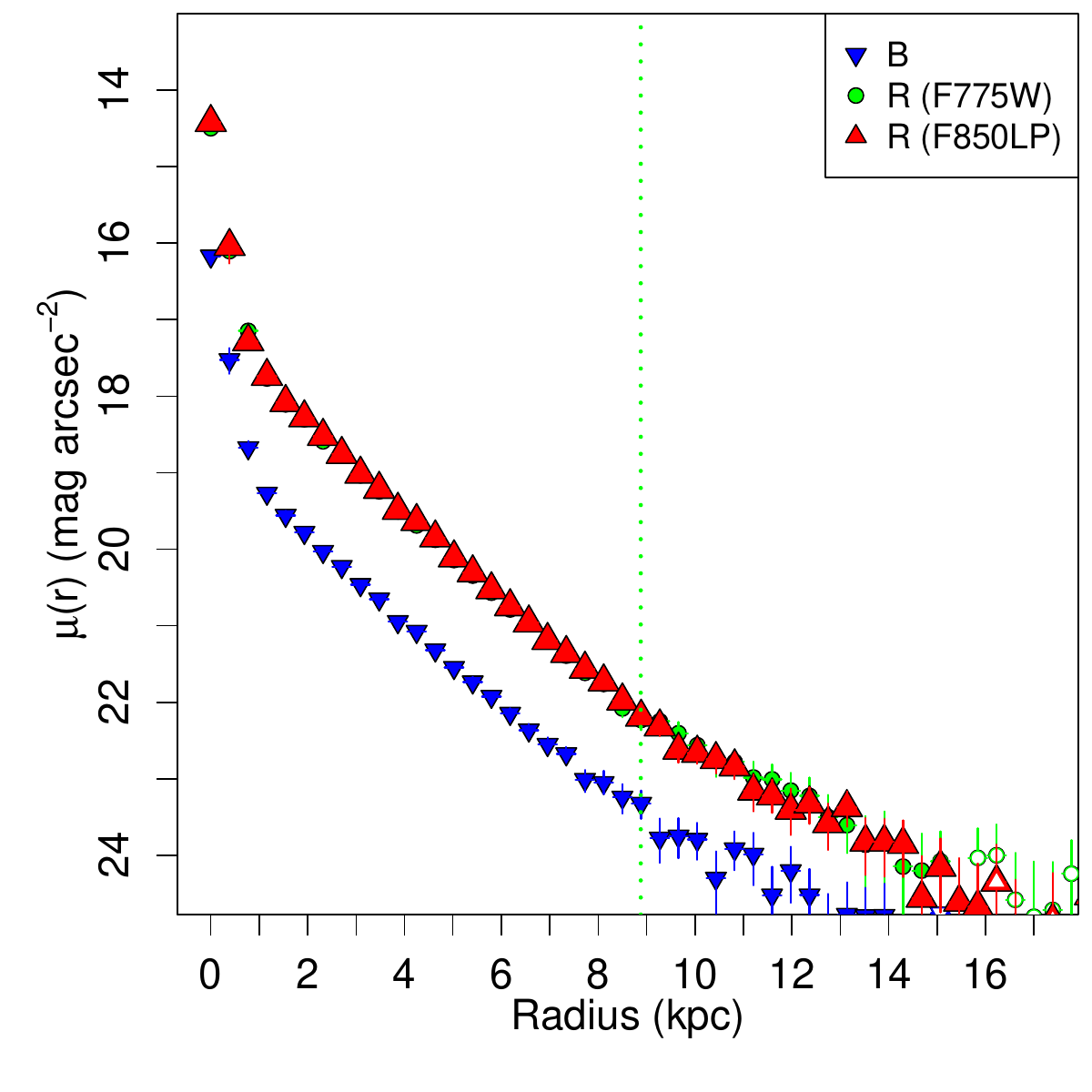}\vspace{-1cm}
\includegraphics[clip,width=0.45\textwidth,page=5]{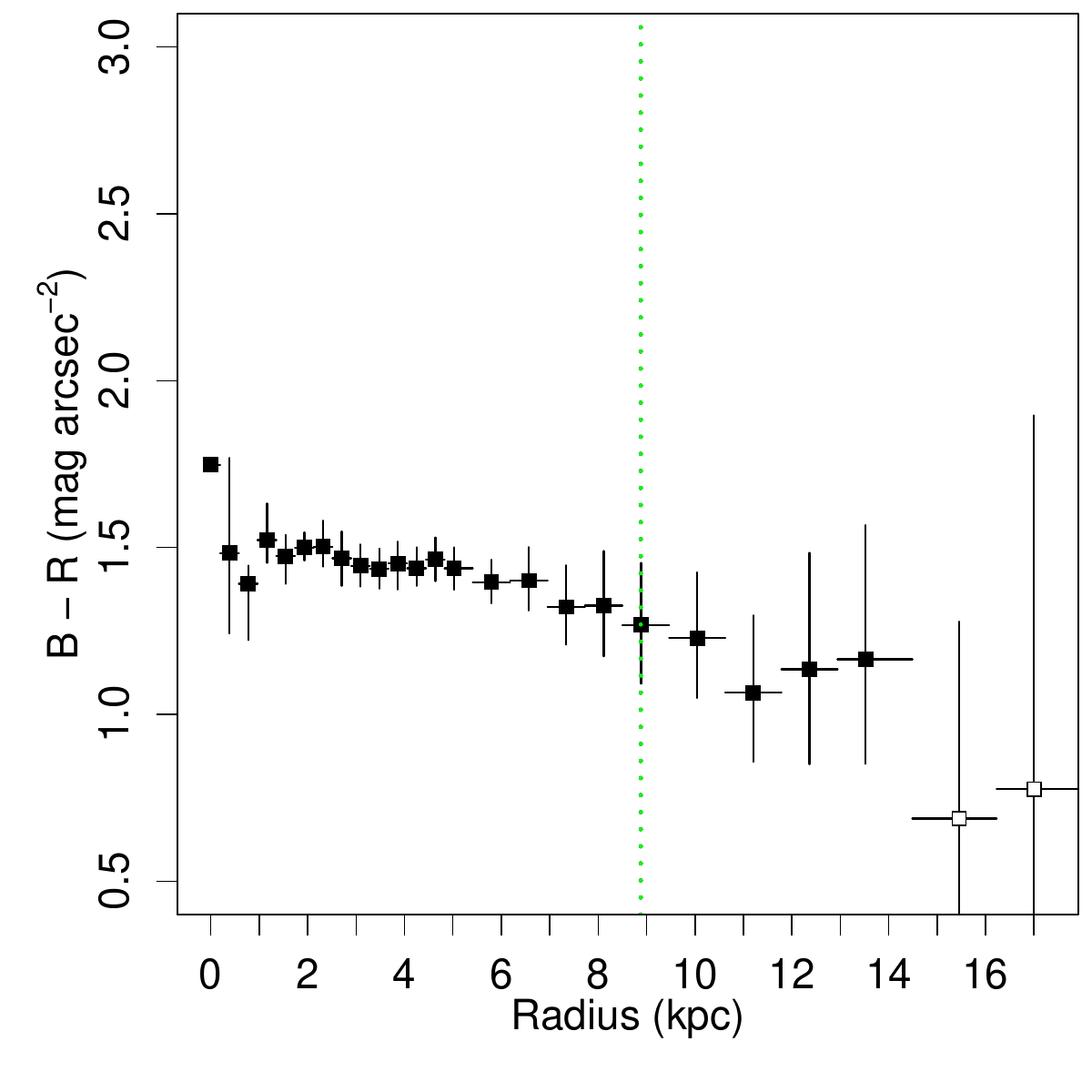}
\end{minipage}%
\vspace{0.5cm}
\caption[]{See caption of Fig.1.}         
\label{fig:img_final}
\end{figure}


\begin{figure}[!h]
\textbf{11) SHARDS10009610:}\\
\begin{minipage}{\textwidth}
\includegraphics[clip,width=0.45\textwidth,page=4]{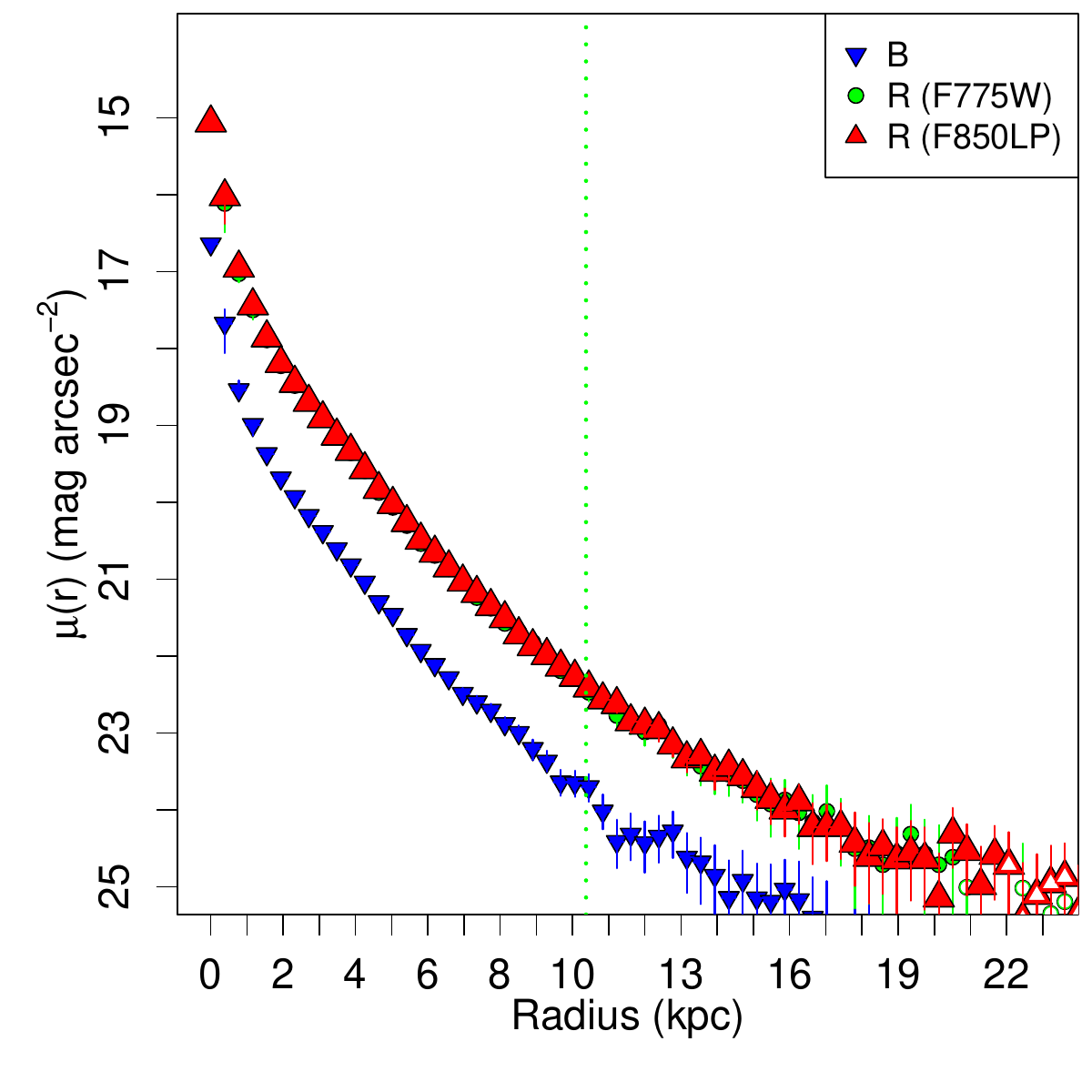}\vspace{-1cm}
\includegraphics[clip,width=0.45\textwidth,page=5]{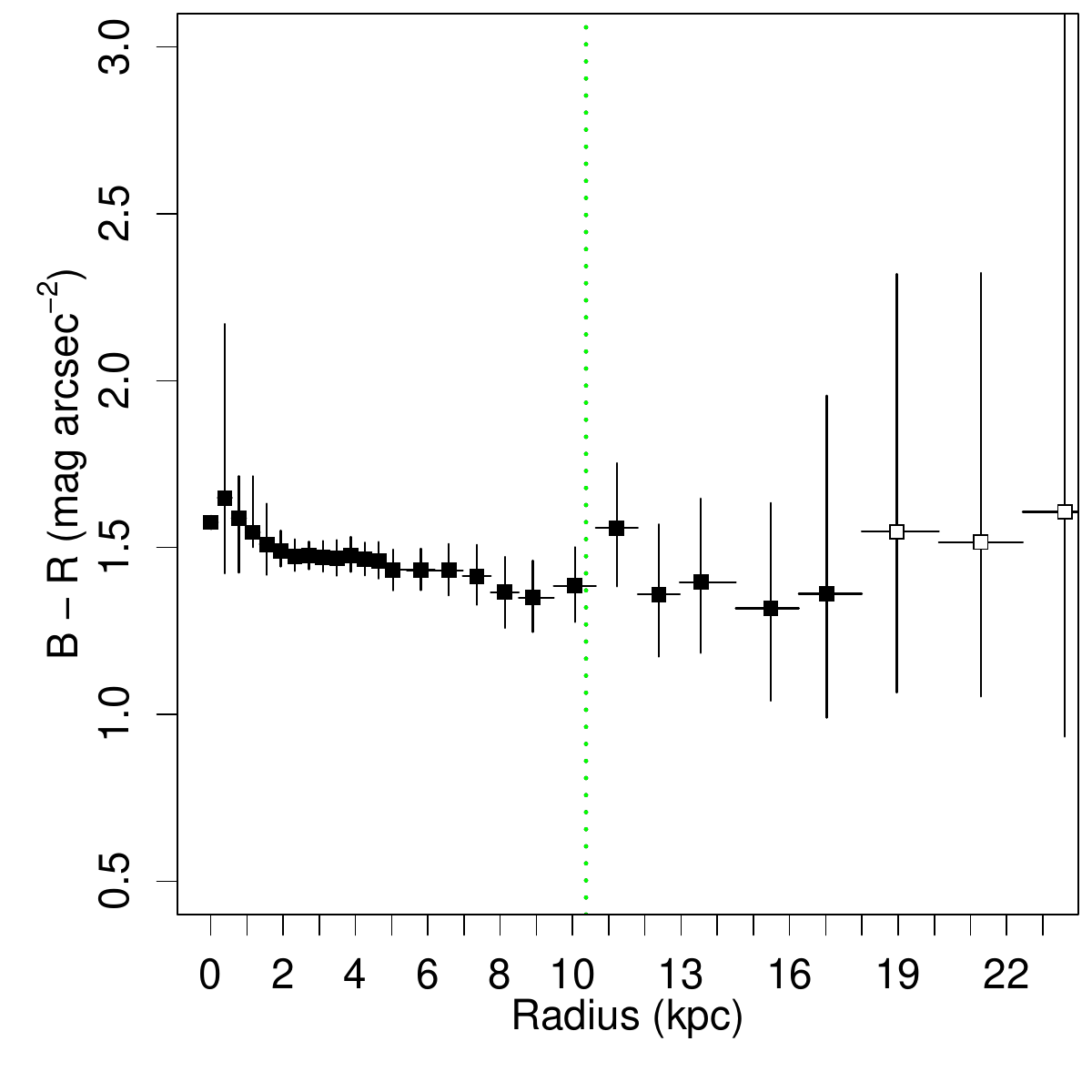}
\end{minipage}%
\vspace{0.5cm}
\caption[]{See caption of Fig.1.}         
\label{fig:img_final}
\end{figure}


\begin{figure}[!h]
\textbf{12) SHARDS20000593:}\\
\begin{minipage}{\textwidth}
\includegraphics[clip,width=0.45\textwidth,page=4]{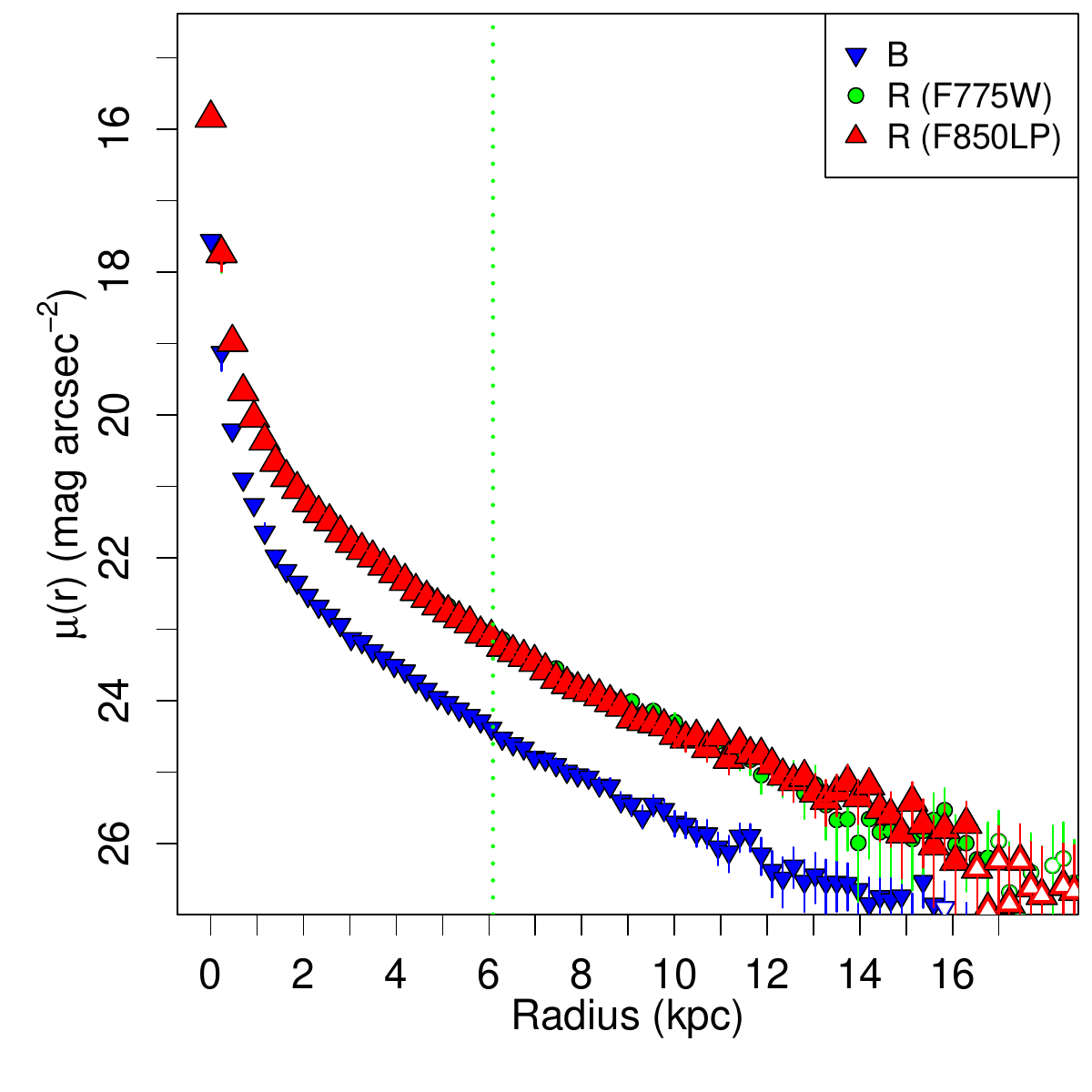}\vspace{-1cm}
\includegraphics[clip,width=0.45\textwidth,page=5]{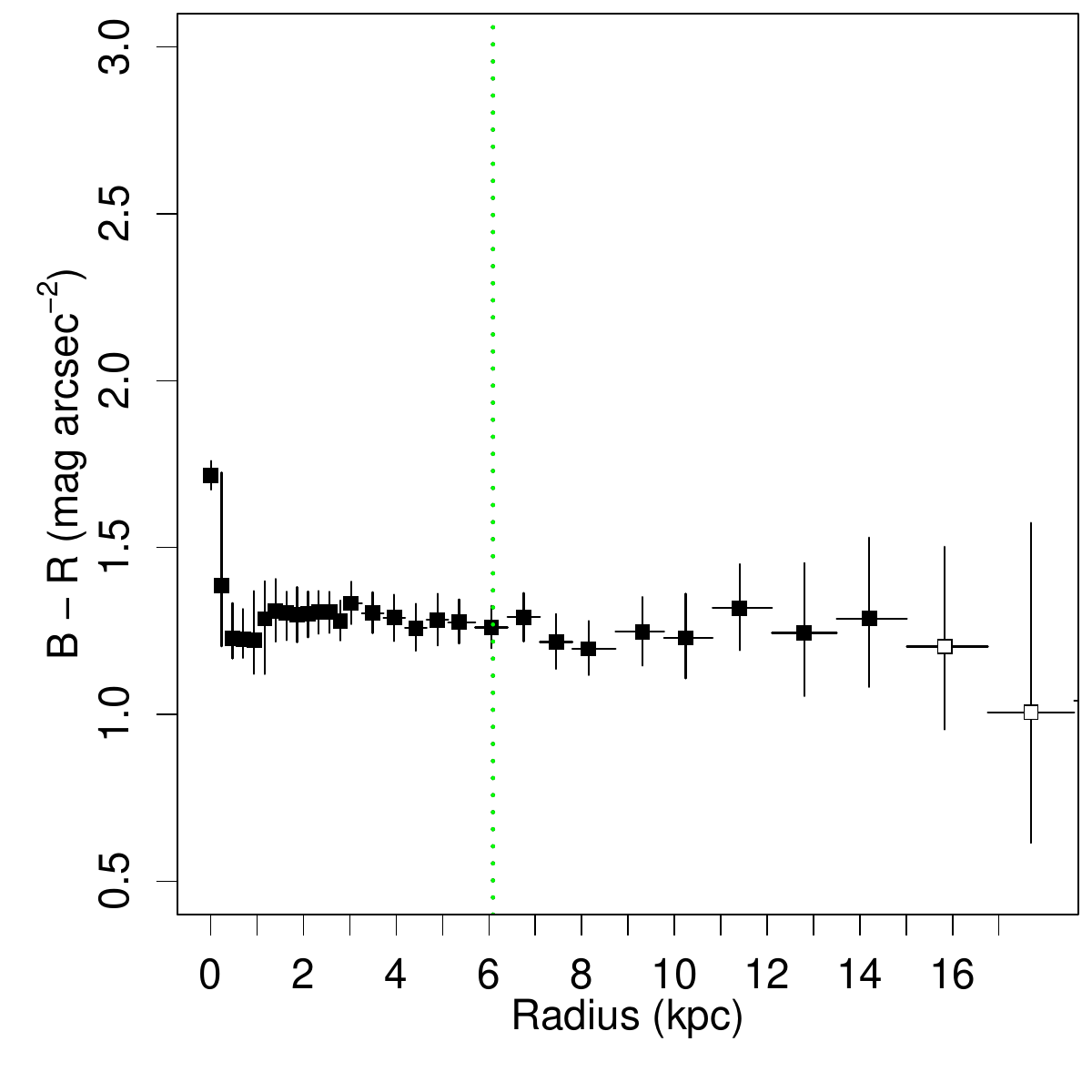}
\end{minipage}%
\vspace{0.5cm}
\caption[]{See caption of Fig.1.}         
\label{fig:img_final}
\end{figure}


\begin{figure}[!h]
\textbf{13) SHARDS20000827:}\\
\begin{minipage}{\textwidth}
\includegraphics[clip,width=0.45\textwidth,page=4]{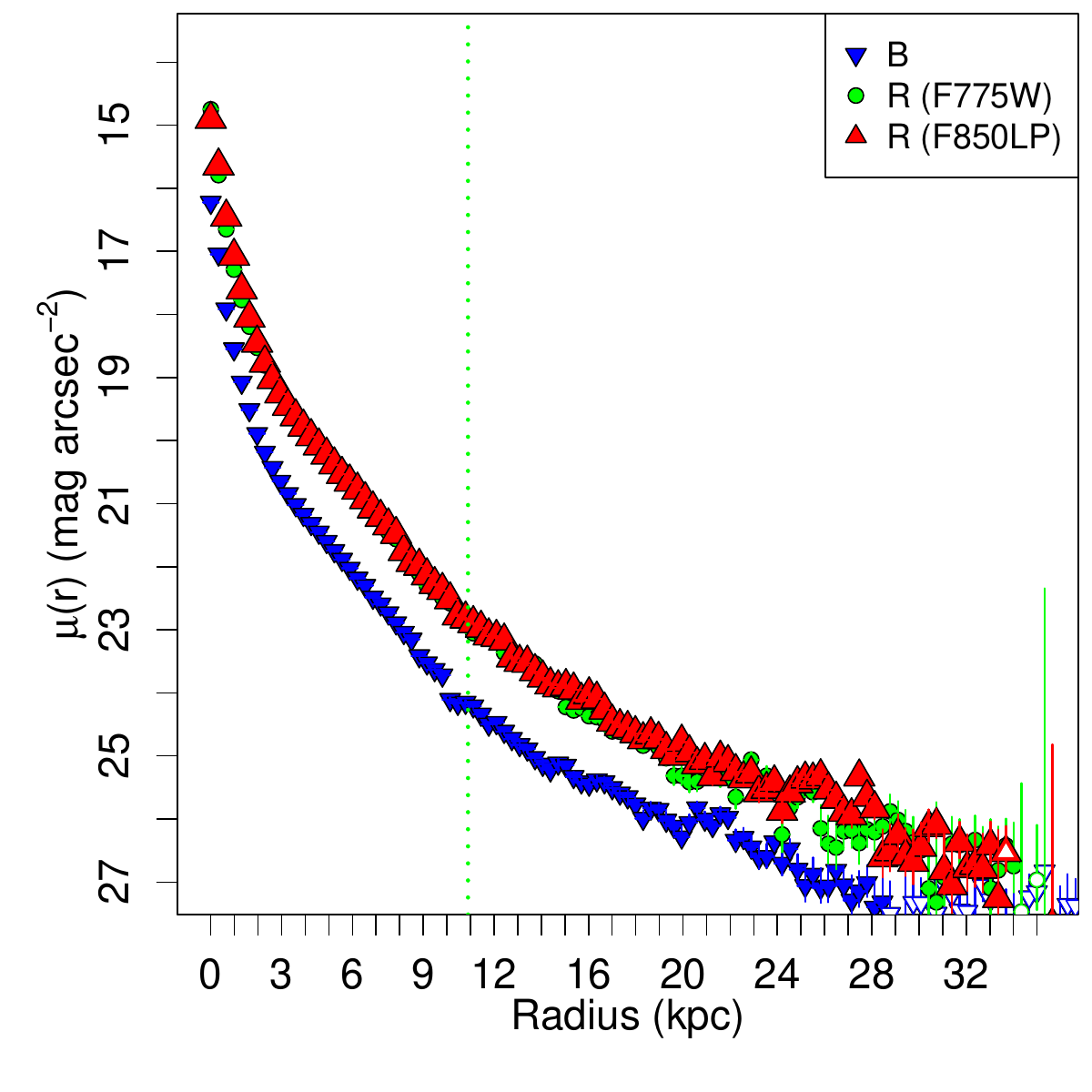}\vspace{-1cm}
\includegraphics[clip,width=0.45\textwidth,page=5]{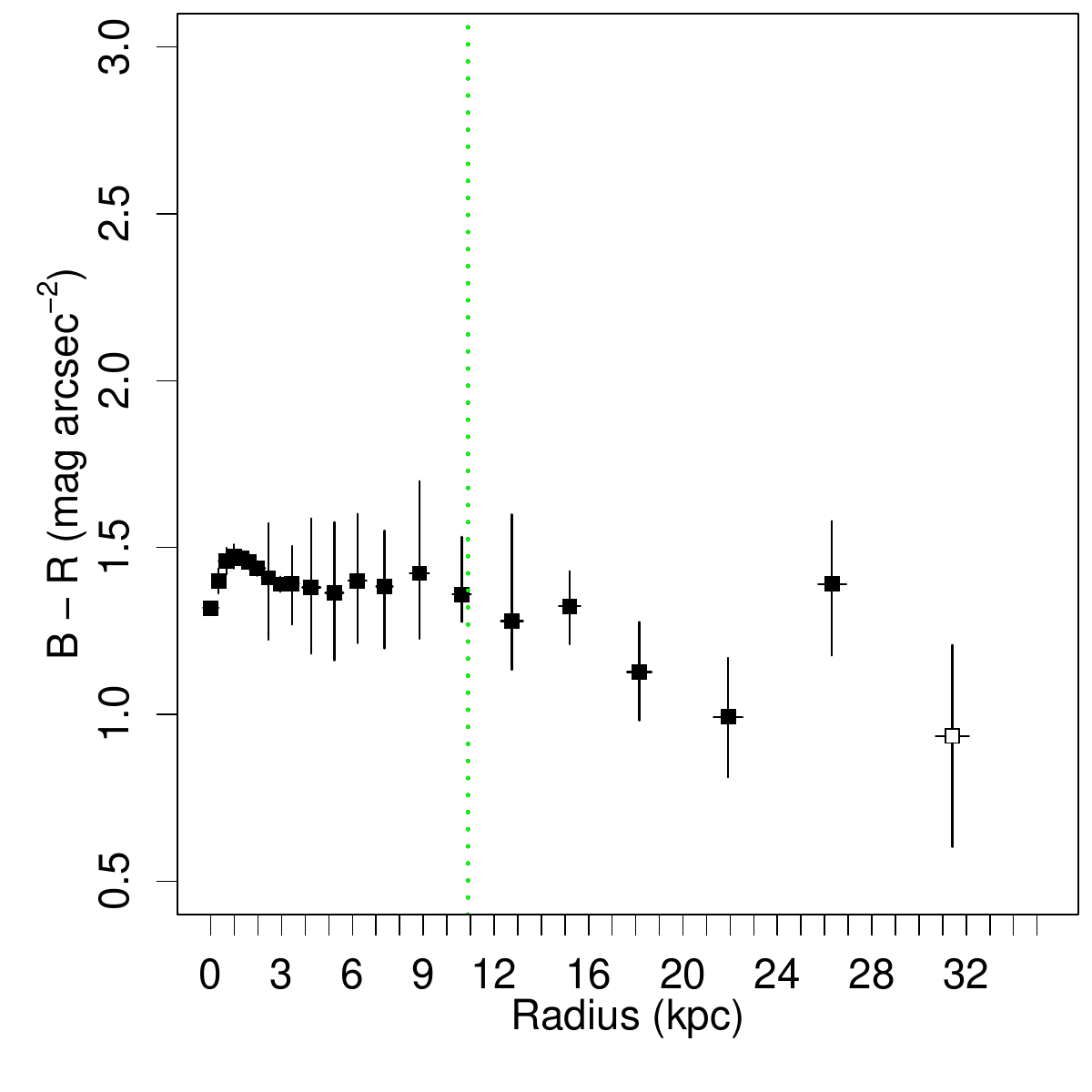}
\end{minipage}%
\vspace{0.5cm}
\caption[]{See caption of Fig.1.}         
\label{fig:img_final}
\end{figure}


\begin{figure}[!h]
\textbf{14) SHARDS20003210:}\\
\begin{minipage}{\textwidth}
\includegraphics[clip,width=0.45\textwidth,page=4]{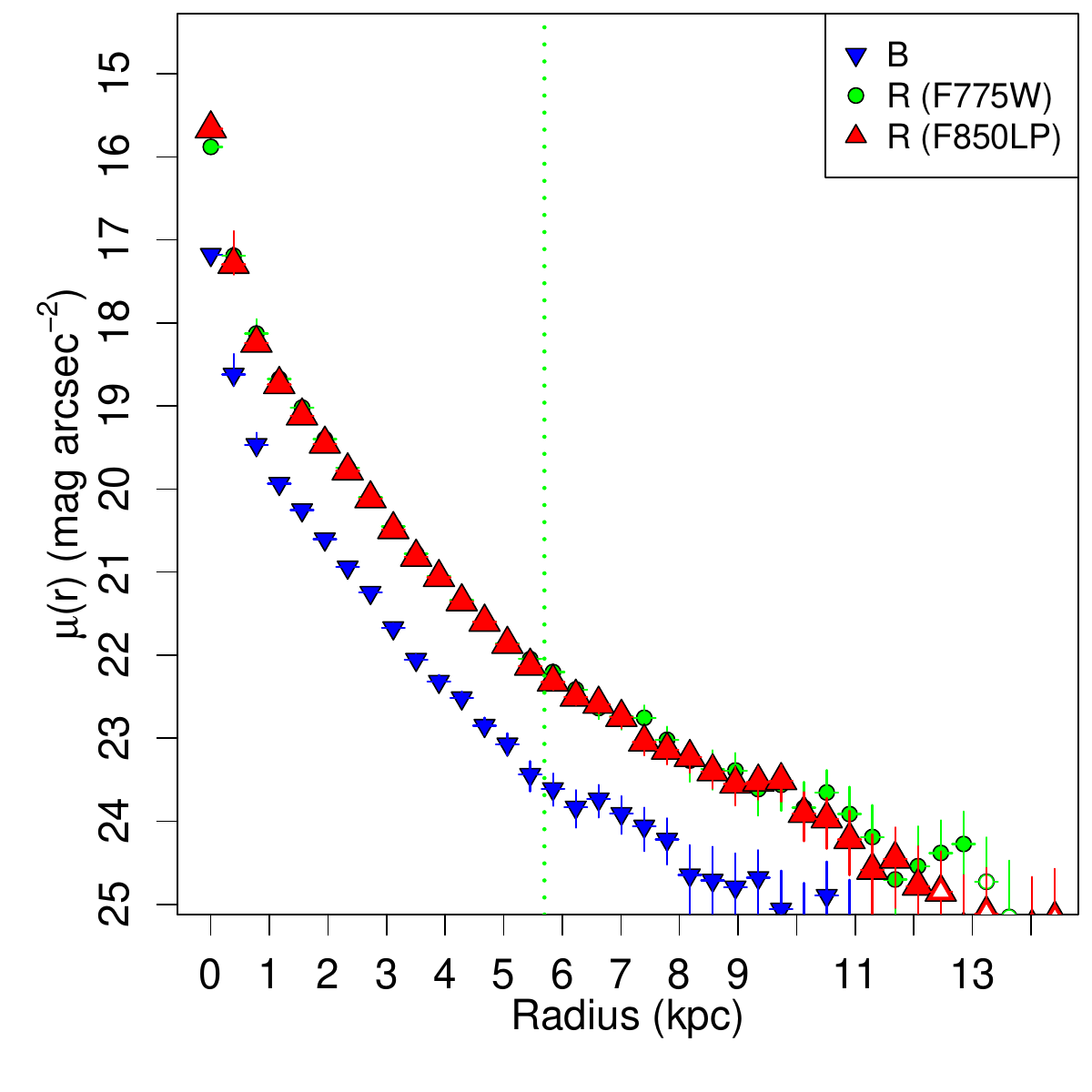}\vspace{-1cm}
\includegraphics[clip,width=0.45\textwidth,page=5]{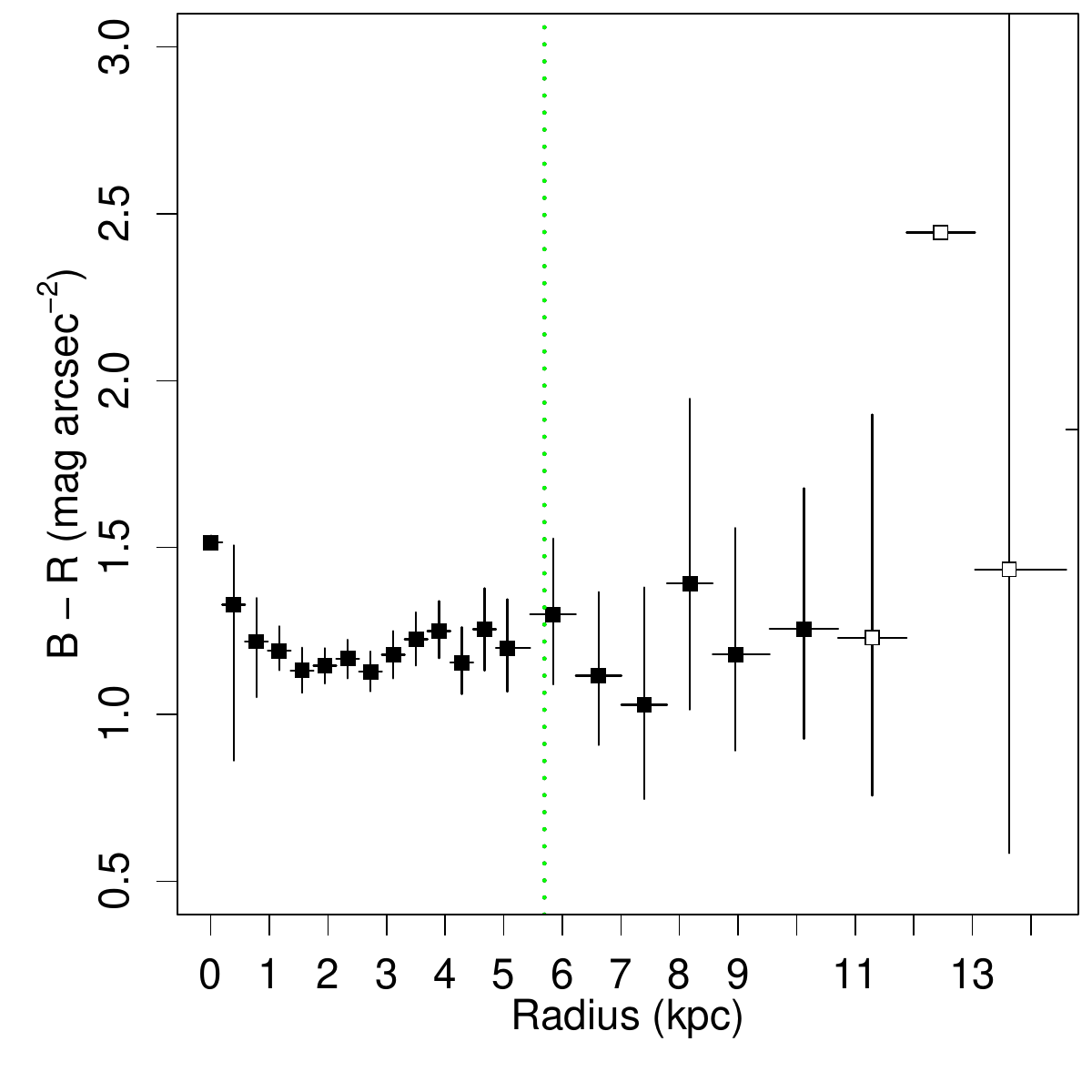}
\end{minipage}%
\vspace{0.5cm}
\caption[]{See caption of Fig.1.}         
\label{fig:img_final}
\end{figure}

\end{appendix}
\clearpage
\newpage
\twocolumn
\small  

\bibliographystyle{aa}
\bibliography{borlaff_14.bib}{}

\end{document}